\documentclass[usenatbib,12pt,preprint]{aastex6}
\usepackage{natbib}

\newcommand{\Msol}{M$_{\odot}$}

\newcommand{\mum}{${\rm \mu m}\ $}
\newcommand{\None}{NGC~2023/2024}
\newcommand{\Ntwo}{NGC~2068/2071}
\newcommand{\Lcrit}{L$_{crit}$}
\defcitealias{Gutermuth09}{G09}
\defcitealias{Kirk11}{KM11}

\begin{document}
\title{The JCMT Gould Belt Survey: Dense Core Clusters in Orion~B}
\author{
H. Kirk\altaffilmark{1}, 
D. Johnstone\altaffilmark{1, 2, 3}, 
J. Di Francesco\altaffilmark{1, 2}, 
J. Lane\altaffilmark{2},
J. Buckle\altaffilmark{4, 5}, 
D.S. Berry\altaffilmark{3}, 
H. Broekhoven-Fiene\altaffilmark{2}, 
M.J. Currie\altaffilmark{3}, 
M. Fich\altaffilmark{6}, 
J. Hatchell\altaffilmark{7}, 
T. Jenness\altaffilmark{3, 8}, 
J.C. Mottram\altaffilmark{9,10}, 
D. Nutter\altaffilmark{11}, 
K. Pattle\altaffilmark{12}, 
J.E. Pineda\altaffilmark{13, 14, 15}, 
C. Quinn\altaffilmark{10}, 
C. Salji\altaffilmark{4, 5}, 
S. Tisi\altaffilmark{6}, 
M.R. Hogerheijde\altaffilmark{9}, 
D. Ward-Thompson\altaffilmark{12}, 
and The JCMT Gould Belt Survey team\footnote{The full members of the JCMT Gould Belt Survey Consortium are 
P. Bastien, D. S. Berry, D. Bresnahan, H. Broekhoven-Fiene, J. Buckle, H. Butner,
M. Chen, A. Chrysostomou, S. Coude, M. J. Currie, C. J. Davis, E. Drabek-Maunder, A. Duarte-Cabral,
J. Di Francesco, M. Fich, J. Fiege, P. Friberg, R. Friesen, G.A. Fuller,
S. Graves, J. Greaves, J. Gregson, J. Hatchell, M.R. Hogerheijde, W. Holland, T. Jenness,
D. Johnstone, G. Joncas, J.M. Kirk ,H. Kirk, L.B.G. Knee, S. Mairs, K. Marsh, B.C. Matthews,
G. Moriarty-Schieven, J.C. Mottram, C. Mowat, K. Pattle, J. Rawlings, J. Richer, D. Robertson, 
E. Rosolowsky, D. Rumble, S. Sadavoy, N. Tothill, H. Thomas, S. Viti, D. Ward-Thompson,
G.J. White, J. Wouterloot, J. Yates, and M. Zhu} 
}

\altaffiltext{1}{NRC Herzberg Astronomy and Astrophysics, 5071 West Saanich Rd, Victoria, BC, V9E 2E7, Canada}
\altaffiltext{2}{Department of Physics and Astronomy, University of Victoria, Victoria, BC, V8P 1A1, Canada}
\altaffiltext{3}{Joint Astronomy Centre, 660 N. A`oh\={o}k\={u} Place, University Park, Hilo, Hawaii 96720, USA}
\altaffiltext{4}{Astrophysics Group, Cavendish Laboratory, J J Thomson Avenue, Cambridge, CB3 0HE, UK}
\altaffiltext{5}{Kavli Institute for Cosmology, Institute of Astronomy, University of Cambridge, Madingley Road, Cambridge, CB3 0HA, UK}
\altaffiltext{6}{Department of Physics and Astronomy, University of Waterloo, Waterloo, Ontario, N2L 3G1, Canada  }
\altaffiltext{7}{Physics and Astronomy, University of Exeter, Stocker Road, Exeter EX4 4QL, UK}
\altaffiltext{8}{LSST Project Office, 933 N. Cherry Ave, Tucson, AZ 85719, USA}
\altaffiltext{9}{Leiden Observatory, Leiden University, PO Box 9513, 2300 RA Leiden, The Netherlands}
\altaffiltext{10}{Max-Planck Institute for Astronomy, K{\"o}nigstuhl 17, 69117 Heidelberg, Germany}
\altaffiltext{11}{School of Physics and Astronomy, Cardiff University, The Parade, Cardiff, CF24 3AA, UK}
\altaffiltext{12}{Jeremiah Horrocks Institute, University of Central Lancashire, Preston, Lancashire, PR1 2HE, UK}
\altaffiltext{13}{European Southern Observatory (ESO), Garching, Germany}
\altaffiltext{14}{Jodrell Bank Centre for Astrophysics, Alan Turing Building, School of Physics and Astronomy, University of Manchester, Oxford Road, Manchester, M13 9PL, UK}
\altaffiltext{15}{Current address: Max Planck Institute for Extraterrestrial Physics, Giessenbachstrasse 1, 85748 Garching, Germany}

\slugcomment{\today}

\begin{abstract}
The JCMT Gould Belt Legacy Survey obtained SCUBA-2 observations of dense cores within
three sub-regions of Orion~B: LDN~1622, \None, and \Ntwo, all of which contain
clusters of cores.  
We present an analysis of the clustering properties of these cores, including the
two-point correlation function and Cartwright's Q parameter.
We identify individual clusters of dense cores across all three regions using a minimal 
spanning tree
technique, and find that in each cluster, the most massive cores tend to be centrally located.
We also apply the independent $M$--$\Sigma$ technique and find a strong correlation between
core mass and the local surface density of cores.  These two lines of evidence jointly suggest
that some amount of mass segregation in clusters has happened already at the dense core
stage.  
\end{abstract}

\section{INTRODUCTION}
Most stars begin their lives within a clustered environment
\citep[e.g.,][]{Lada03,Porras03}, highlighting the importance of understanding this mode
of star formation.
Over the past decade, significant effort has been made to characterize the clustering
properties of the youngest stellar clusters to aid in constraining theories of
cluster formation and evolution.  Studies of nearby young stellar clusters
span a range of properties from very small and sparse systems \citep[e.g.,][]{Kirk11}
to denser and more populous systems \citep[e.g.,][]{Gutermuth09,Feigelson13,Kuhn14}.
Combining cluster catalogues with detailed modelling can inform
a variety of topics including the timescale of the initial cluster formation
\citep[e.g.,][]{Tan06,Parmentier14}, 
the presence of multiple generations of cluster formation and
their geometries \citep[e.g.,][]{Ellerbroek13},
the role of early sub-cluster merging in the appearance of
present-day clusters \citep[e.g.,][]{Moeckel09}, and the effects of gas explusion and
feedback on cluster formation and early evolution \citep[e.g.,][]{Pfalzner13,Krumholz14}.

A complementary approach to measuring the properties of young stellar systems 
is characterizing the properties of the dense gas and
dust prior to and during the initial stages of cluster formation.
A variety of studies also exist in this regime as well, 
characterizing the properties and stability of dense cores, the kinematic properties of 
core and cluster gas, the influence of outflows, and heating of cluster cores
\citep[see, for example][]{Tafalla06,Kirk07,Wang08,Friesen09,Foster09,Maruta10,Csengeri11,Palau13,Pattle15}.
In terms of characterizing the dust continuum properties of dense cores in clusters (such as
distributions of their masses and positions), the James Clerk Maxwell Telescope (JCMT) Gould 
Belt Survey (GBS) can make a strong contribution.  The JCMT GBS
performed a uniform large-scale mapping of thermal dust emission at 850~\mum\ and 450~\mum\ 
of nearby molecular clouds (the Gould Belt) visible from the northern hemisphere
\citep{WardThomp07}.
These maps include catalogues of dense cores to be identified around many
nearby cluster-forming regions including Orion~A (Salji et al. 2015 and Mairs et al., in prep),
Orion~B \citep{Kirk16}, and Ophiuchus \citep{Pattle15}.  This extensive set of
dense core catalogues allows for a uniform analysis of the clustering properties of dense
cores, which can then provide other constraints on the initial conditions of clusters
and the physical processes shaping their evolution.  Our focus in the present analysis
is on the clustered population within Orion~B, a region known to harbour several active
cluster-forming regions \citep[e.g.,][]{Meyer08}.

One particularly controversial aspect of cluster formation is
mass segregation, namely is it present, and if so, is it primordial?
Difficulties arise in the measurement of mass segregation from a host of challenges
including the definition of a cluster, the definition of mass segregation, and
the effects of observational biases.  The determination of 
primordiality is also fraught with additional complications which again include how
clusters are defined, cluster age determination, and whether or not the present day 
cluster is the product of an earlier merger of smaller systems that has since
dynamically relaxed.  
While ongoing innovative work can
help to overcome these challenges, it is also possible to address the question from
the other side of star formation: if some degree of mass segregation is apparent in dense 
cores before the onset of star formation, then this lends support to the idea of
primordial mass segregation in young stellar clusters.

In this paper, we use two independent techniques to look for the presence of mass
segregation within clusters of dense cores in the Orion~B molecular cloud.  We follow
the initial characterization of dense cores in this region by \citet{Kirk16},
which is summarized in Section~2.  In Section 3, we introduce the Minimal Spanning Tree,
and with it we find a general tendency
for the most massive dense core(s) of a cluster to lie near the cluster centre.  In Section~4, we
introduce the $M$--$\Sigma$ technique, and with it we find more massive cores have a tendency to lie
in zones of higher core-core surface density.  In Section~5, we present additional measures
of the clustering properties of the dense cores, including the two-point correlation
function.  In Section~6, we discuss our results and their implications for the broader
picture of cluster formation.

\section{OBSERVATIONS}
\label{sec_obs}

Three separate regions within Orion~B were observed with SCUBA-2 \citep{Holland13} 
at 850~\mum and 450~\mum as part of the JCMT Gould Belt Survey \citep{WardThomp07}:
LDN~1622, \None, and \Ntwo.  NGC~2024, NGC~2068, and NGC~2071 are the most vigorous
sites of ongoing star formation in Orion~B, representing 60\% to 90\% of its current YSOs
\citep[e.g.,][]{Lada91,Meyer08}.  The SCUBA-2 observations span a generous
area around these three clustered star-forming regions, as well as several less 
prominent ones, with total areal coverages of 0.6~deg$^2$, 2.1~deg$^2$, and 1.7~deg$^2$ 
around L1622, \None, and \Ntwo, respectively.
Analysis of the SCUBA-2 maps was first presented in Kirk et al. (2016, hereafter K16),
using the GBS Legacy Release 1 reduction methodology \citep[see, e.g.,][]{Mairs15}.  
In brief, individual observations
were reduced using the JCMT's standard 
Starlink software \citep[see][for a description of \it{makemap}]{Chapin13}, 
and then mosaicked together.  Smaller-scale sources (i.e., those of size below roughly 2.5\arcmin)
are generally well-recovered in the final map, while larger-scale sources can be subject
to some filtering (see K16 for more details).  All of the Orion~B maps and associated
data products are available at https://doi.org/10.11570/16.0003

K16 identified dense cores in the 850~\micron\ map -- using the {\it FellWalker} 
algorithm \citep{Berry15}, finding a total of 29 dense cores in L1622, 546 in \None, 
and 322 in \Ntwo.
{\it FellWalker} effectively identifies cores based on local peaks of emission,
defining their extents based on the locations at which local gradients in flux are
directed toward the peak.  As such, dense core boundaries are irregular, and each pixel
in the map is assigned to a maximum of one core.  
Some of the dense cores identified are elongated -- 12\% have ratios of 2 or higher
in comparing their vertical and horizontal extents -- but there is no indication in K16 that
their other basic properties differ from the rest of the cores.
In our present analysis, we treat all cores identically, regardless of
their elongation.
The robustness of individual
cores is ensured through {\it FellWalker} criteria specifying a minimum spatial
and flux separation between closely-spaced peaks, a minimum core size larger than the
telescope beam, and a minimum gradient and flux level for all pixels associated with 
the peak.
Of the dense cores identified in the three regions, 5, 25, and 34, respectively, 
were classified as protostellar, based on
a comparison with the {\it Spitzer}-based YSO catalogue of \citet{Megeath12} and the
{\it Herschel}-based YSO catalogue of \citet{Stutz13}, noting that the latter catalogue 
covers a deep but limited area within Orion~B.
The minimum flux for a dense core to be identified was 7.4~mJy, which corresponds to
a mass of $\sim 0.01$~\Msol, assuming a distance of 415~pc, a temperature of 20~K,
and a dust opacity of 0.0125~g~cm$^{-2}$ at 850~\micron\ (see K16 for details).
For a star-forming region at a single distance with an invariant population
of dust grains, the observed 850~\mum flux is proportional to both the total 
amount of material present in the beam and its temperature.
The assumption of a constant 20~K temperature is reasonable
for the starless cores, but is likely to cause the mass to be overestimated for
the protostellar cores by a factor of a few.  For this reason,
we separate the two populations of cores for some of our analysis here.

Comparing the mass within lower column density material, as measured by 
{\it Herschel} and {\it Planck} by \citet{Lombardi14}, K16 find that the 
\None\ and \Ntwo\ regions are likely in the process of forming several
hundred protostars each, while L1622 harbours a more modest number of
protostars and lacks the necessary dense material to be able to form 
significantly more.

In relating the results of our clustering analysis below to the initial
conditions for protostellar clusters, we make several reasonable assumptions,
which are discussed in more detail in Section~6.2 and Appendix~B.2.  While it
is not necessary to assume that each core forms a single star, we do assume that similar
cores will form similar groups of stars, regardless of their locations in the 
larger environment.  Similarly, while we do not assume a one-to-one relationship between
the masses of a dense core and a protostar, we do assume that the most massive cores are
most likely to form the most massive protostars.  It is possible that some of
the dense cores in the K16 catalogue will eventually disperse without forming any
stars.  These cores, however, are likely to be the least massive ones, which we find are 
the least likely to be associated with clustering.  We note that K16 estimated
that the majority of starless dense cores in Orion~B are presently bound, 
due to a combination of self-gravity and
pressure from the external weight of the ambient cloud material.

\section{Minimal Spanning Tree Analysis}
\label{sec_clusters_mst}
A visual examination of the three regions mapped in Orion B reveals clusters of dense cores,
especially in the larger and more active \None\ and \Ntwo\ regions.  We use a minimal spanning tree
(MST) to identify these clusters to analyze their properties.  MSTs have been used
previously to identify clusters of stars \citep[e.g.,][]{Gutermuth09,Kirk11,Billot11,Masiunas12}, 
and as a tool
to compare simulations with observations \citep[e.g.,][]{Kirk14}.  An MST basically 
defines a structure in which all sources are connected by their minimum possible separations,
i.e., branches.  Within an MST structure,
clusters are apparent as sets of points connected by short branches.  We follow the method
of \citet[][hereafter G09]{Gutermuth09} to estimate \Lcrit, the critical branch length,
used to distinguish clustered from isolated sources.  G09 found that young stellar clusters
tend to have a characteristic distribution of branch lengths when plotted as cumulative
number versus length: i.e., a steep, nearly linear rise at short branch lengths, followed by
a turn-over to a shallow, nearly-linear slope at the longest branch lengths.  G09 defined
\Lcrit\ as the intersection point between linear fits to the two ends of the distribution
(see Figure~1 of G09).  Sources which remain connected after all branches with lengths
above \Lcrit\ are removed are considered clusters.  In this manner, clusters are easily
identified as local over-densities, rather than relying on a fixed surface density
threshold.  \citet[][hereafter KM11]{Kirk11} followed 
the same procedure as \citetalias{Gutermuth09}, setting the minimum number of YSOs 
for a cluster at 11 since fewer sources 
make the determination of cluster properties such as the centre position difficult.
Here, as in \citetalias{Kirk11}, the cluster centre is defined as the median position
of cluster members.

\begin{figure}[htb]
\plottwo{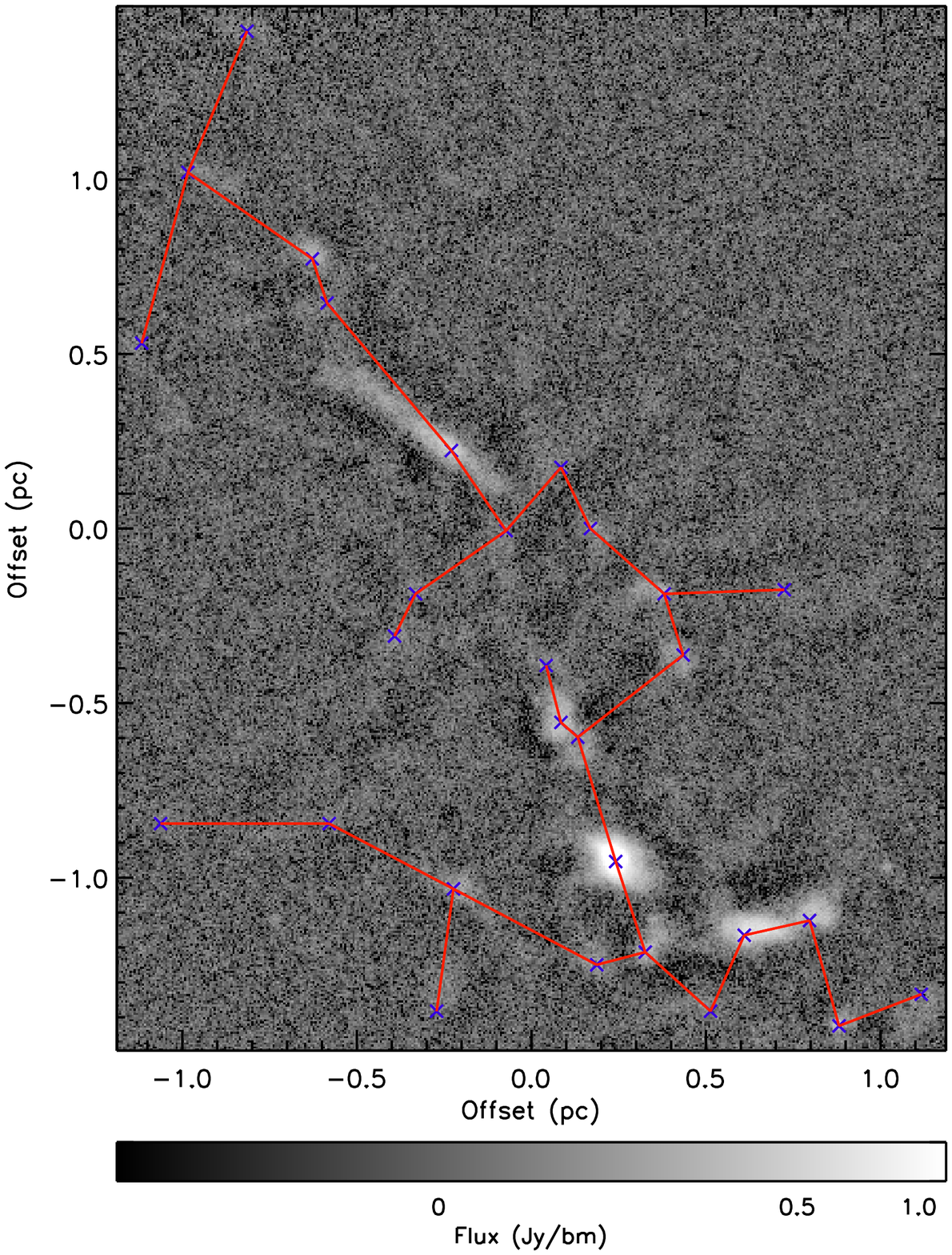}{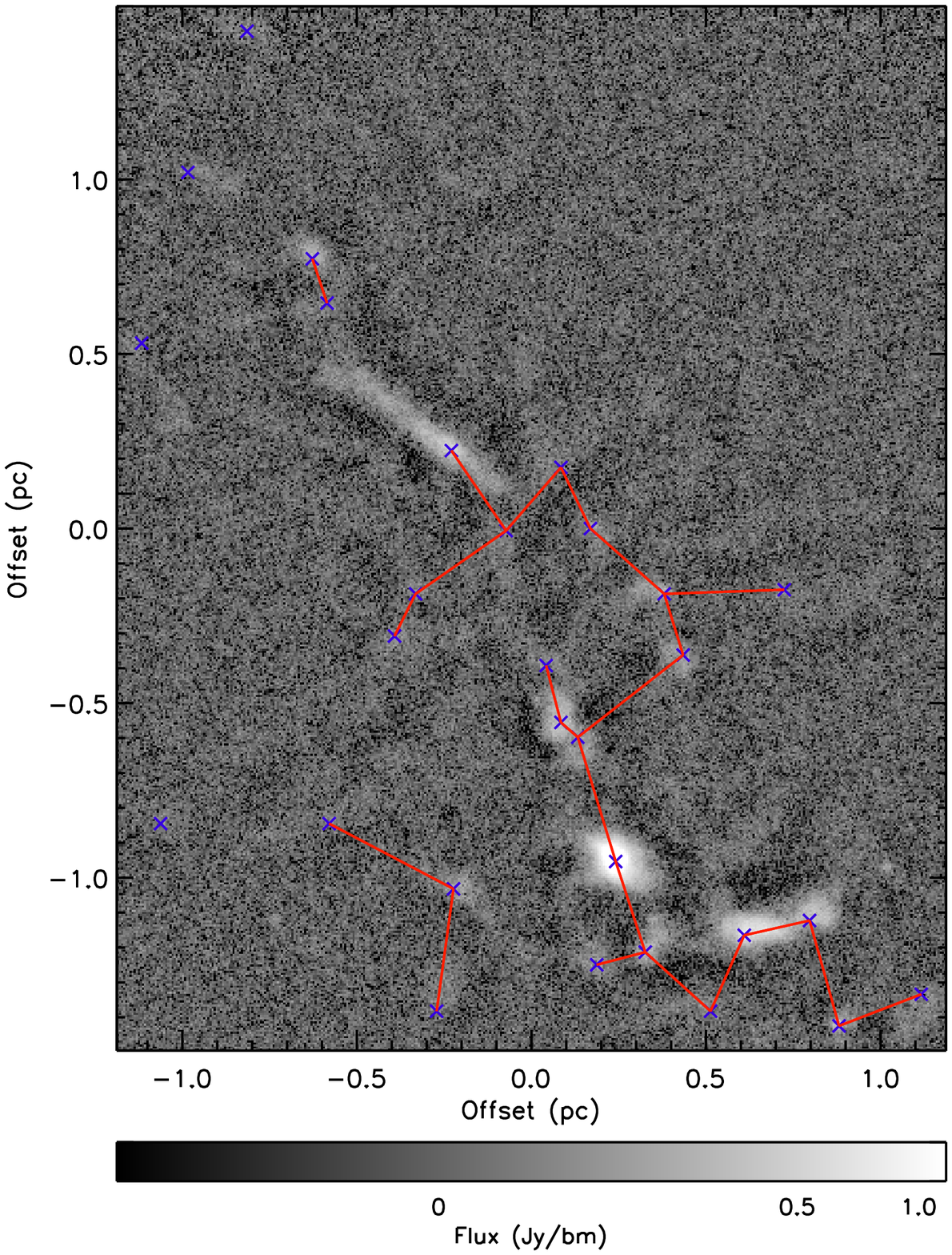}
\caption{The minimal spanning tree structure for L1622.  In both panels, the background
        greyscale image shows the SCUBA-2 850~\mum emission (truncated to the extent over which
        dense cores were identified), while the blue crosses show the dense cores.
        In the left panel, the red lines show the entire original MST structure.
        In the right panel, branches longer than \Lcrit\ (0.42~pc) have been removed.}
\label{fig_full_mst_L1622}
\end{figure}

\begin{figure}[htb]
\plottwo{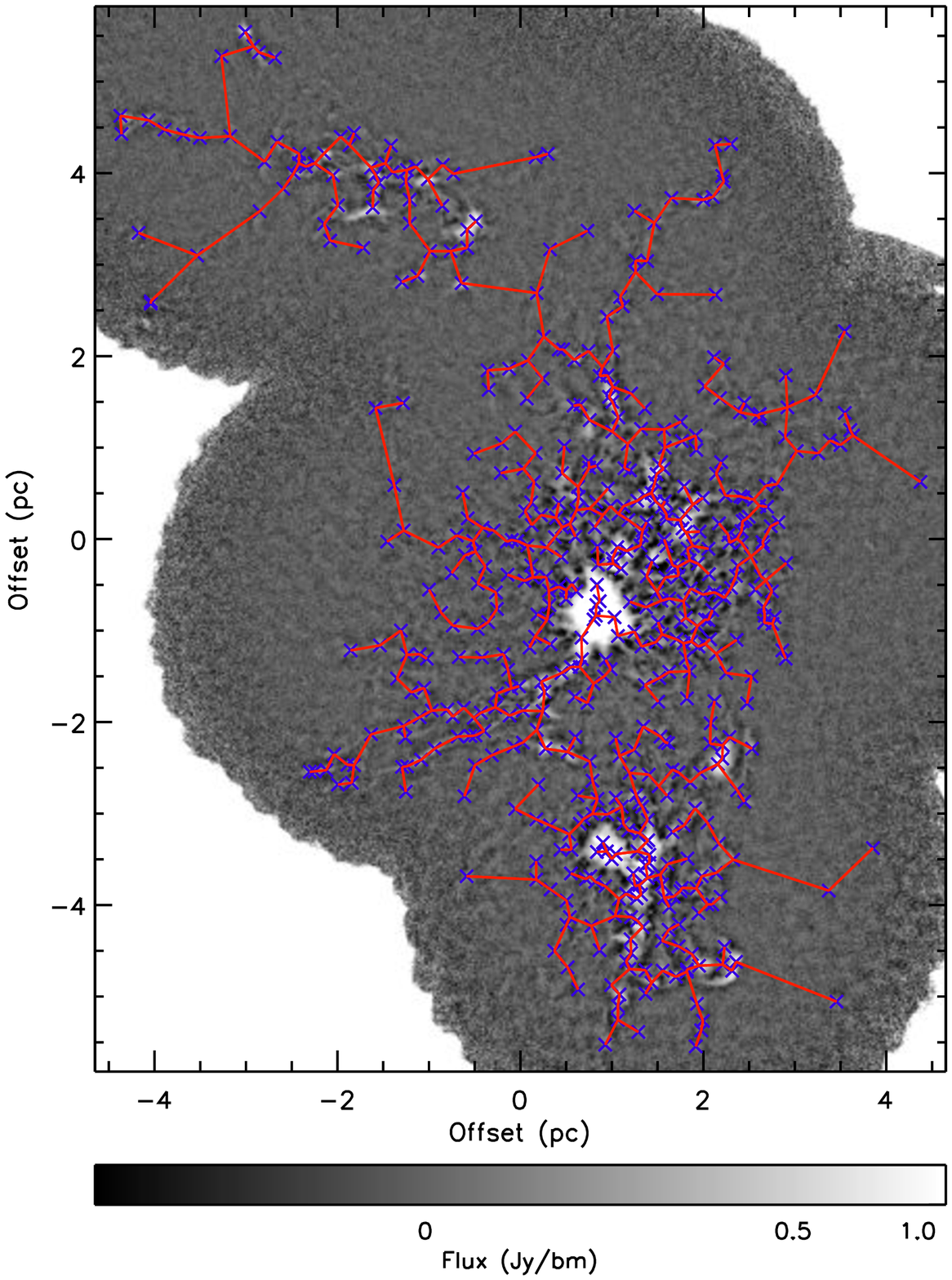}{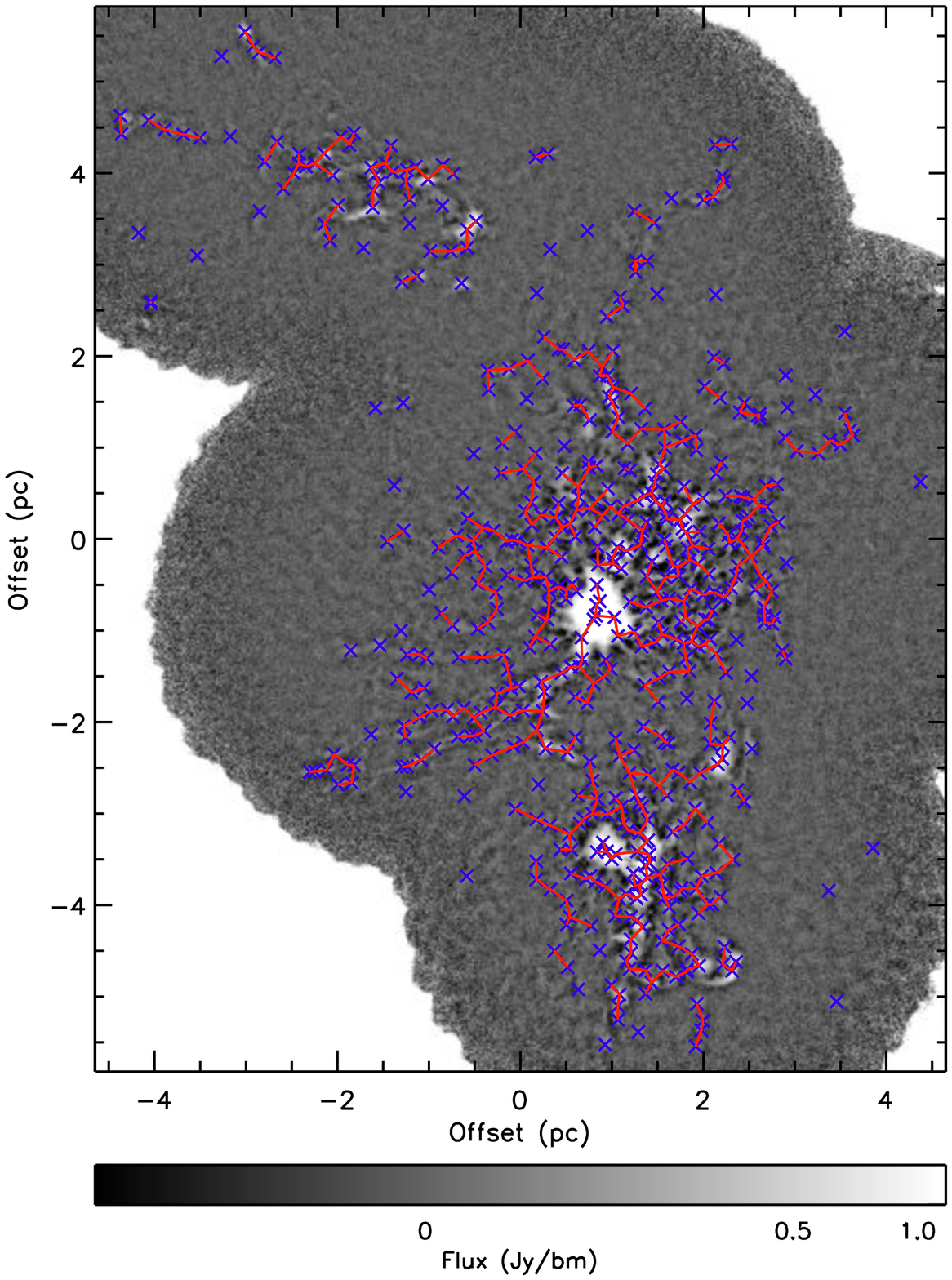}
\caption{The minimal spanning tree structure for \None.  In both panels, the background
        greyscale image shows the SCUBA-2 850~\mum emission (truncated to the extent over which
        dense cores were identified), while the blue crosses show the dense cores.
        In the left panel, the red lines show the entire original MST structure.
        In the right panel, branches longer than \Lcrit\ (0.26~pc) have been removed.}
\label{fig_full_mst_N2023}
\end{figure}

\begin{figure}[htb]
\plottwo{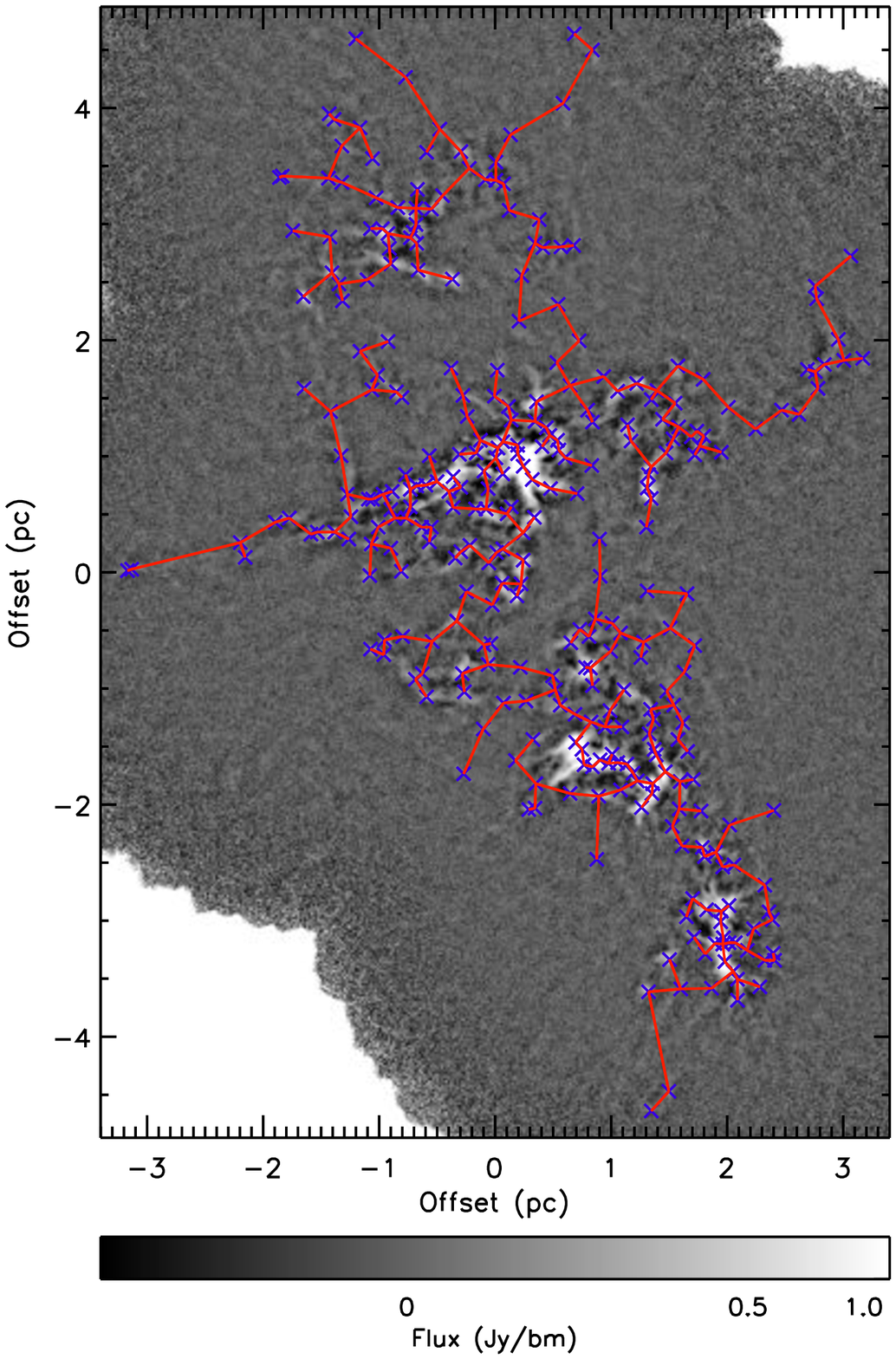}{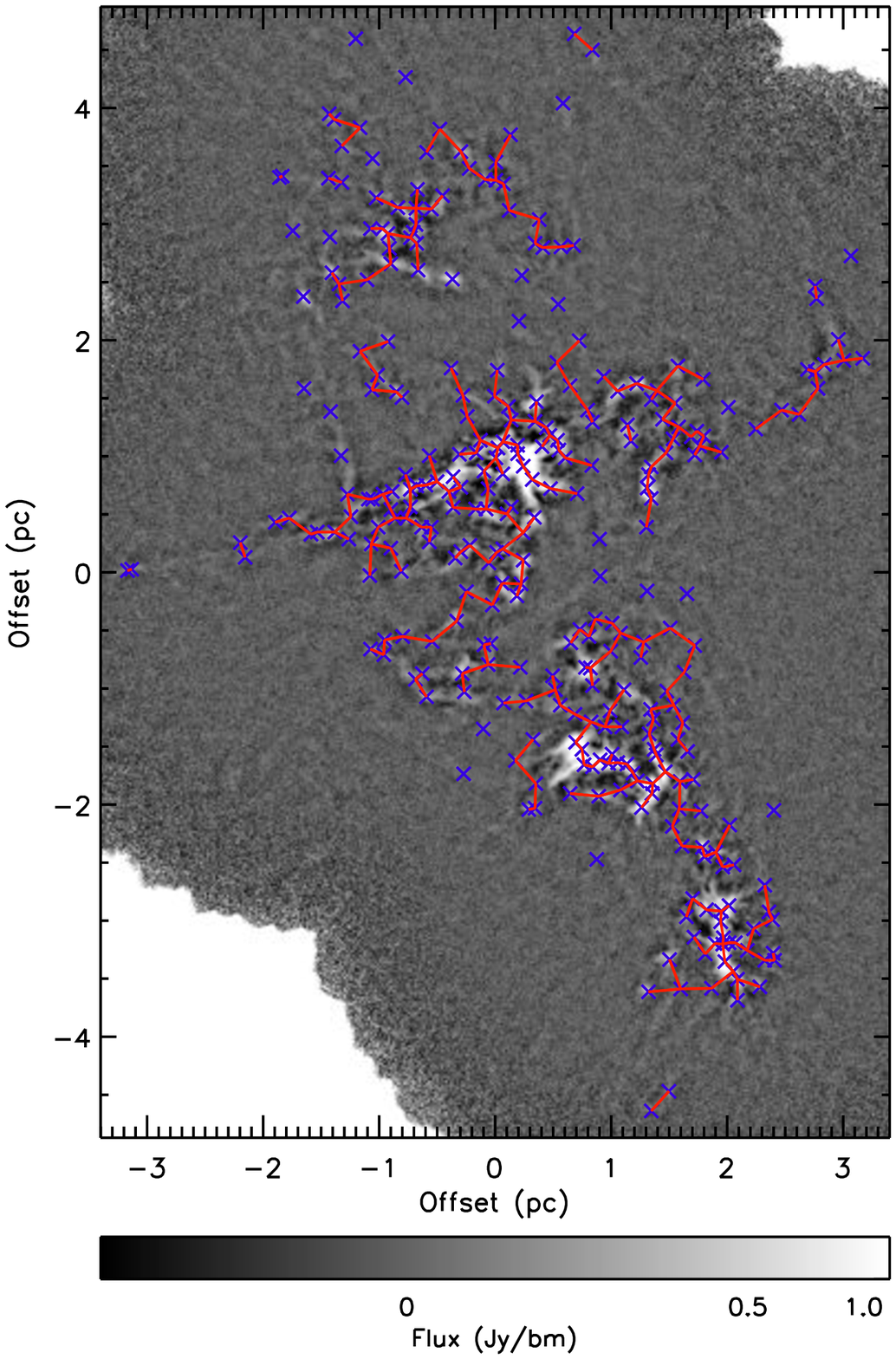}
\caption{The minimal spanning tree structure for \Ntwo.  In both panels, the background
	greyscale image shows the SCUBA-2 850~\mum emission (truncated to the extent over which 
	dense cores were identified), while the blue crosses show the dense cores.
	In the left panel, the red lines show the entire original MST structure.  
	In the right panel, branches longer than \Lcrit\ (0.28~pc) have been removed.}
\label{fig_full_mst_N2068}
\end{figure}

We find that the cumulative branch length distribution of cores in each Orion~B region
follows a similar distribution, and so we apply the same cluster-identification method as in
\citetalias{Gutermuth09} and \citetalias{Kirk11}, including the minimum number of cluster
members (eleven) and centre position used in the latter analysis.  
Figures~\ref{fig_full_mst_L1622} to \ref{fig_full_mst_N2068} show the 
full and chopped MST structures in L1622, \None, and \Ntwo\ using the procedure described
in \citetalias{Kirk11}.  
The main clusters selected by this method 
correspond very well with how the eye would subdivide the region into clusters.
In \None\ and \Ntwo, multiple clusters are identified, while in L1622 only a single cluster 
is identified.  In total, 69\% (20/29), 71\% (389/546), and 75\% (243/322) cores
were associated with a cluster in each of L1622, \None, and \Ntwo. 
We measure best-fit \Lcrit\ values of 0.42~pc (L1622), 0.26~pc (\None), and 
0.28~pc (\Ntwo).  Table~\ref{tab_clusts} summarizes the properties of the clusters identified in
all three regions, and each cluster is also shown in the Appendix in 
Figures~\ref{fig_mst_clusts1} and \ref{fig_mst_clusts2}.  A full examination of the effects
of the uncertainty in \Lcrit\ on the cluster definitions and subsequent results is also
given in the Appendix.

\begin{deluxetable}{cllllllllll}
\tablecolumns{11}
\tablewidth{0pc}
\tabletypesize{\scriptsize}
\tablecaption{Properties of MST-identified clusters\label{tab_clusts}}
\tablehead{
\colhead{Region} &
\colhead{Index} &
\colhead{N\tablenotemark{a}} &
\colhead{$S_{max}$\tablenotemark{b}} &
\colhead{$S_{max,sl}$\tablenotemark{b}} &
\colhead{$S_{med}$\tablenotemark{b}} &
\colhead{$S_{tot}$\tablenotemark{b}} &
\colhead{$O_{max}$\tablenotemark{c}} &
\colhead{$O_{max,sl}$\tablenotemark{c}} &
\colhead{$O_{med}$\tablenotemark{c}} &
\colhead{$f_{proto}$\tablenotemark{d}} \\
\colhead{ } &
\colhead{ } &
\colhead{ } &
\colhead{(Jy)} & 
\colhead{(Jy)} & 
\colhead{(Jy)} &
\colhead{(Jy)} &
\colhead{(pc)} &
\colhead{(pc)} &
\colhead{(pc)} &
\colhead{(\%)} 
}
\startdata
       LDN~1622 &  1 &   20 &   4.89 &  1.71 &  0.17 &  12.23 &  0.48 &  0.83 &   0.63 &   25.0 \\
  NGC~2023/2024 &  1 &  259 & 119.52 &  5.34 &  0.17 & 420.96 &  0.58 &  0.86 &   1.22 &   4.2 \\
  NGC~2023/2024 &  2 &  104 &  11.12 & 11.12 &  0.16 & 103.10 &  0.42 &  0.42 &   0.80 &   8.7 \\
  NGC~2023/2024 &  3 &   15 &   3.00 &  2.71 &  0.19 &   9.36 &  0.39 &  0.43 &   0.27 &   20.0 \\
  NGC~2023/2024 &  4 &   11 &   0.87 &  0.87 &  0.08 &   2.34 &  0.15 &  0.15 &   0.24 &   0.0 \\
  NGC~2068/2071 &  1 &   89 &  46.39 &  2.36 &  0.35 & 109.96 &  0.59 &  0.62 &   0.71 &   12.4 \\
  NGC~2068/2071 &  2 &   30 &   7.71 &  1.94 &  0.29 &  31.93 &  0.15 &  0.06 &   0.34 &   26.7 \\
  NGC~2068/2071 &  3 &   66 &   6.70 &  6.22 &  0.19 &  67.22 &  0.42 &  0.47 &   0.58 &   15.2 \\
  NGC~2068/2071 &  4 &   23 &   2.73 &  2.73 &  0.17 &  12.21 &  0.48 &  0.48 &   0.34 &   4.3 \\
  NGC~2068/2071 &  5 &   20 &   1.24 &  1.24 &  0.26 &   8.82 &  0.27 &  0.27 &   0.28 &   15.0 \\
  NGC~2068/2071 &  6 &   15 &   0.64 &  0.64 &  0.10 &   1.97 &  0.44 &  0.44 &   0.44 &   0.0 \\
\enddata

\tablenotetext{a}{Total number of cluster members.}
\tablenotetext{b}{Maximum core flux, maximum starless core flux, median core flux
	of individual cluster members, and the total flux of all cores in the cluster.  
	Under the assumption of a constant dust temperature and opacity, the ratio 
	of maximum to median flux is equal to the ratio in the mass of the most massive core 
	to the median mass.}
\tablenotetext{c}{Offset from the cluster's centre of the highest flux core and the highest
	flux starless core, and the median offset of all cores.}
\tablenotetext{d}{The fraction of cluster members which are associated with a {\it Spitzer}
	or {\it Herschel} YSO.}

\end{deluxetable}


\subsection{Offset Ratios}

\begin{figure}[htb]
\plottwo{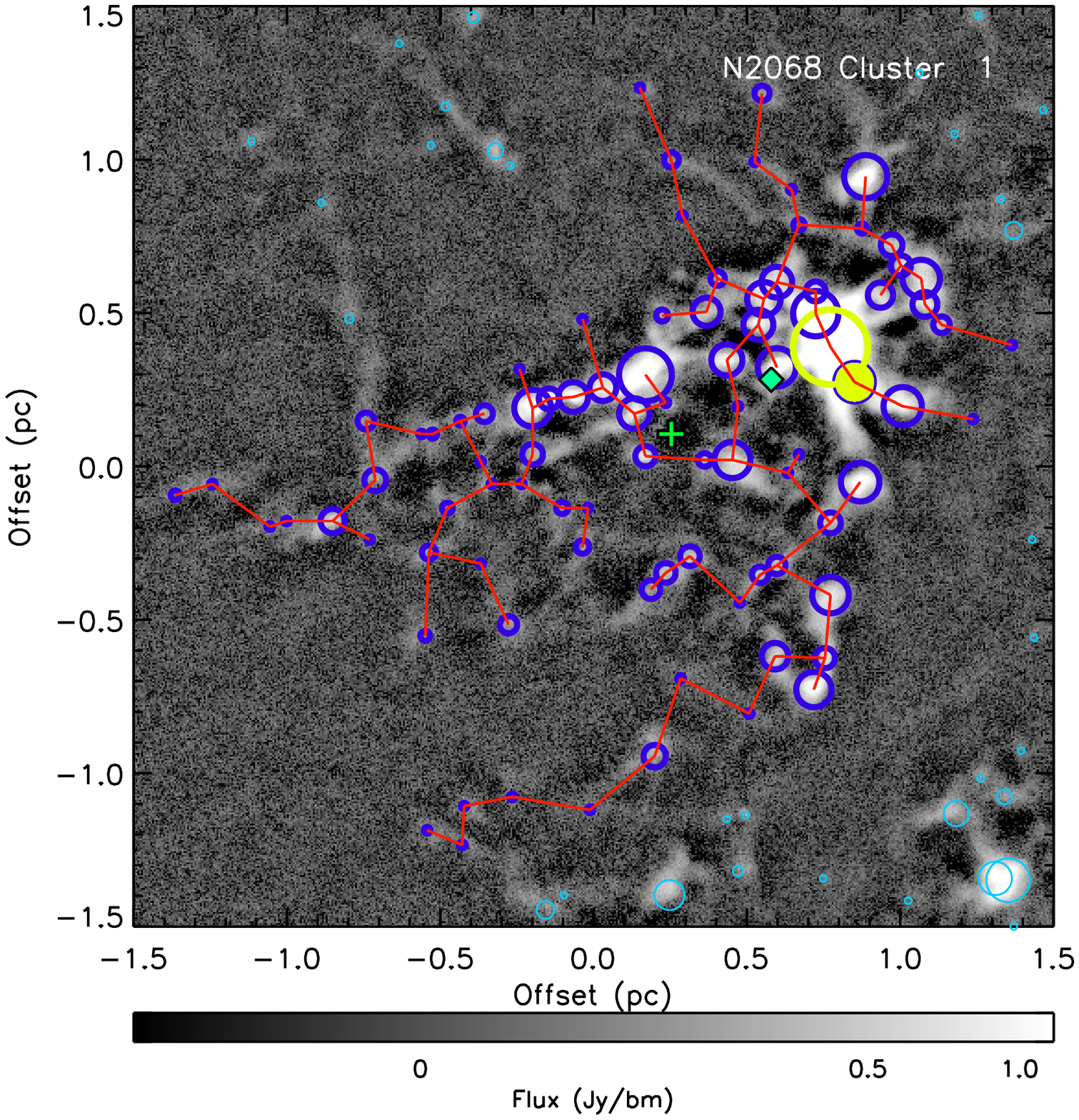}{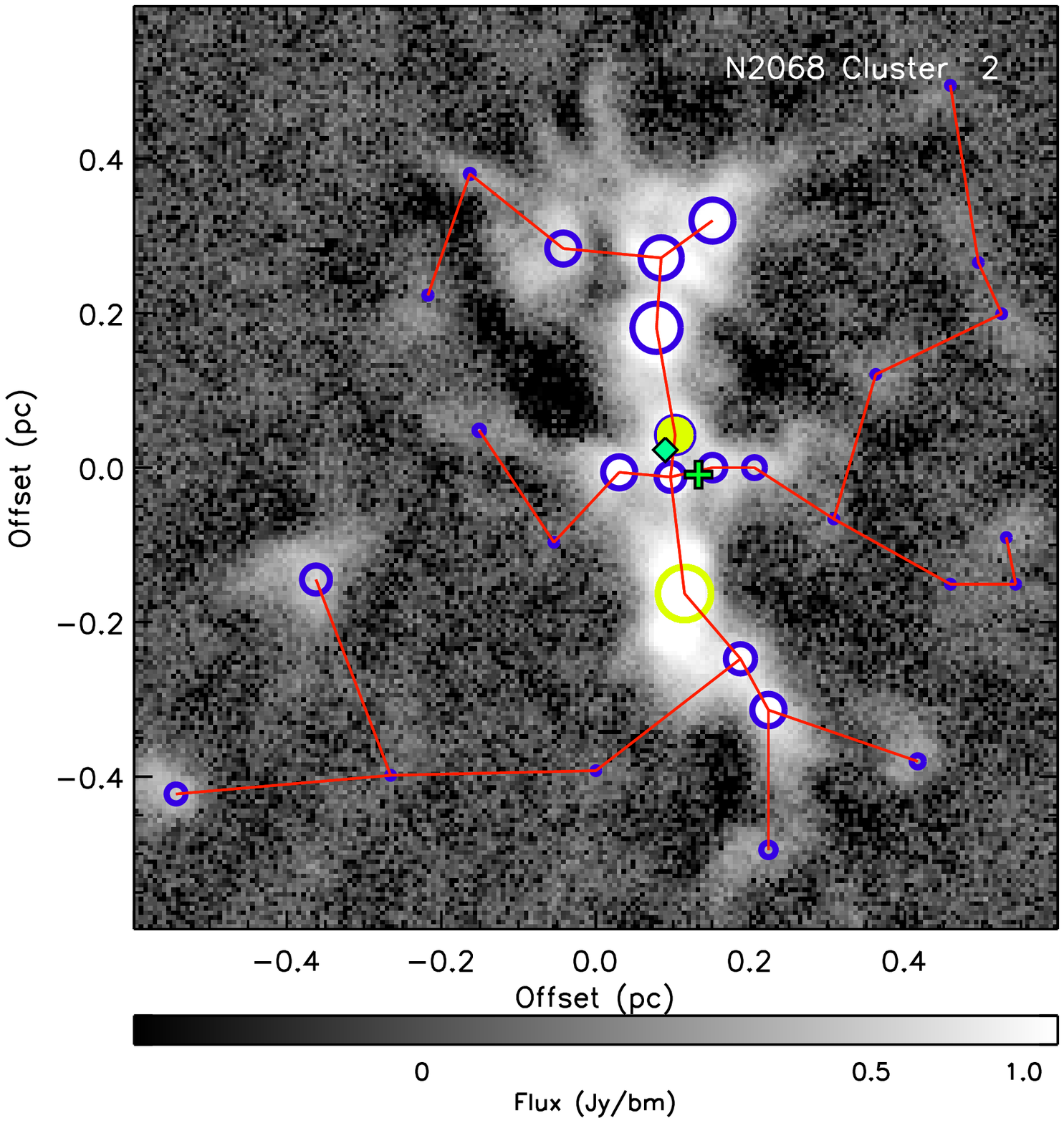}
\caption{A zoomed-in view of the largest two MST-based clusters identified in \Ntwo.  
	The circles indicate the relative locations of dense 
	cores, with the circle size scaling with the total core flux.  Thicker dark blue circles
	denote cluster members, while thin light blue circles indicate non-members in the vicinity.
	The yellow open and filled circles indicate the locations of the highest flux
	cluster member and highest flux starless core cluster member respectively.
	The red lines indicate the MST structure after branches longer than \Lcrit\ have been
	removed.  The green plus sign indicates the cluster centre (median position),
	while the filled turquoise diamond indicates the flux weighted mean core position.
	}
\label{fig_small_mst}
\end{figure}

\begin{figure}[htb]
\plottwo{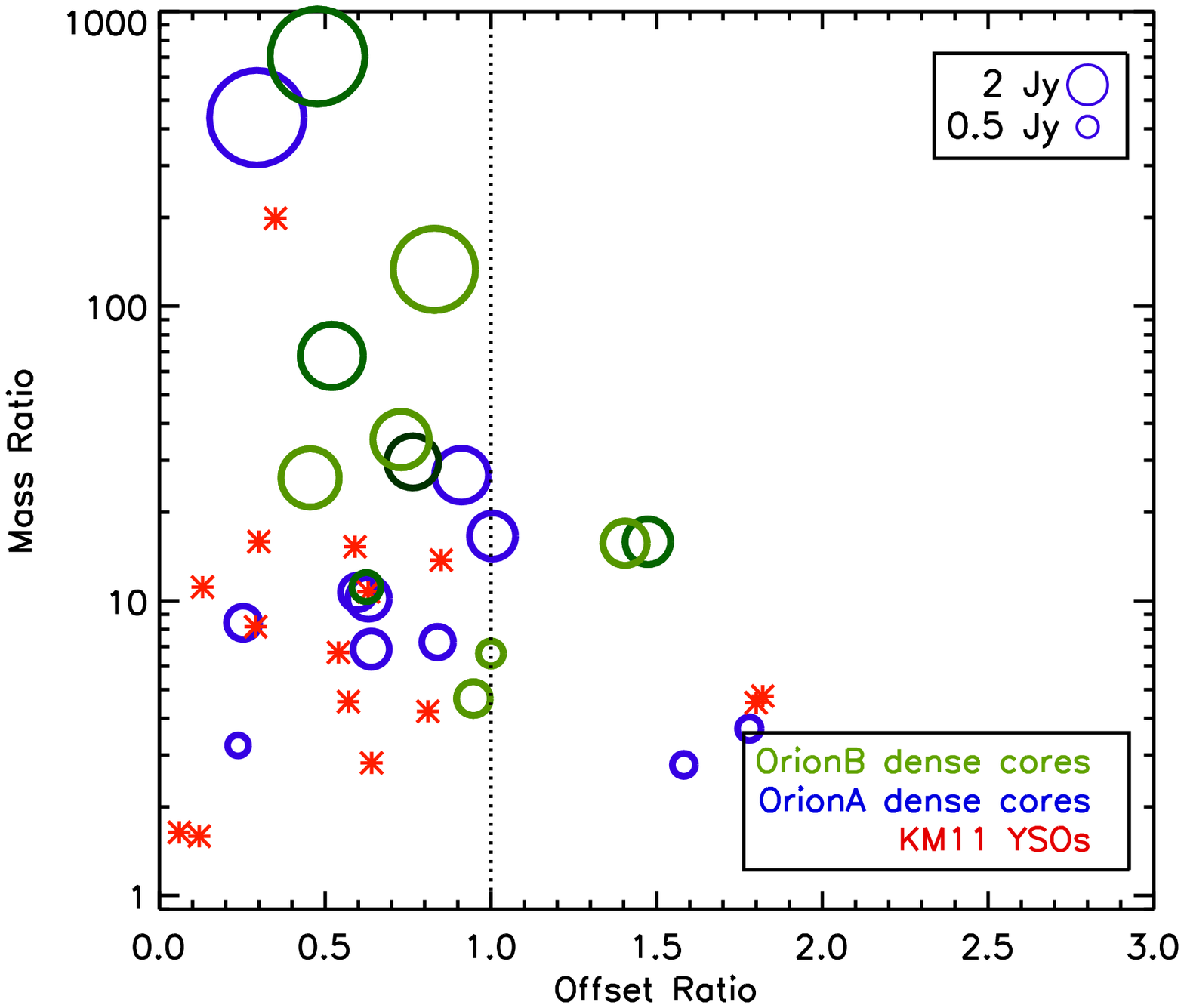}{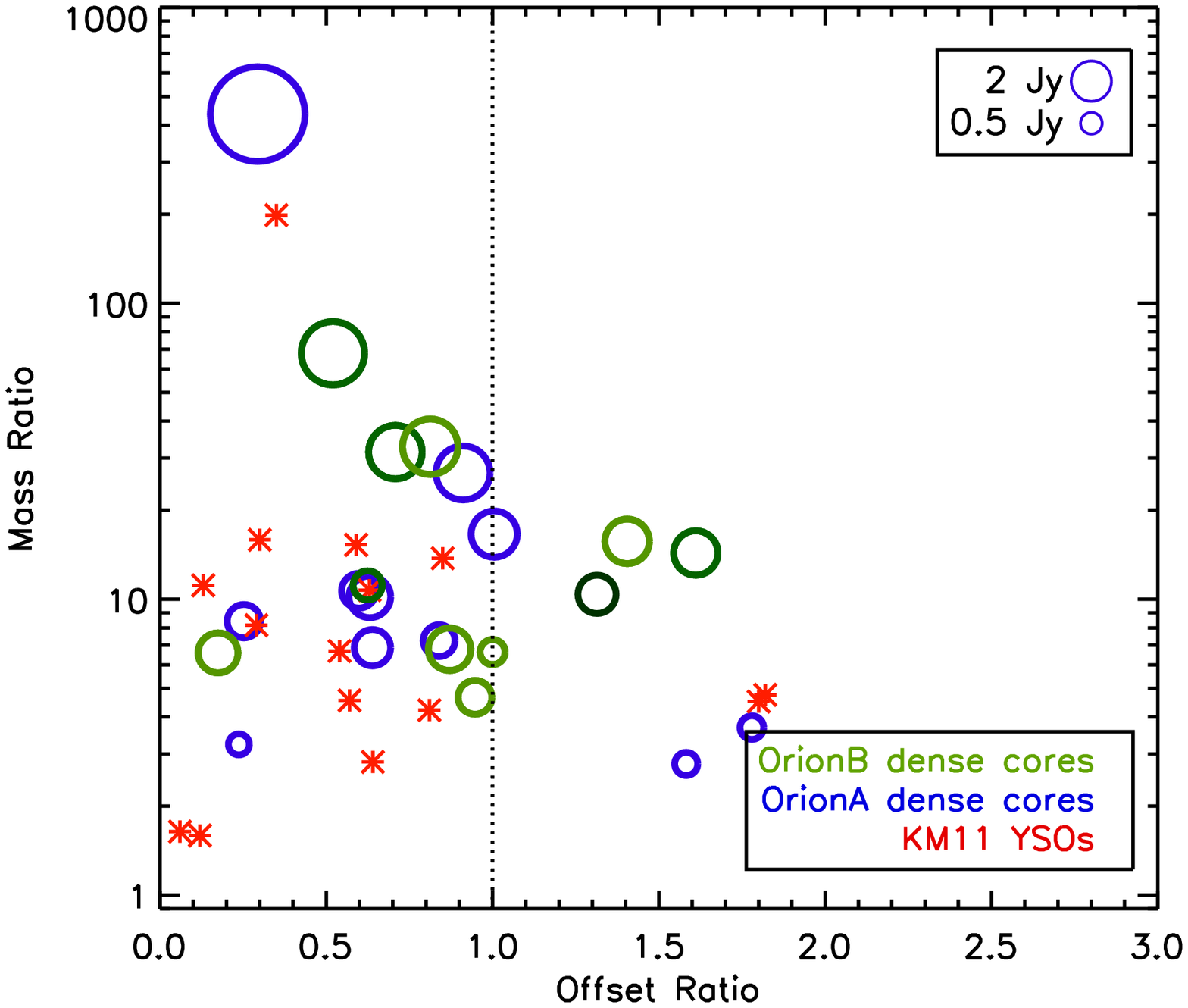}
\caption{A comparison of mass and offset ratios for the dense core clusters in Orion~B (green
	circles).  The size of the circles scale with the total flux of the most massive dense
	core in the cluster, and the circle shading indicates the region (L1622, \None, and
	\Ntwo\ correspond to darker through lighter shades of green).  For comparison, the
	blue circles indicate dense core clusters measured using the same technique by
	Lane et al. (in prep), and the red asterisks show the YSO clusters analyzed in
	\citetalias{Kirk11}.  The left panel shows the results considering the full population
	of dense cores in each cluster, while the right panel shows the results for the
	highest flux / most massive {\it starless} dense core in each cluster.
	}
\label{fig_offset_ratio}
\end{figure}

Figure~\ref{fig_small_mst} shows a zoomed-in view of the largest two clusters identified
in \Ntwo.  In the figure, the size of the circle scales with the total flux of the dense 
core.  The highest flux core (open yellow circle) tends to be relatively
centrally located in both of the examples shown in 
Figure~\ref{fig_small_mst}.  We can quantify
this tendency further, using a technique introduced by \citetalias{Kirk11}.  For each core in a 
cluster, we calculate its offset from the cluster centre.  The ratio of the offset
of the highest flux core to the median value of all of the core offsets
gives an indication of how relatively close the core is to the cluster centre,
with offset ratios less than one indicating a centrally-located core.  As might be naively
expected, \citetalias{Kirk11} showed that clusters with randomly located most massive
members tend to have equal incidences of offset ratios above and below one.
In Figure~\ref{fig_offset_ratio}, we show the distribution of offset ratios found for
dense core clusters in Orion B, as well as their `mass ratios'.  Here, we define  
`mass ratio' as the ratio between the flux of the highest flux cluster member and the 
median cluster member flux.  For a constant conversion factor between flux and mass
(see the discussion below), this ratio in fluxes is identical to
that of the masses.  The mass ratio is a very rough proxy for the degree to which
the most massive cluster member dominates the system gravitationally, namely ratios near one
indicate that all cluster members have similar fluxes (and hence masses), 
while high ratios indicate a diversity in fluxes and masses.  We note that we do not 
perform any quantitative analysis using the mass ratio and use it in 
Figure~\ref{fig_offset_ratio} purely for illustrative purposes.  Appendix~B
discusses in detail the sources of potential bias in our measurement of the offset
ratio, and shows that if any bias is present, it serves to slightly 
increase the offset ratios measured here.

Figure~\ref{fig_offset_ratio} also shows the mass and offset ratios for clusters of dense cores
identified in Orion~A by Lane et al. (in prep).  There, 
they use a different core-identification technique,
but the same procedure for identifying clusters of cores and measuring offset ratios.
In both Orion~B and Orion~A, it is clear that the majority of clusters have centrally-located
highest flux members.

We also compared the offset ratio distributions to those expected from random core locations.
We created 10,000 synthetic clusters with either a uniform 2D (circular) or uniform 3D
(spherical) distribution, randomly assigning one of the cluster
members to be the most massive one.  From these clusters, we calculated the nominal 
cluster centres and offset ratios
using the same procedure as for the observations.
We tried this test for clusters with 15, 25, and 50 
members.  For the Orion~B clusters alone, a two-sided Kolmogorov-Smirnov test \citep{Conover99}
gives a probability between 11\% to 17\% of the observed offset ratios being drawn from any 
of the random distributions we tested.
Combining the offset ratios for both Orion~A and B clusters, the probability of 
being a random
distribution drops to between 3\% and 4\%.
The decreasing probability is primarily attributable to the increased sample size, as
the fraction of clusters with offset ratios above one is similar in both samples (much less
than 0.5), whereas the random samples all have fractions of clusters with 
offset ratios above 1 which are very close to 0.5.
A direct comparison of the Orion B and A offset ratios with a two-sided KS test 
yields a probability of 74\% of similarity, which is not
statistically significant.

The offset ratios measured for the dense core clusters in Orion~B and Orion~A also bear a striking
resemblance to the offset ratios \citetalias{Kirk11} measured for small, nearby, YSO 
clusters (red asterisks on Figure~\ref{fig_offset_ratio}).  
A direct comparison between the offset ratios in \citetalias{Kirk11} and those of
Orion~B (and Orion~A), however, requires two assumptions.  First, the flux-ranking is the same as
the mass-ranking of the dense cores, and, second, the most massive protostar tends to form 
out of the most massive core.

Regarding the first assumption,
we know that it will not always hold, as hotter dense cores will appear brighter for 
the same intrinsic mass.  Protostellar
cores in particular can be expected to be hotter.  Indeed, several protostellar cores in
Orion~B are known to have temperatures of order 50~K, which would imply masses
about 3.3 times smaller than those obtained for a 20~K temperature (see discussion
in K16).  We therefore re-ran our analysis of the Orion~B clusters 
selecting the highest flux {\it starless}
dense core in each cluster, to account roughly for the potential bias introduced in
flux measurements in the protostellar cores.  The highest flux starless core in 
each cluster is also noted in Figures~\ref{fig_small_mst}, \ref{fig_mst_clusts1}, 
and \ref{fig_mst_clusts2}.  In some clusters, the highest flux core is actually starless, so
our derived offset ratio is unchanged.  In the remaining clusters, while the individual
offset ratio measures change, the overall distribution of values remains fairly
similar, as is illustrated in the right panel of Figure~\ref{fig_offset_ratio}.
This test indicates that our measurement of typically small offset ratios is true for both
the highest flux core and the most massive core.

A two-sided KS test comparing the Orion B and A dense core cluster offset ratios with the 
\citetalias{Kirk11} YSO cluster offset ratios yields a probability of 24\% of similarity,
also inconclusive due to the small sample sizes.  
The fraction of offset ratios greater than one is
somewhat smaller for the YSO cluster sample (15\%) than for the combined dense core sample (22\%).
If the most massive YSO is formed from the most massive core,
this difference in offset ratios could imply evolution in the position of the most 
massive cluster object between the
dense core and YSO stage.  \citet{Evans09} estimate a lifetime of 0.54 -- 0.7~Myr for the Class 0 
plus I phases of YSOs \citep[while][give an updated value of 0.54~Myr]{Heiderman15}, 
which, for cluster velocity dispersions of 
0.5 -- 0.9~km~s$^{-1}$ \citep{Foster15},
would correspond to a maximum motion of 0.3 -- 0.6~pc.  
A second estimate of the typical velocity dispersion within clusters can be obtained
through a combination of the YSO-to-core velocity dispersion, estimated to be about 
0.1 -- 0.2~km~s$^{-1}$ based on the offsets of embedded YSOs from the parent dense core
\citep[e.g.,][Mairs et al. 2015, in prep.]{Jorgensen07,Jorgensen08} with the core-to-core velocity
dispersion within a clustered environment, estimated to be several times larger.
For example, the line-of-sight velocity dispersion between cores in clustered environments 
is 0.3 -- 0.8~km~s$^{-1}$ in Perseus \citep{Kirk07}.  The total YSO-to-YSO 
velocity dispersion using this
second technique is therefore similar to the \citet{Foster15} result.  Over the typical
lifetime of an embedded YSO, the amount of motion possible
is large enough to have a significant
impact on the offset ratios of the smaller clusters in our sample.
It is also possible, however, that Orion~B and A represent a different cluster-forming environment
than the \citetalias{Kirk11} sample (e.g., KM11's sample includes Taurus YSOs which typically
have lower source-source surface densities than found in Orion~B -- see Section~4), 
so that the two samples cannot be thought of as direct
correspondents at different ages.  Further complicating the direct evolutionary picture, 
as noted in Section~2, is the fact that different cores may exhibit differing amounts of 
fragmentation, and the least massive cores might dissipate without forming any protostars.

\subsection{Beyond the Offset Ratio}
\begin{figure}[htbp]
\begin{tabular}{cc}
\includegraphics[width=2.7in]{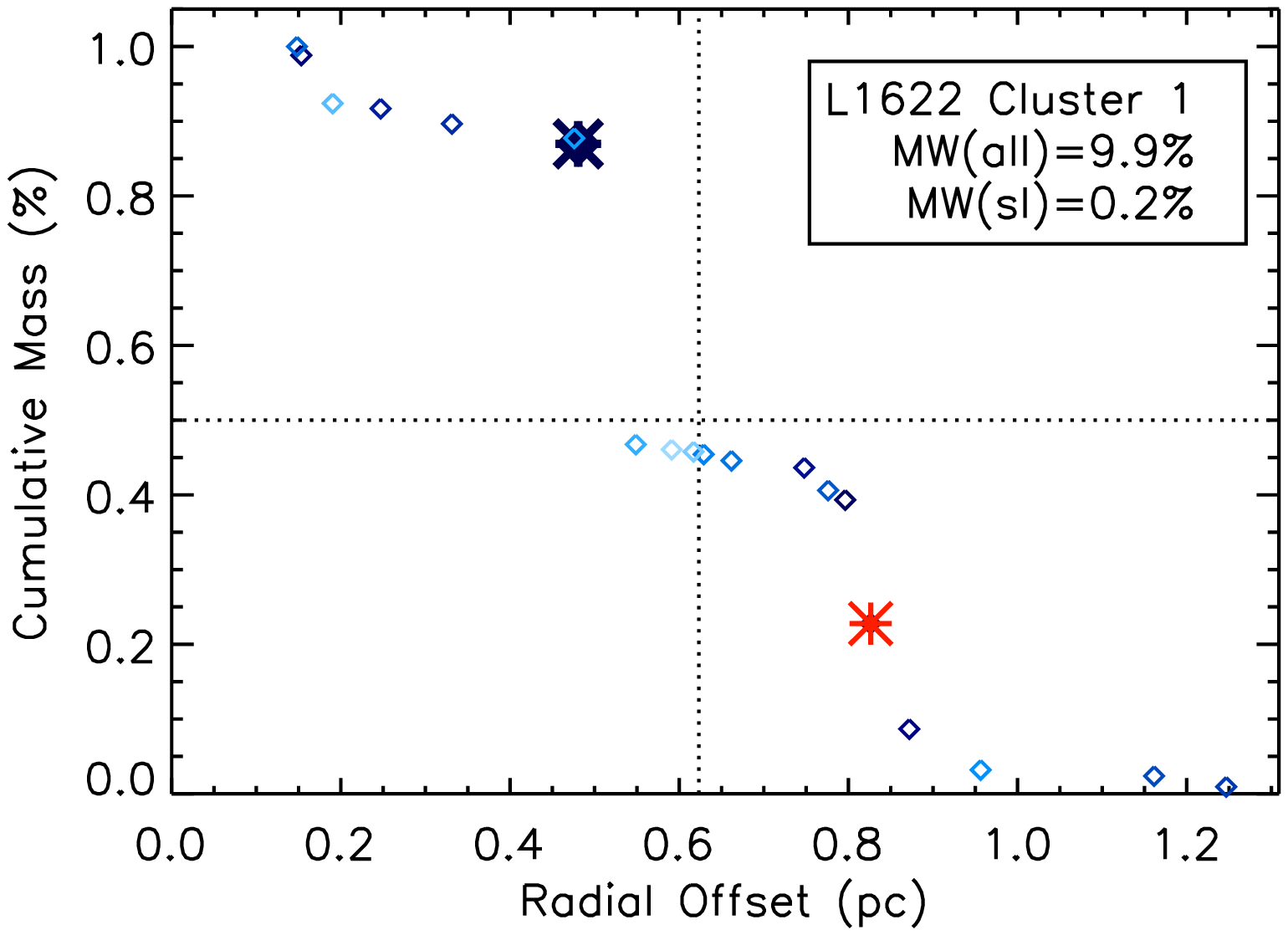} &
\\
\includegraphics[width=2.7in]{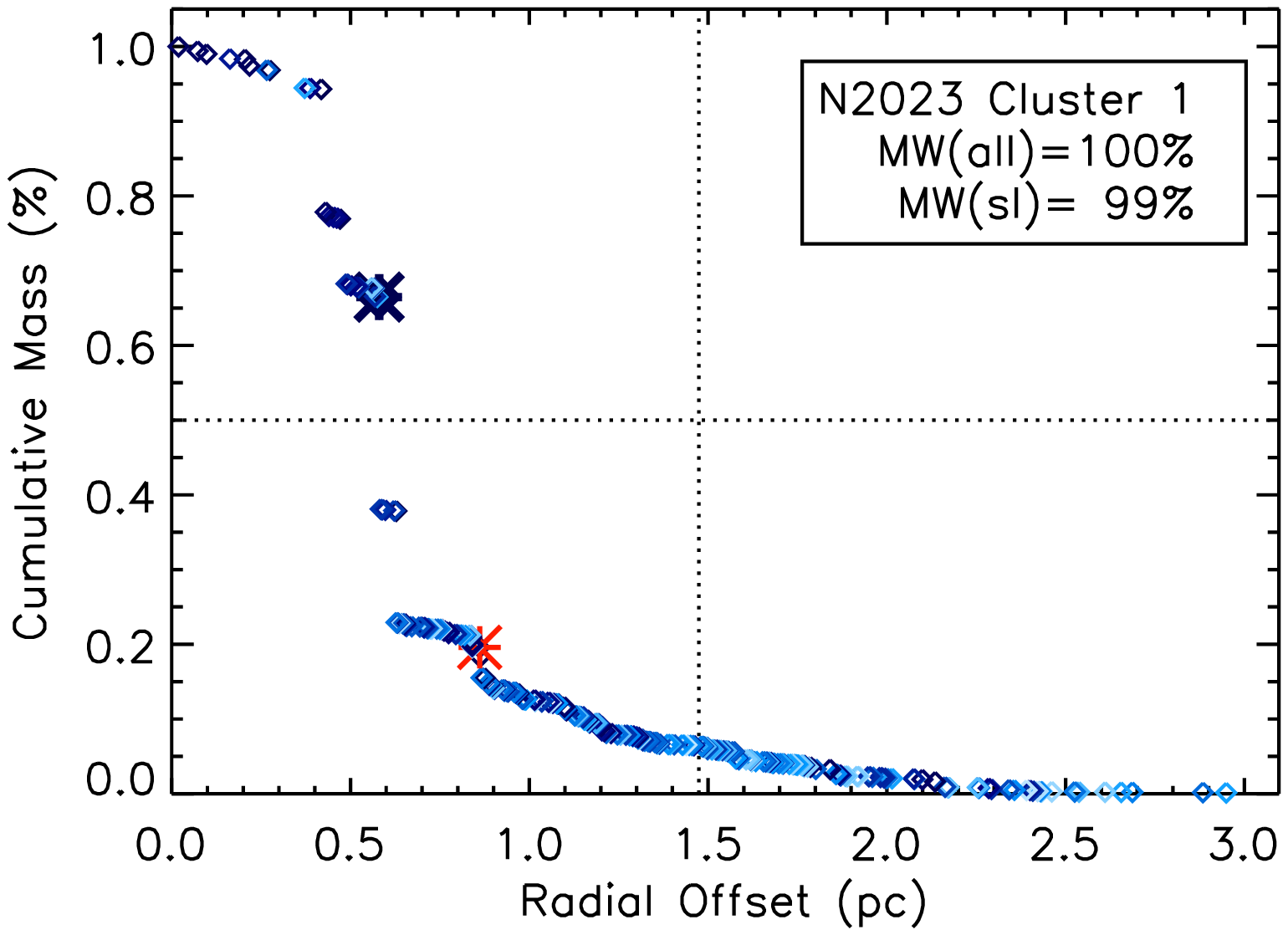} &
\includegraphics[width=2.7in]{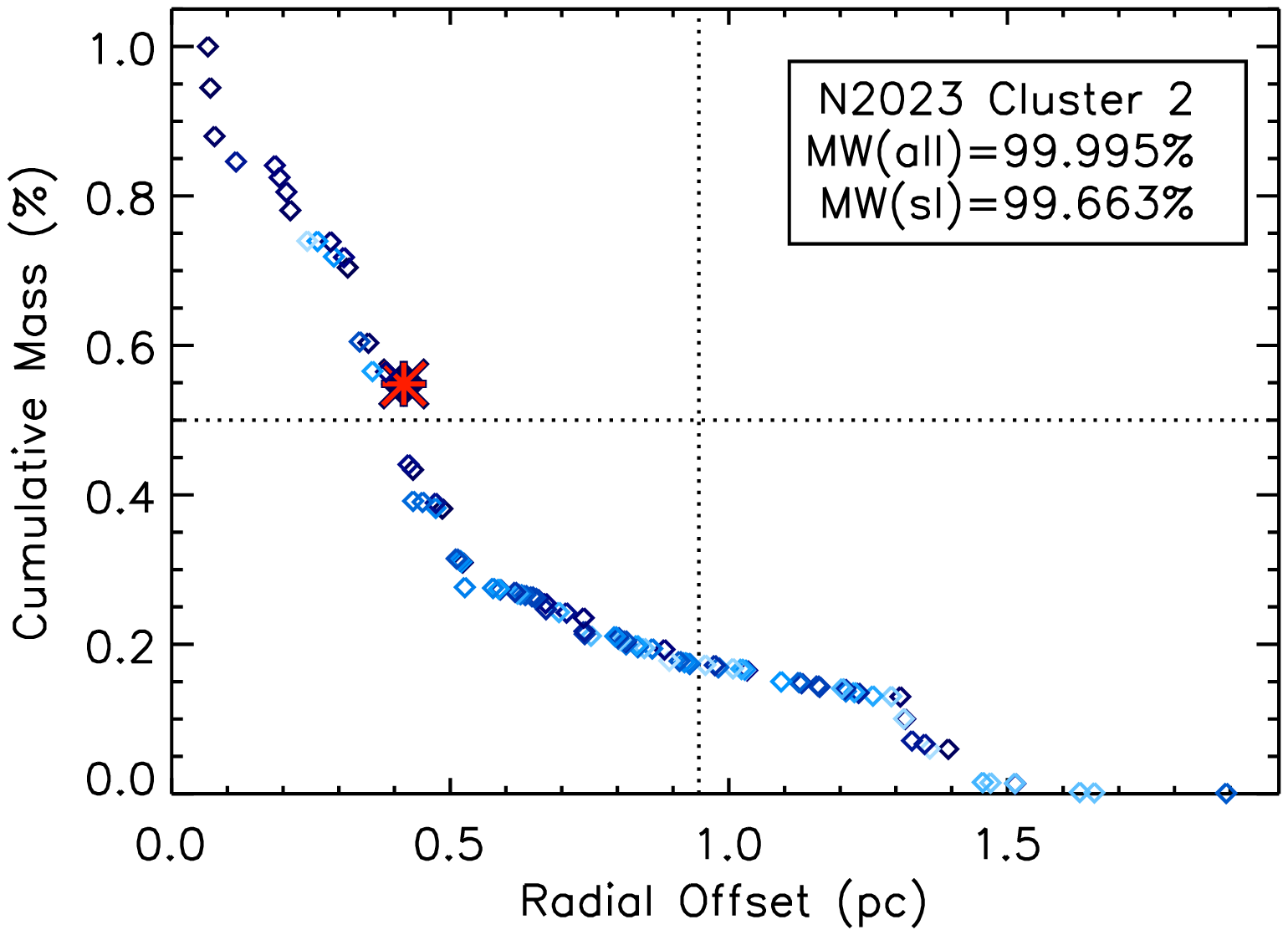} \\
\includegraphics[width=2.7in]{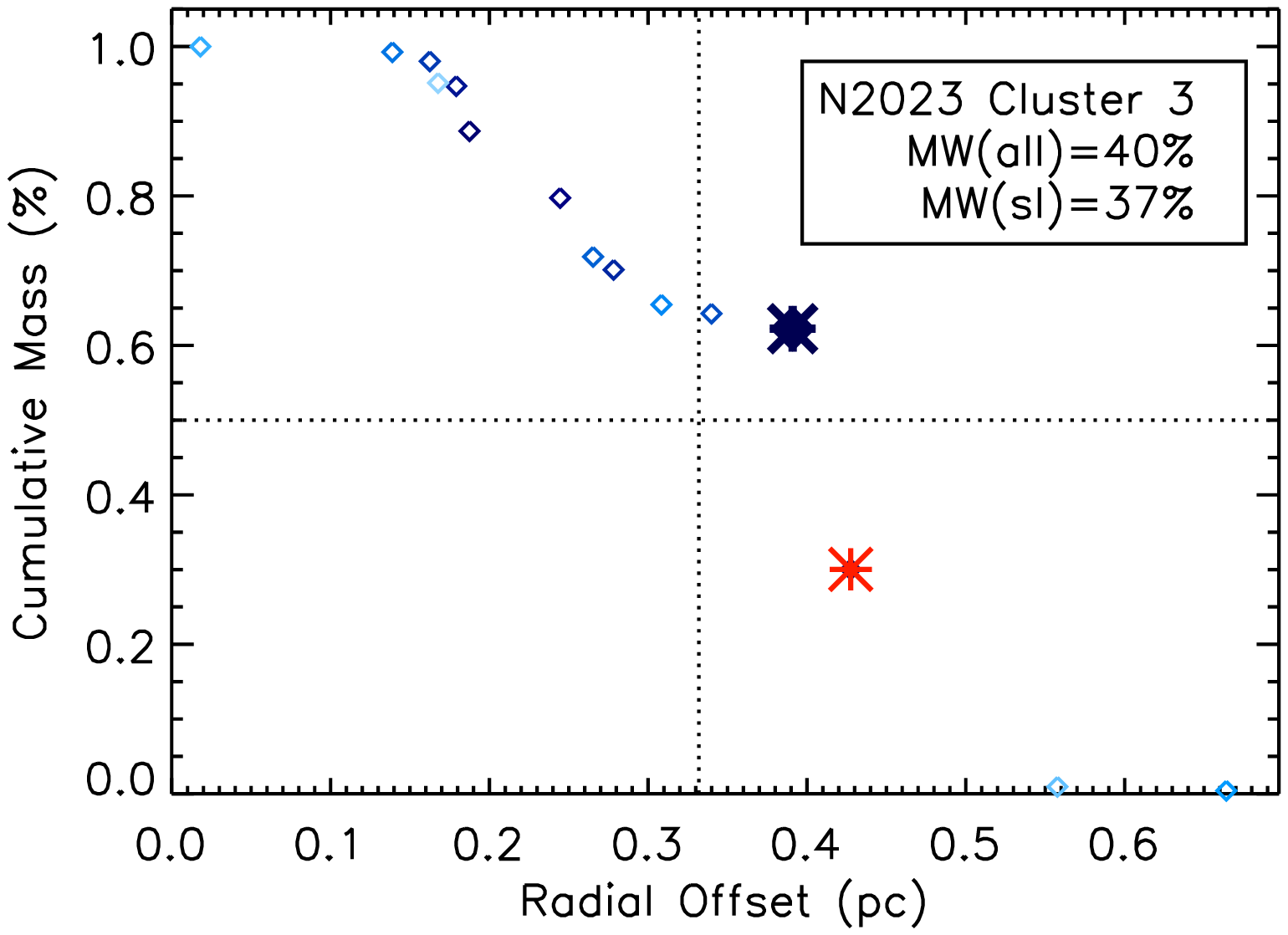} &
\includegraphics[width=2.7in]{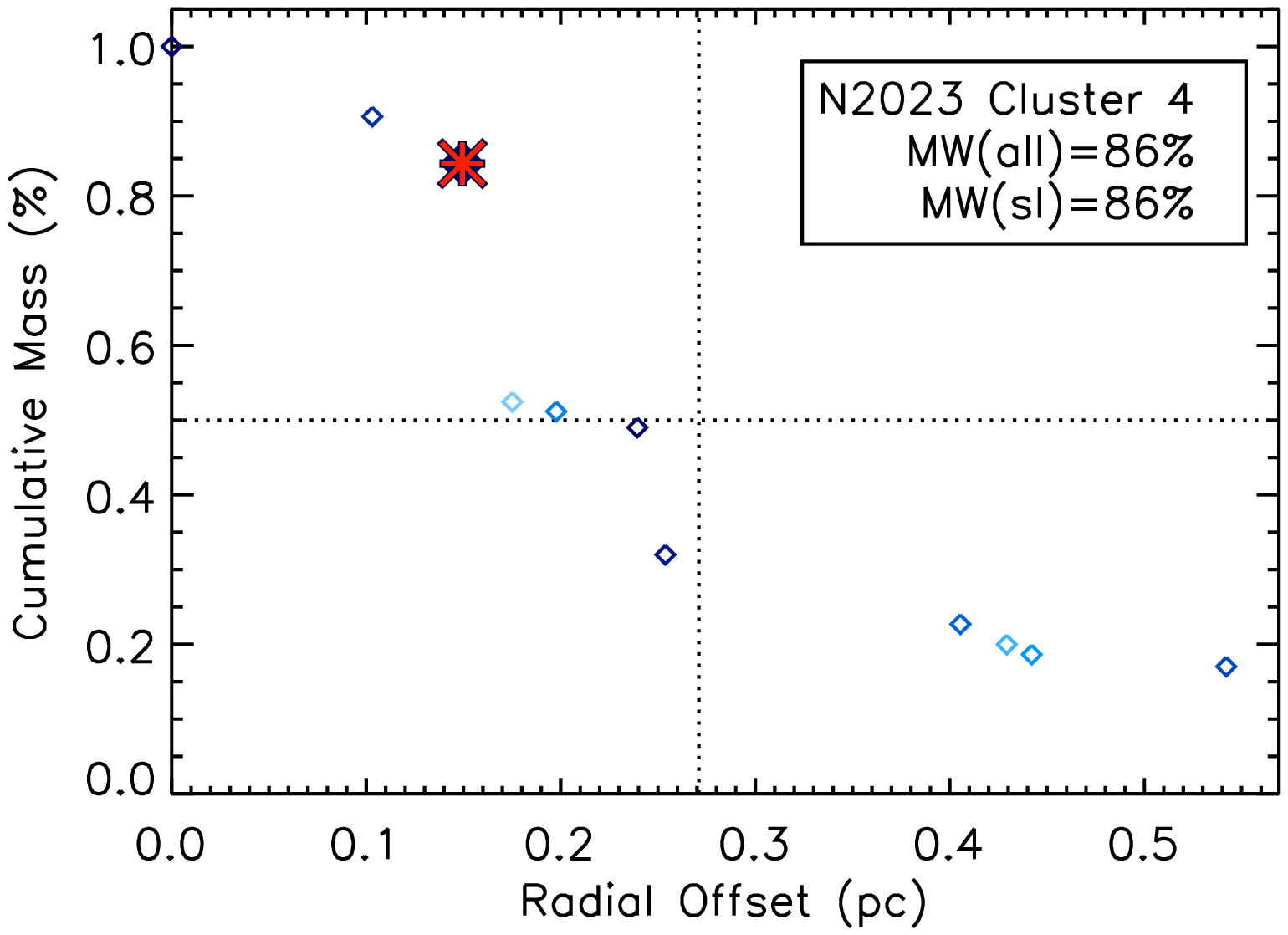} \\
\end{tabular}
\caption{The cumulative mass of sources at a given offset or larger from their MST-based
	cluster centre.  The flux / mass rankings of the cores are indicated by the shading
	with darker indicating higher flux, the large dark blue asterisk indicating the 
	highest flux member of each cluster, while the bright red asterisk denotes the
	highest flux starless core within each cluster.  
	The horizontal dotted line indicates a cumulative mass
	fraction of 50\%, while the vertical dotted line indicates half of the maximum
	offset.  The Mann-Whitney probability that the inner half of the dense 
	cores have typically higher fluxes than the outer half of cores is given in the
	upper right corner of each panel, for both the entire cluster population (`all') 
	as well as only the starless cores (`sl').  
	This figure shows the clusters in L1622 (top row) 
	and \None\ (middle and bottom rows).
	}
\label{fig_mass_offset1}
\end{figure}

\begin{figure}[htbp]
\begin{tabular}{cc}
\includegraphics[width=3.1in]{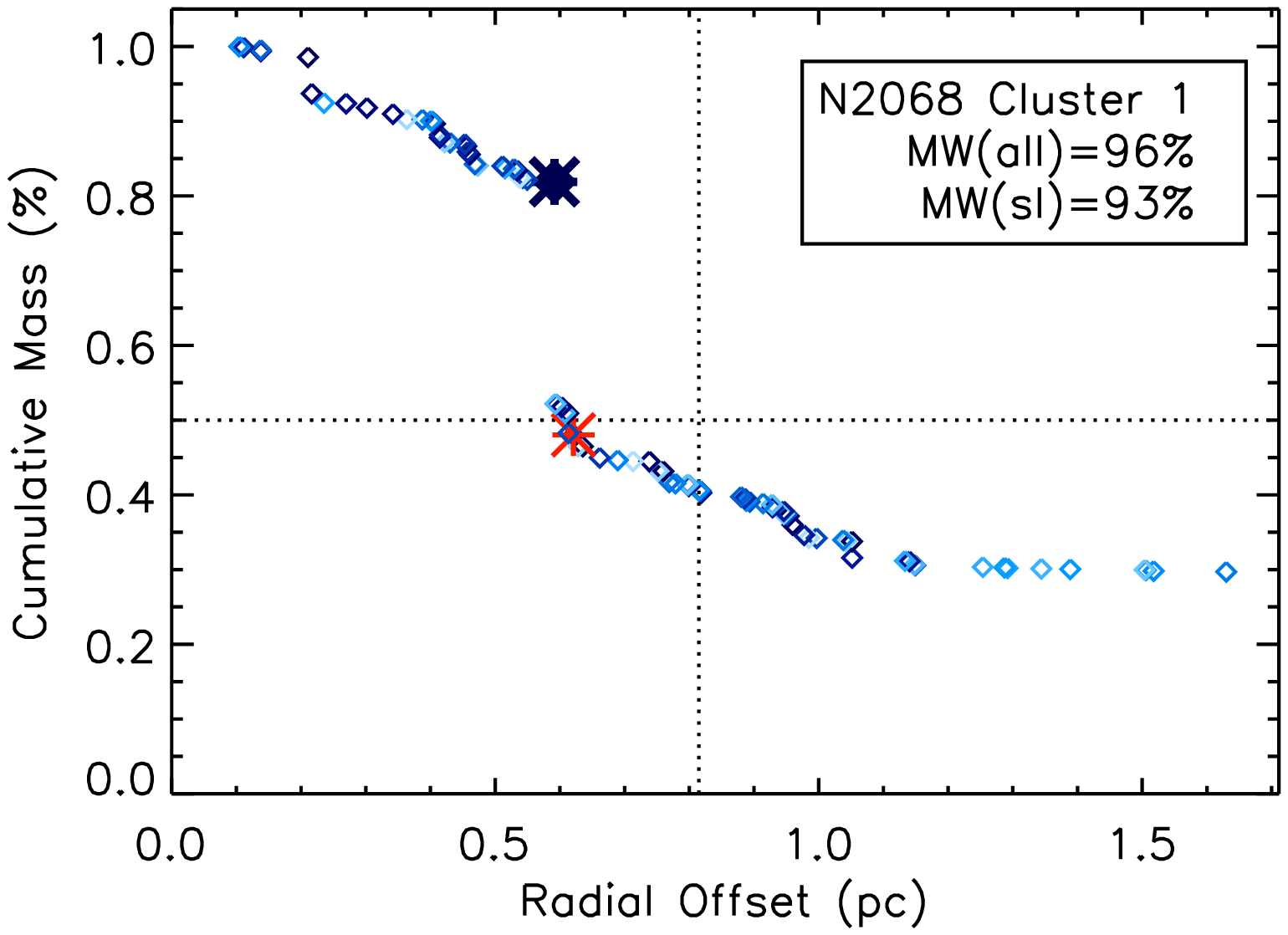} &
\includegraphics[width=3.1in]{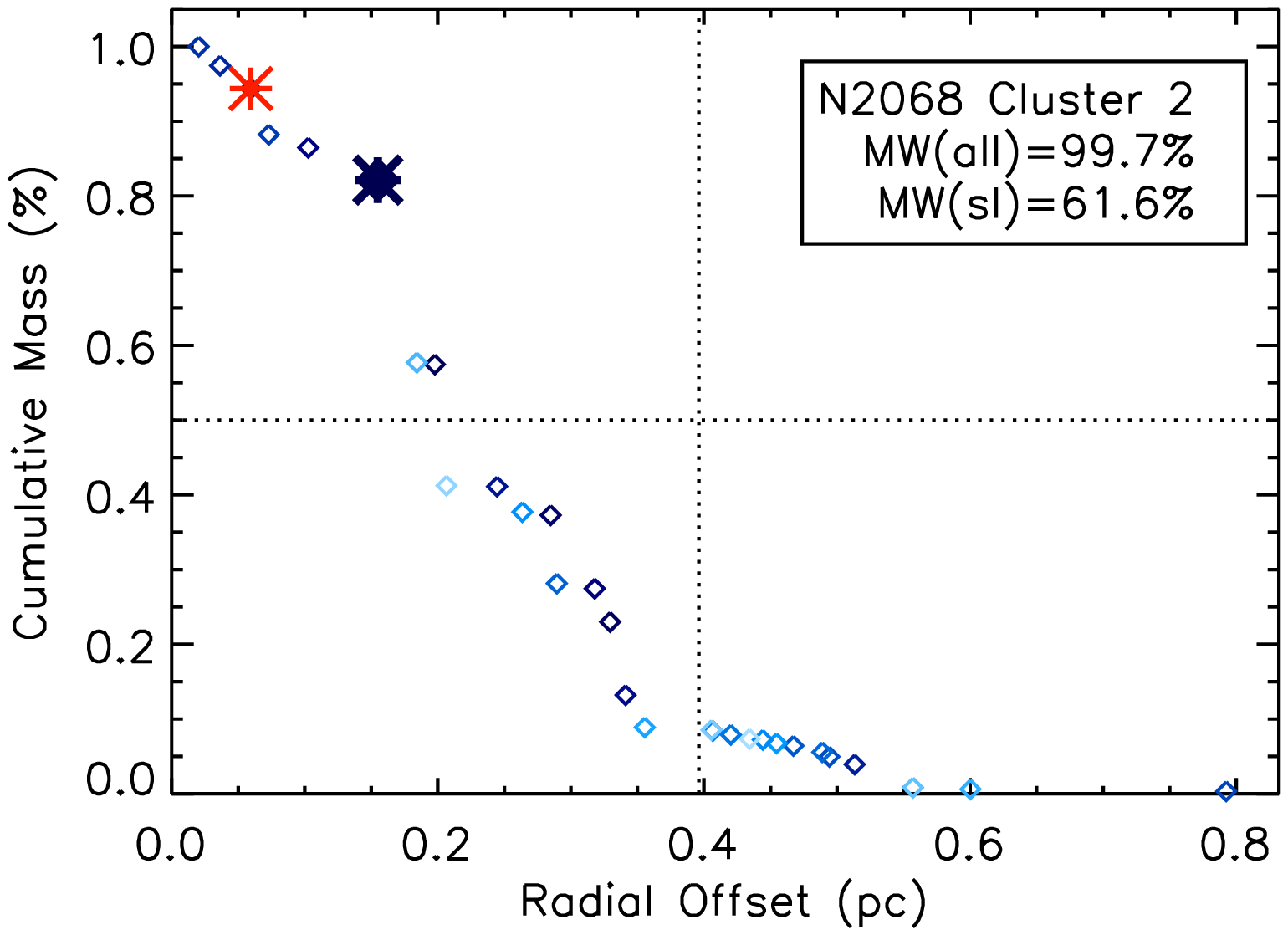} \\
\includegraphics[width=3.1in]{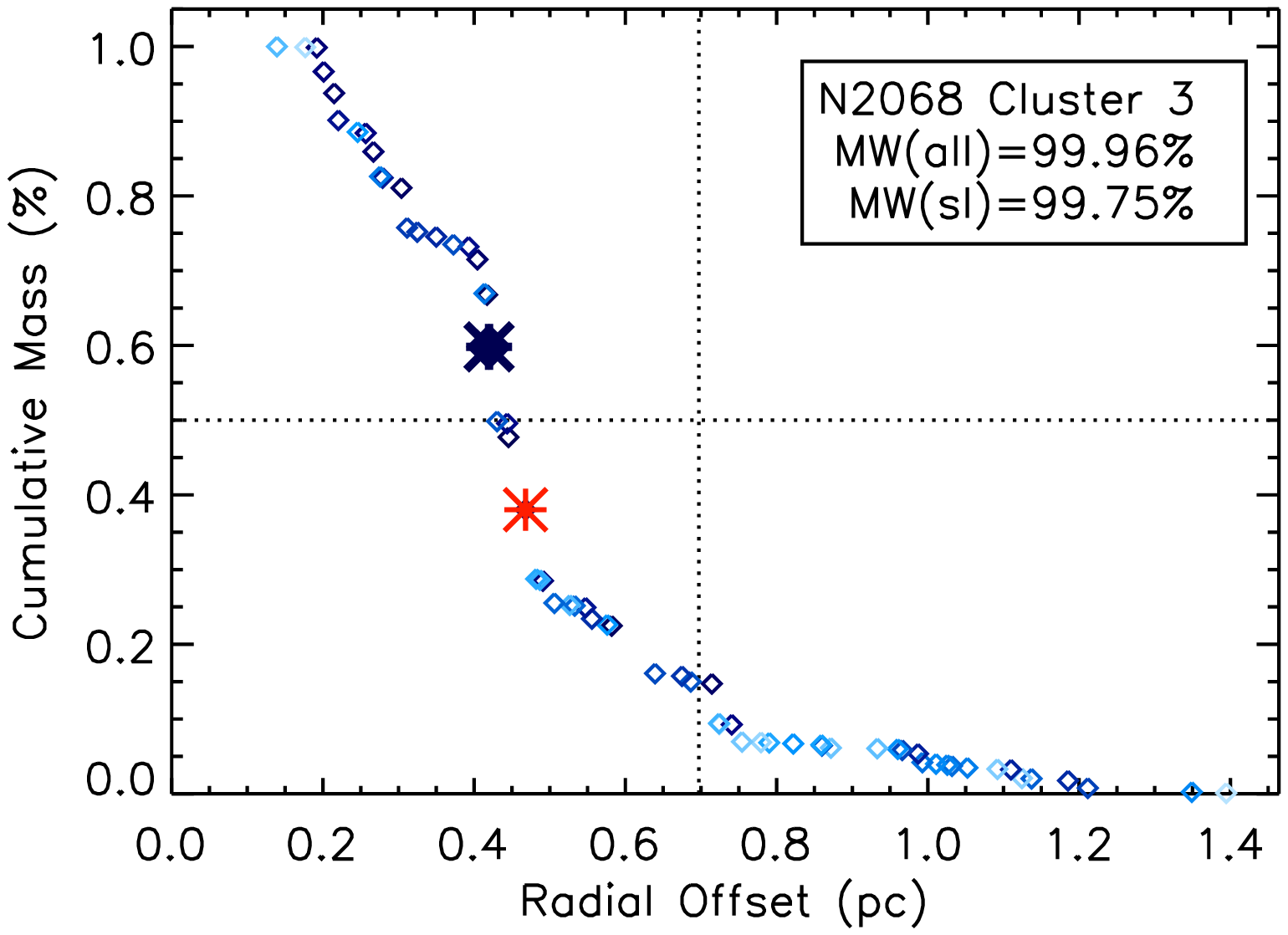} &
\includegraphics[width=3.1in]{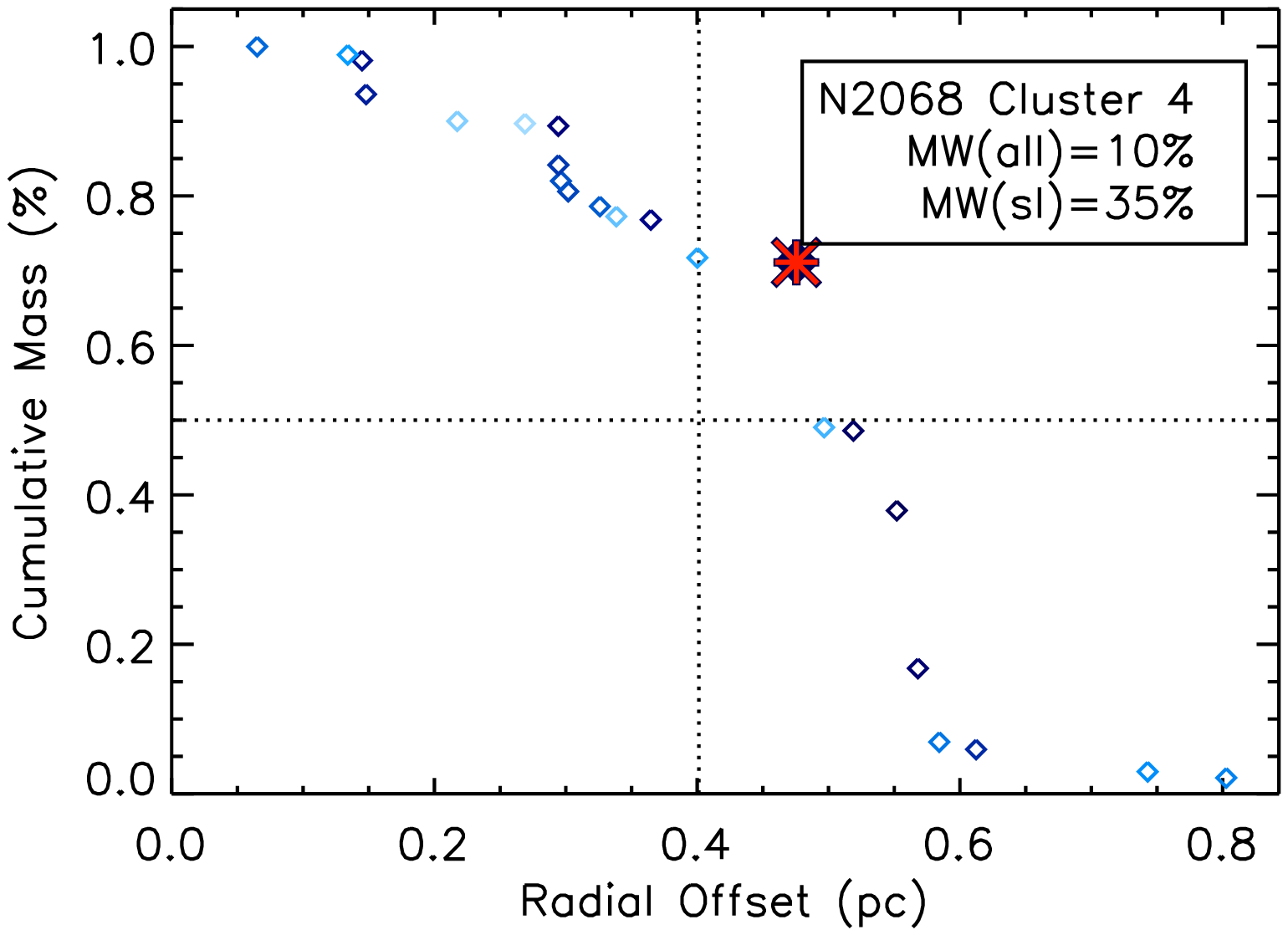} \\
\includegraphics[width=3.1in]{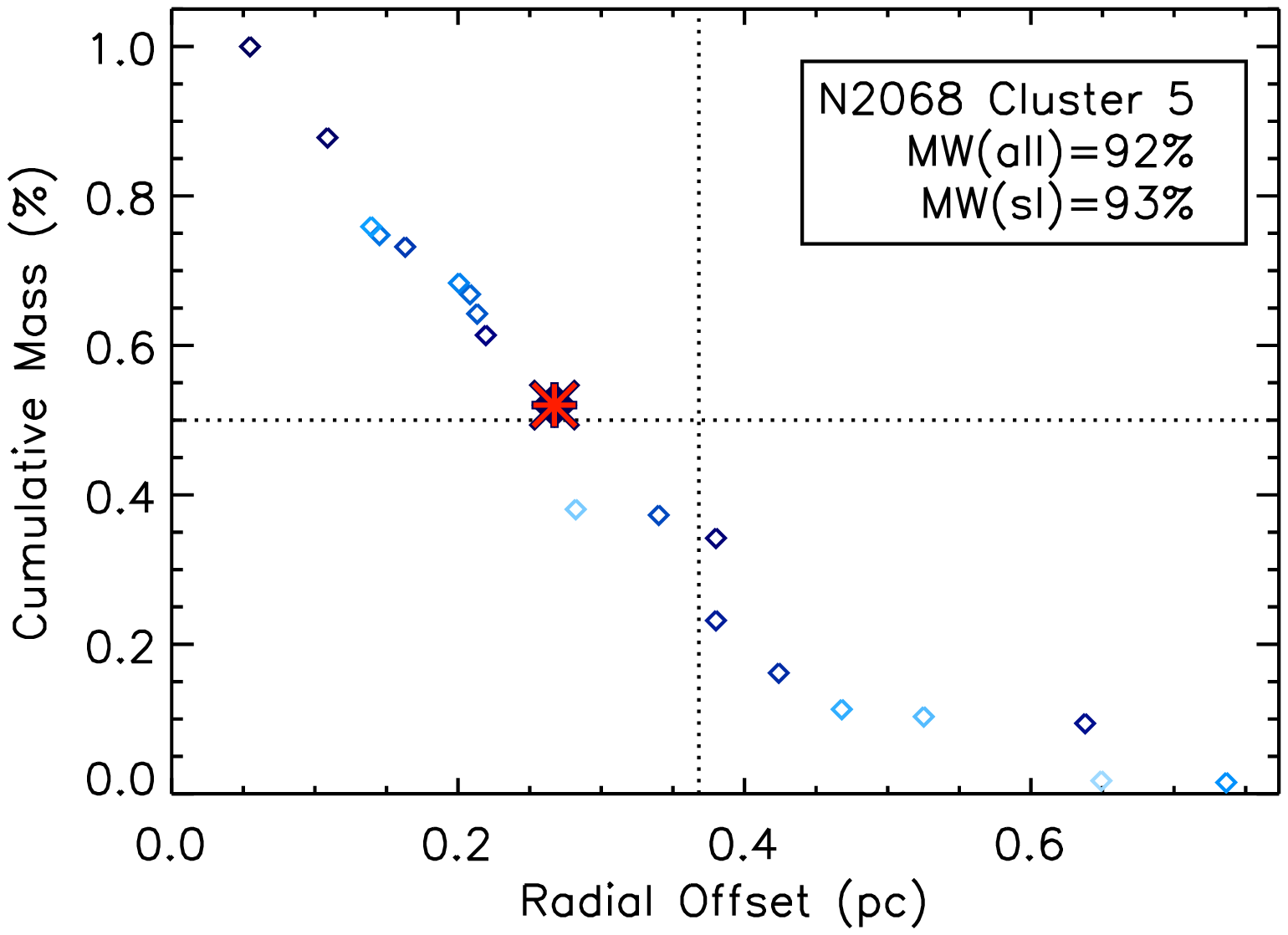} &
\includegraphics[width=3.1in]{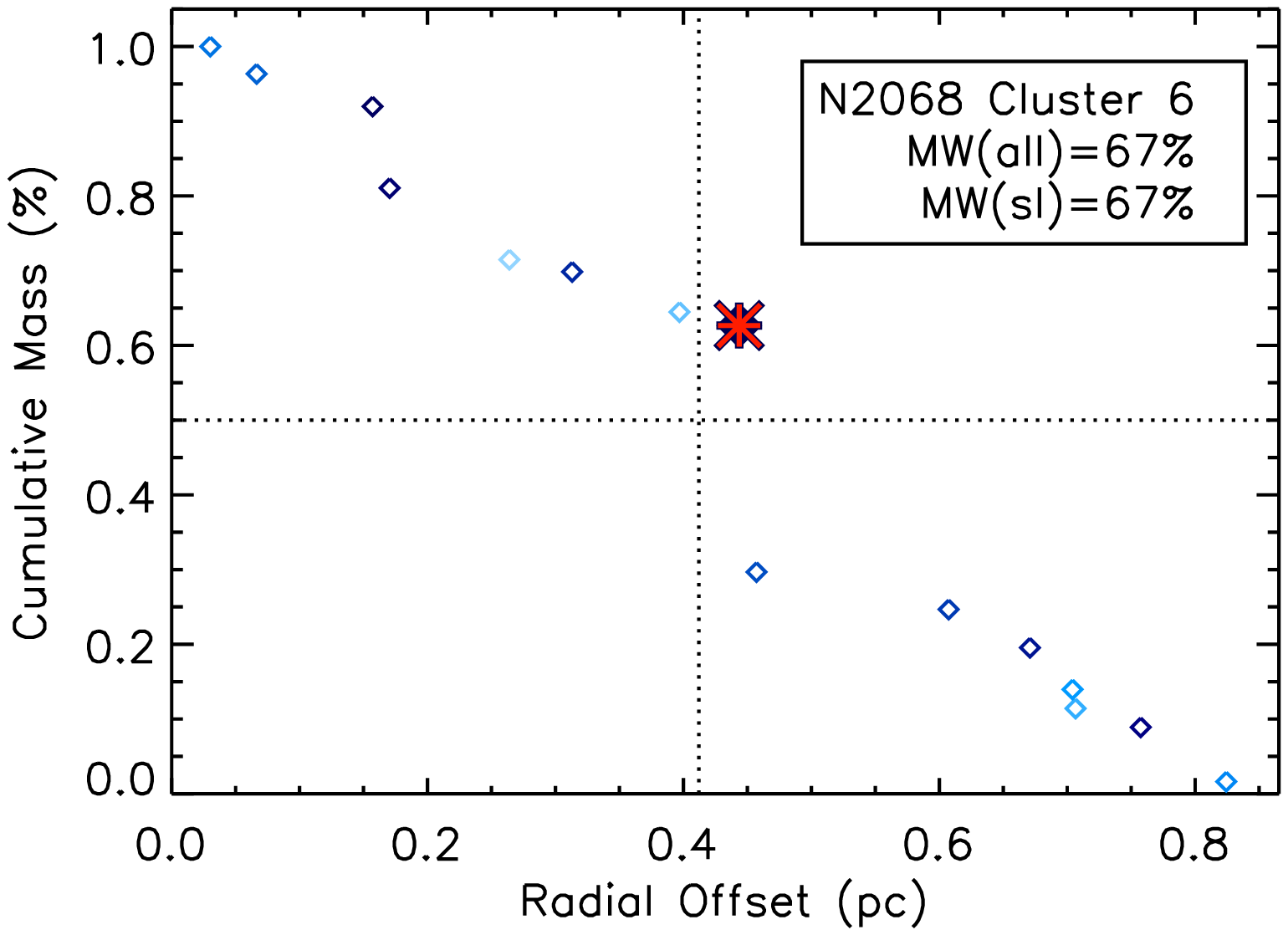} \\
\end{tabular}
\caption{The cumulative mass fraction of sources at a given offset or larger from their MST-based
        cluster centre, for clusters identified in \Ntwo, showing the relative flux / mass
	rankings of the cores.  See Figure~\ref{fig_mass_offset1} for the plotting conventions 
	used.} 
\label{fig_mass_offset2}
\end{figure}

In addition to measuring the offset ratio as a proxy for mass segregation, we briefly
take a more qualitative look at the general tendency of higher flux cores 
to be located
closer to cluster centres.  Figures~\ref{fig_mass_offset1} and \ref{fig_mass_offset2} show
the cumulative distributions of dense core masses as a function of offset from the cluster 
centre for dense core clusters in Orion~B.
Visually, many of the clusters appear to have a higher
fraction of high flux cores closer to the cluster centre.  We test this hypothesis
by running a Mann-Whitney, or Wilcoxon, test \citep{Conover99}, also used by
\citet{Kirk12} on the dense core fluxes in the inner and outer
halves of the clusters.
The Mann-Whitney (MW) test compares the rank order of
values within the two sub-samples to calculate the probability that one sub-sample has
typically larger or smaller values than the other.  The probability that the inner half of the 
dense cores have higher fluxes than the outer half are reported in the legend of each
panel in Figures~\ref{fig_mass_offset1} and \ref{fig_mass_offset2}.  We ran this
test for both the full cluster population, and examining only the starless cores.
Indeed, several of the clusters do show strong signs of having more massive cores 
closer to the centre.  For example,
\None\ clusters 1 and 2 and \Ntwo\ clusters 1, 2, and 3 all have probabilities above 95\%
of higher flux members in the inner half (although \Ntwo\ cluster 2 has a lower
probability for only the starless core population),
while several other clusters have somewhat high probabilities.  L1622 has a notably low
probability, indicating a tendency for higher fluxes
of dense cores in the {\it outer} half, i.e., a probability
of 0.2\% that the inner starless cores have higher fluxes and correspondingly a 99.8\% 
probability that
the outer cores have higher flux.  Cluster 4 in \Ntwo\ also has a low, but 
less statistically significant probability.

\section{Local Surface Densities}

\label{sec_clusters_msig}
A second complementary method for measuring mass segregation that also works well in regions
with substructure is presented in \citet{Maschberger11}.  Their method does not explicitly
subdivide sources into clusters, like the MST technique, but instead uses the local surface
density of sources to distinguish between cluster centres and outskirts.  The local surface
density of sources is estimated using the separation to the Nth nearest neighbour,
\begin{equation}
\sigma_{N} = \frac{N}{\pi r_N^2} ,
\end{equation}
where $r_N$ is the distance to the Nth nearest neighbour.  This estimate has a fractional
uncertainty of $N^{-0.5}$ \citep{Casertano85,Gutermuth09}\footnote{We note that these formulae 
are written
with N-1 in the references listed, because there, the Nth nearest neighbour is calculated at 
every location in an image, not only for a pre-defined list of sources.  In the former case, 
finding the Nth nearest neighbour at the position of a source would include the source itself, 
i.e., the second nearest neighbour would be the closest source to the one in question.
Since in most other contexts,
the Nth nearest neighbour doesn't count the source itself for N, we adopt this terminology here.
In the previous example, the source would be called the first nearest neighbour.}.
\citet{Maschberger11} used N=6 for their analysis and we tried both N=5 and N=10, and find similar
results.  Only the N=5 results are shown here.

\begin{figure}[htbp]
\begin{tabular}{cc}
\includegraphics[width=2.9in]{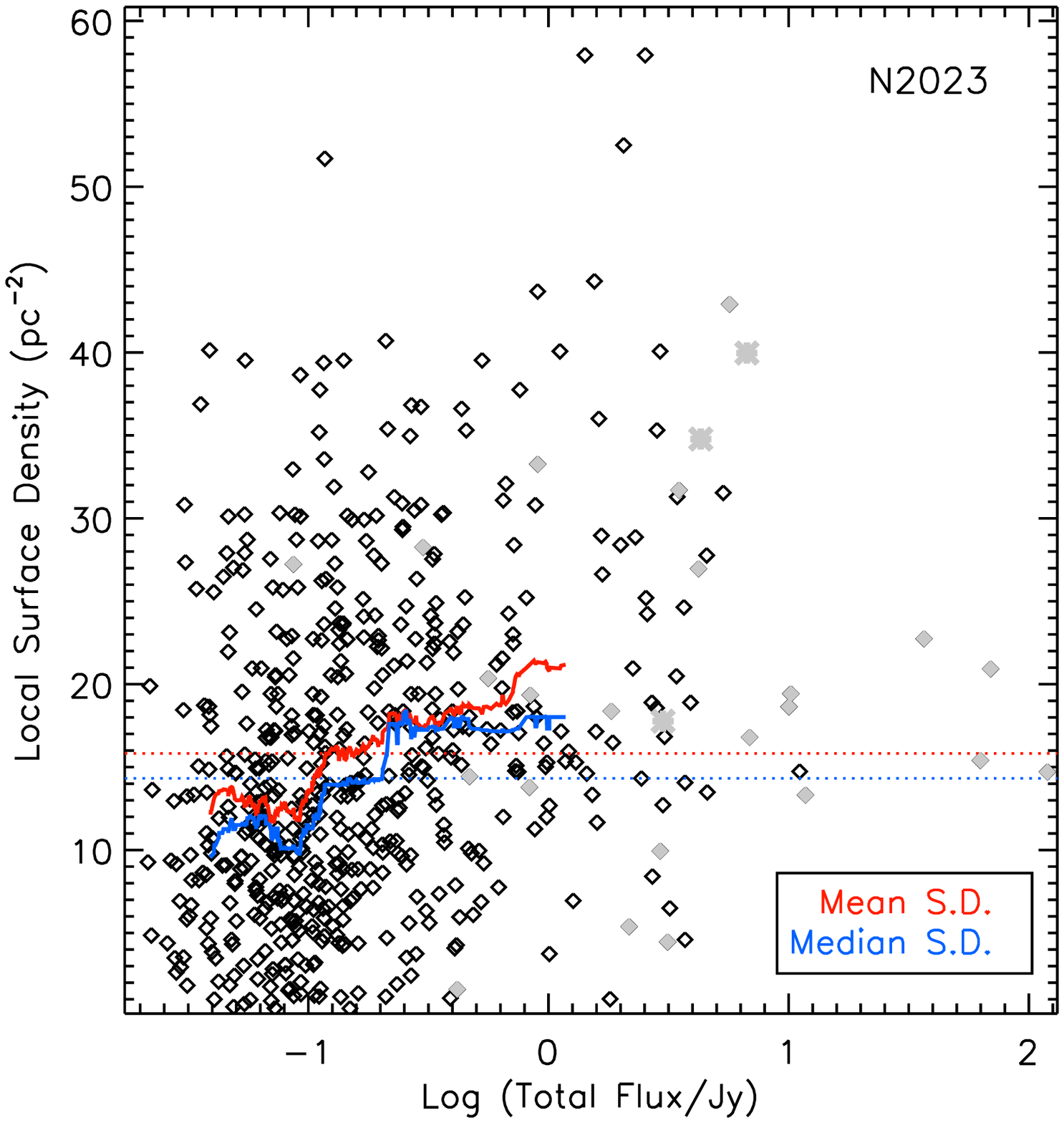}&
\includegraphics[width=2.9in]{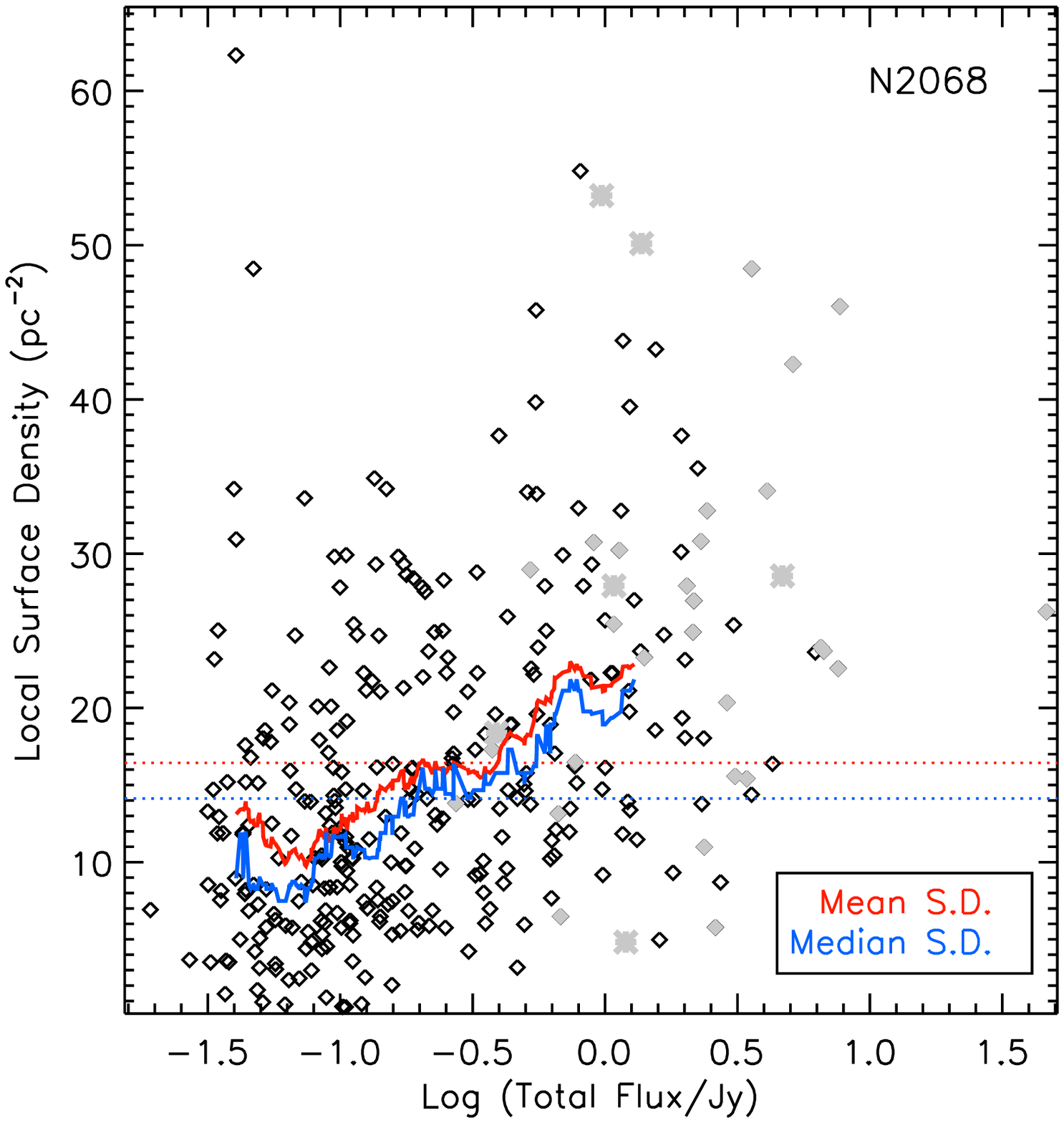} \\
\includegraphics[width=2.9in]{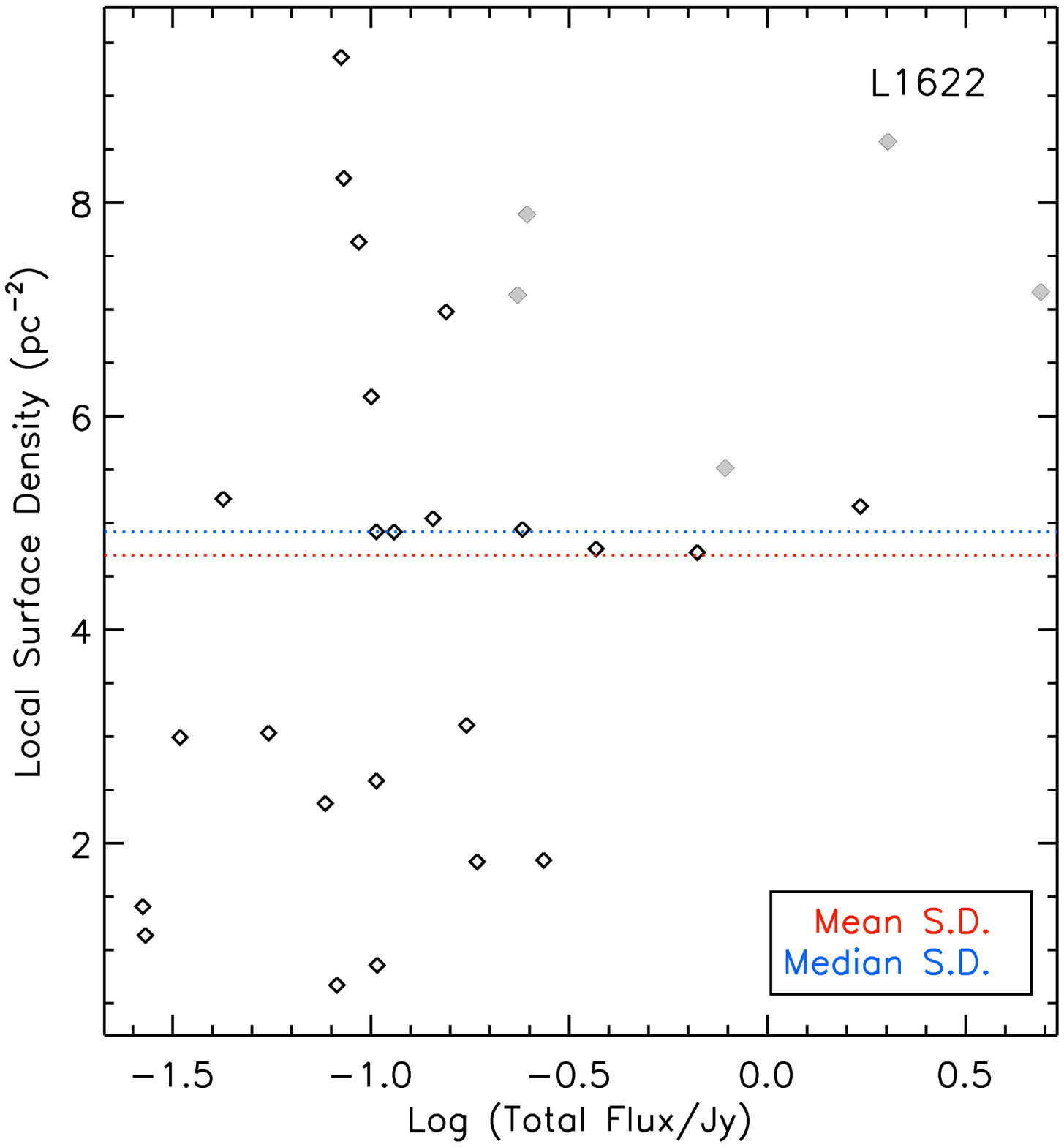} &
\\
\end{tabular}
\caption{A comparison of the local surface densities of cores and their total fluxes.  The top
	row shows \None\ and \Ntwo\ (left and right), while the bottom row shows L1622.  In each
	plot, the five nearest neighbours were used to determine each core's local surface
	density (see text for details).  Protostellar cores are denoted as the grey
	solid diamonds \citep[using the][{\it Spitzer} catalogue]{Megeath12} and grey 
	asterisks \citep[using the][{\it Herschel} catalogue]{Stutz13}.
	The solid red and blue lines in the top row show co-moving
	mean and median surface density values for the starless cores; a co-moving window of 
	40 starless cores was 
	used for \None\ while a window of 20 starless cores was used for \Ntwo.  
	The dotted red and blue
	lines show the global mean and median surface density.  In L1622, there were insufficient
	cores to calculate co-moving window values.
	}
\label{fig_mass_vs_surfdens}
\end{figure}

Once a source's local surface density has been estimated, mass segregation, if present,
should manifest itself as a positive trend between source mass and local surface density.
In all three maps of Orion~B, we find that more massive (higher total flux) dense cores
tend to inhabit areas with higher core surface densities.  Figure~\ref{fig_mass_vs_surfdens}
shows the dense core surface densities and total fluxes for each of the three regions.
Protostellar cores, which may have higher fluxes due to their elevated temperatures, 
are shown in light grey.
Although the correlation is not tight, it is clear that all three regions have a relative absence
of cores in the top left (high surface density, low total flux) and bottom right (low surface
density, high total flux) corners.  Following \citet{Maschberger11}, we calculated the
co-moving mean and median of the core surface densities as a function of total flux.  We
used a window full width of 40 starless cores in \None\ and 20 starless cores in \Ntwo. 
The results are
shown as the red and blue solid lines in Figure~\ref{fig_mass_vs_surfdens} and clearly highlight
the trend for higher surface densities around higher flux cores.  L1622, however, has too few 
cores for the co-moving mean and median analysis to be useful.  In \None\ and \Ntwo,
the overall trend is unchanged with the window width adopted -- we used the largest width
feasible in order to best highlight the overall trend while decreasing random variations.  
Independent of the co-moving windows, we also used a two-sided KS test
to see whether or not the apparent trends were significant.  In each region, we split the cores
into lower and higher flux subsets for both the full sample of dense cores and only
the starless dense cores, and calculated the probability that the corresponding
two sets of surface densities were drawn from the same parent sample.  
The resulting probabilities confirm what was visually suggested 
in Figure~\ref{fig_mass_vs_surfdens}, namely that \None\ and \Ntwo\ show strong signs of dense core
mass segregation, with probabilities of less than 10$^{-5}$ regardless of how the high and low
flux samples are split and using either N=5 or 10 for the surface density calculation.
L1622 has suggestive hints of
the same trend when the protostellar cores are included in the analysis, but again has 
too few cores for this result to be statistically significant on
its own (probabilities larger than a few percent).

\section{Other Clustering Properties}
Moving away from mass segregation measurements, we also characterize the general clustering
properties of each region.  One commonly-used measure of such properties
is the two-point correlation function or
the closely related surface density of companions \citep[e.g.,][]{Gomez93,Larson95,Simon97}.
The two-point correlation function, $w$, is defined as
\begin{equation}
w(\theta) = N_p(\theta) / N_r(\theta) - 1
\end{equation}
where $N_p$ is the number of pairs of sources at a given angular separation, $\theta$ and
$N_r$ is the mean number of pairs of a random set of sources distributed over the same
observed area \citep[e.g.,][]{Gomez93}.  The surface density of companions, $\Sigma(\theta)$,
can be calculated as 
\begin{equation}
\Sigma(\theta) = (N_{tot}/A) \times (1 + w(\theta))
\end{equation}
where $N_{tot}$ is the total number of sources and $A$ the total area observed \citep{Simon97}.

\begin{figure}[htb]
\begin{tabular}{ccc}
\includegraphics[width=2.05in]{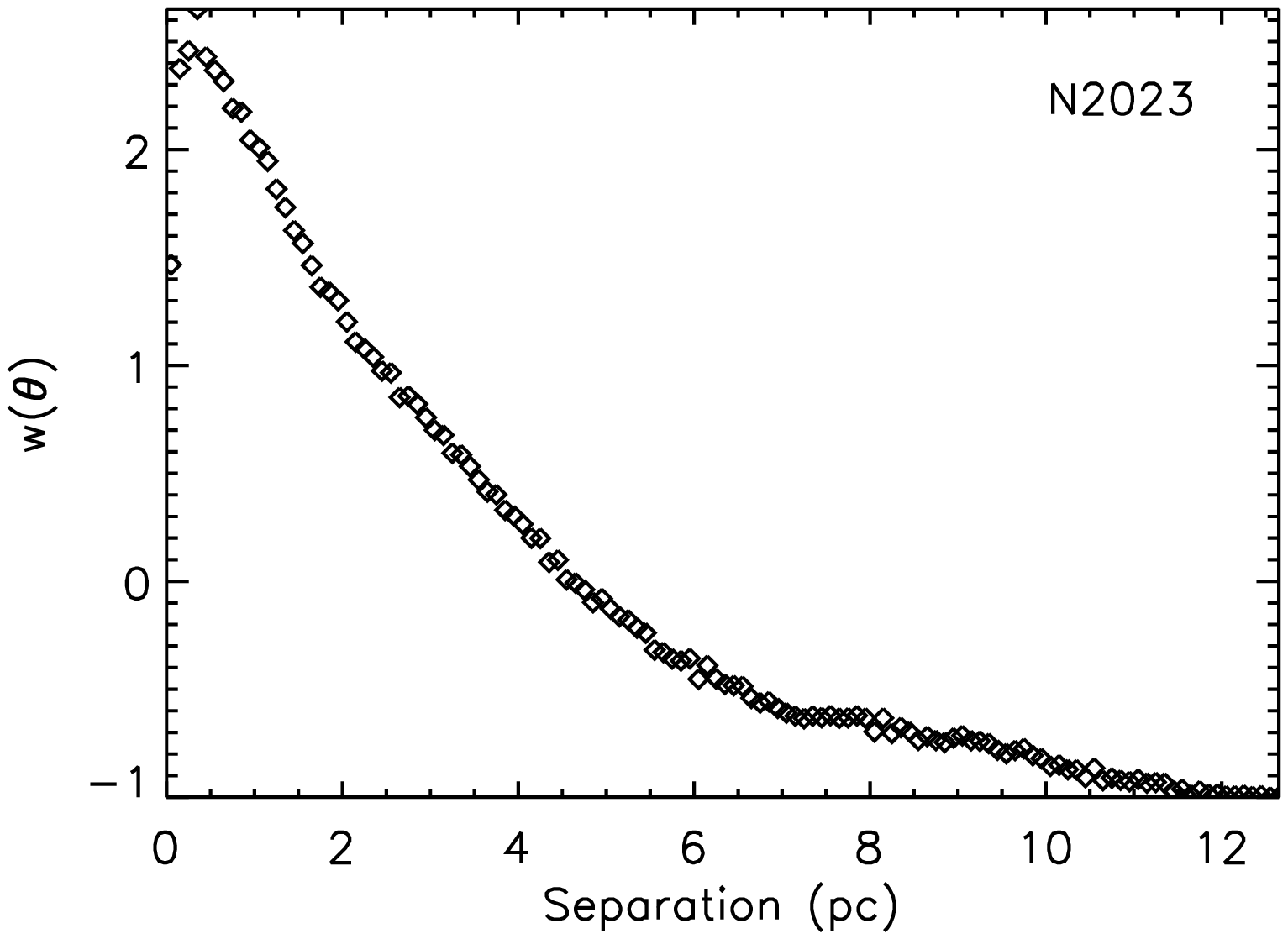} &
\includegraphics[width=2.05in]{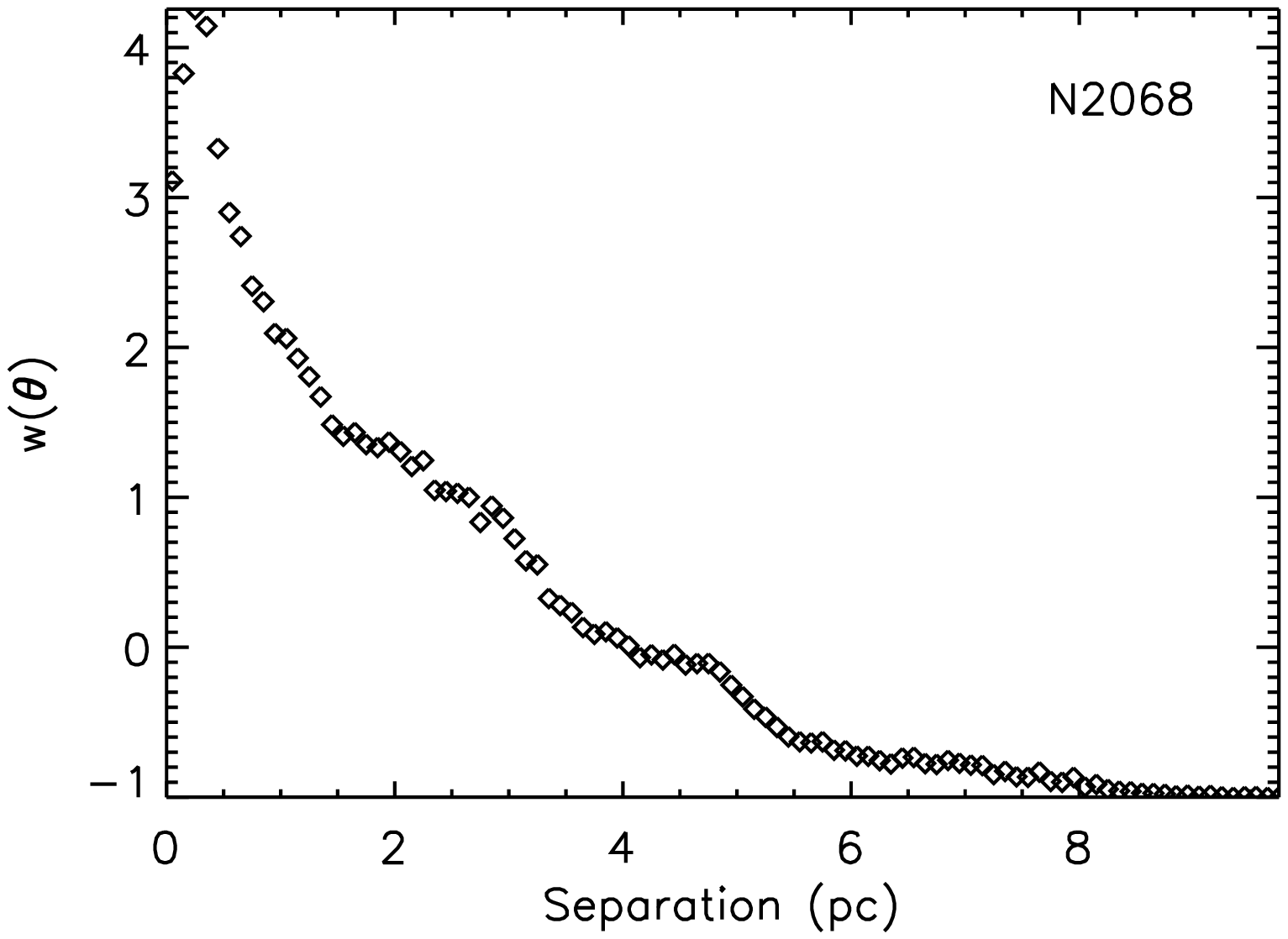} &
\includegraphics[width=2.05in]{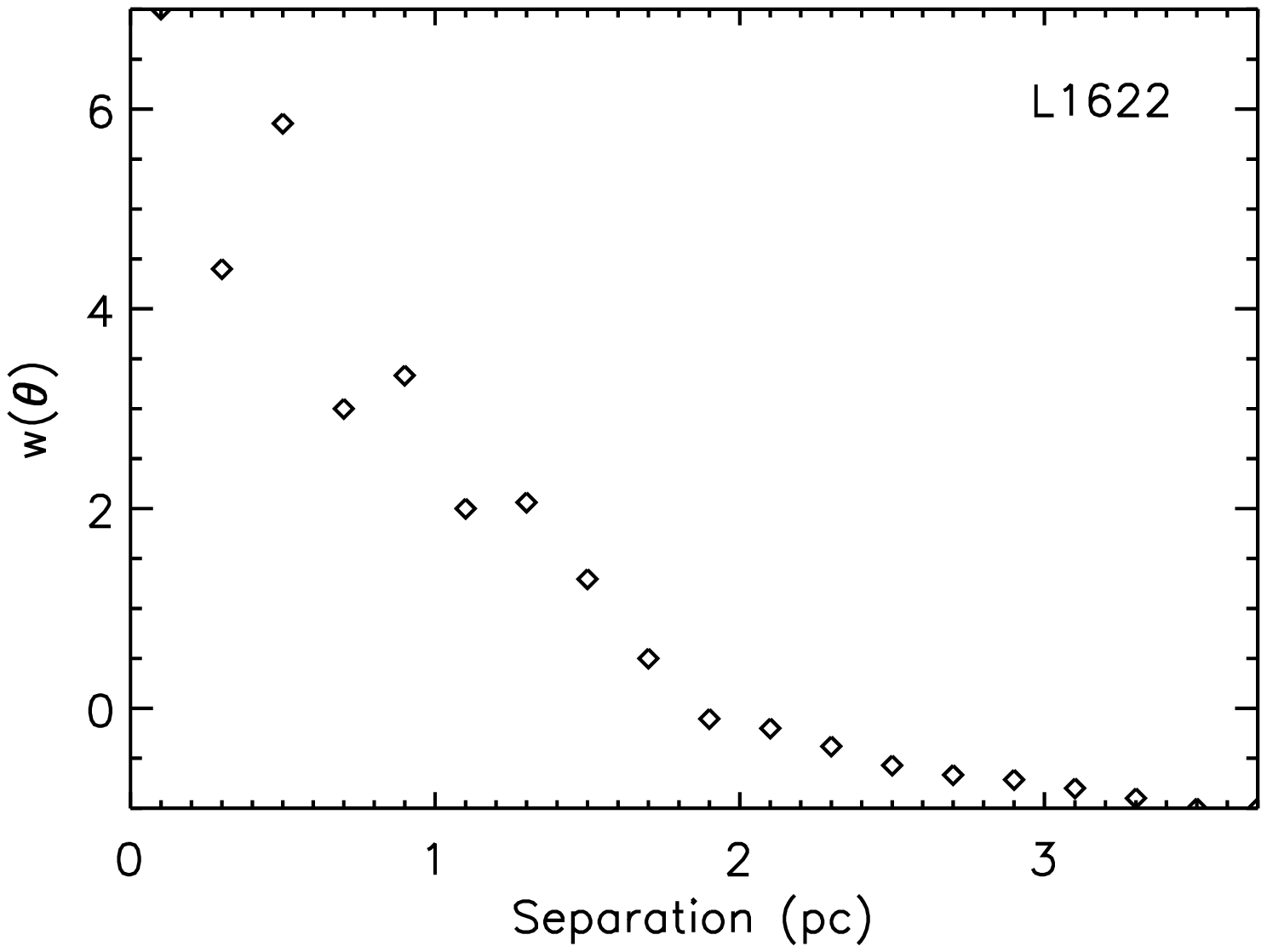} \\
\includegraphics[width=2.05in]{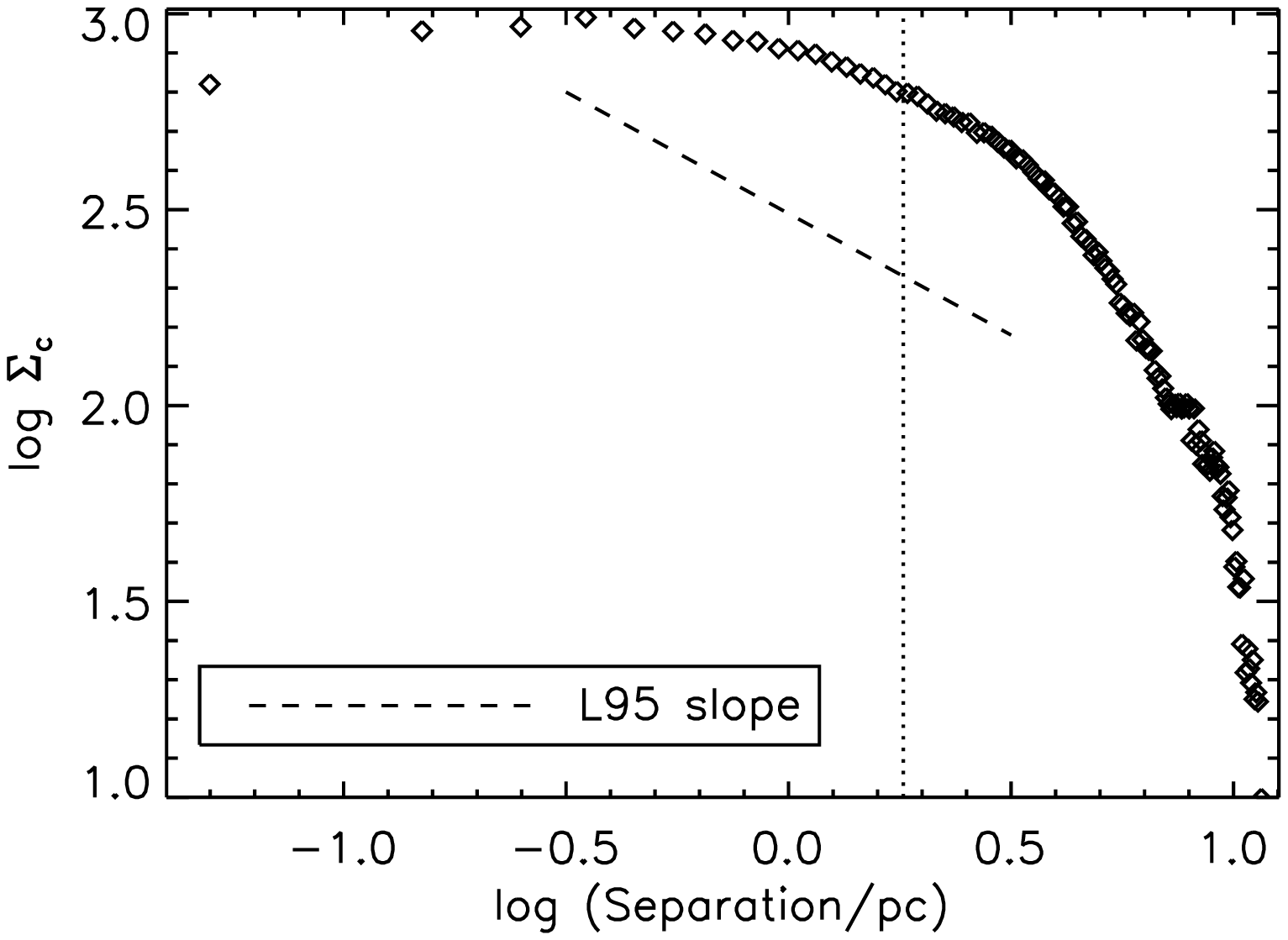} &
\includegraphics[width=2.05in]{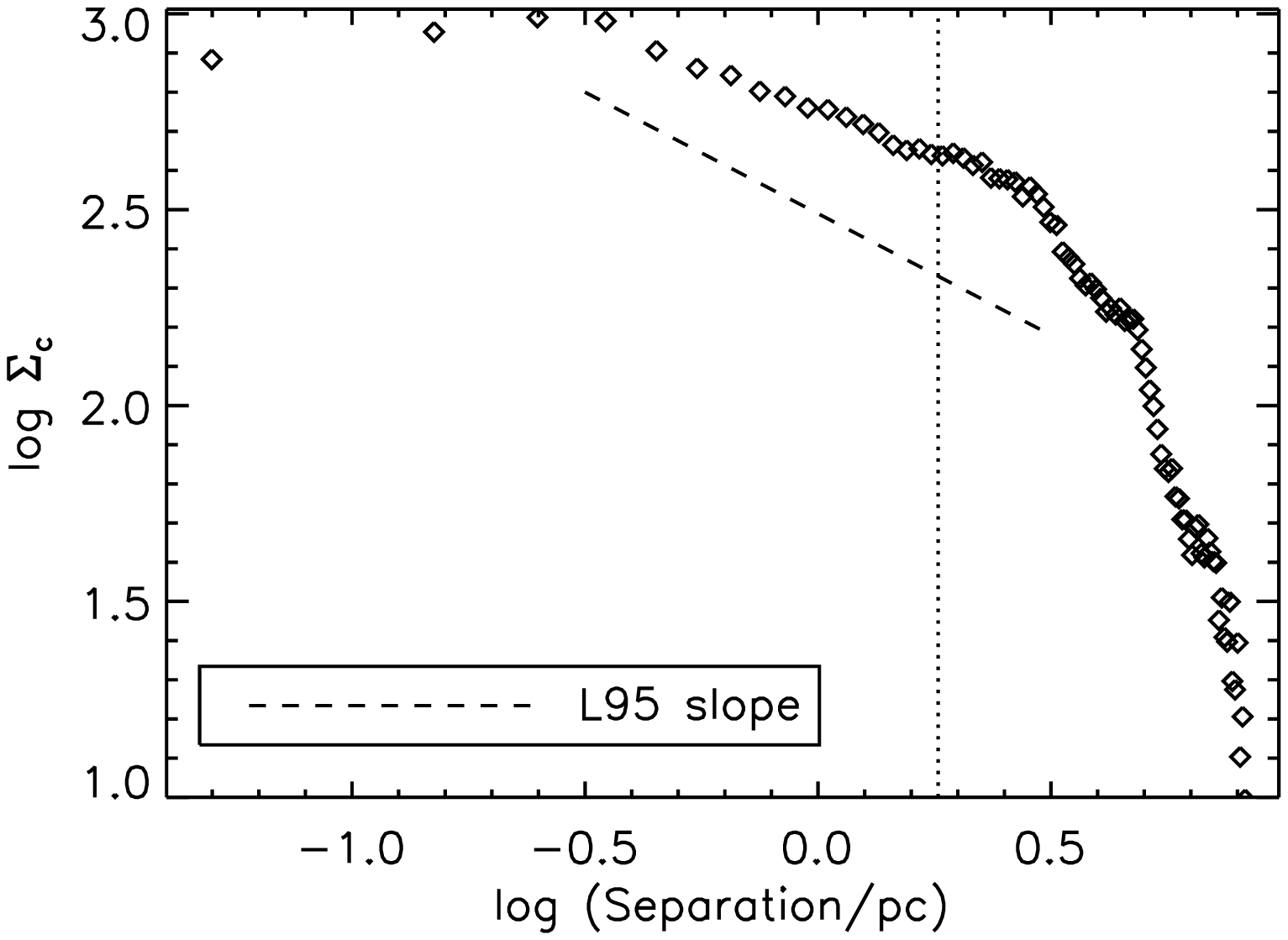} &
\includegraphics[width=2.05in]{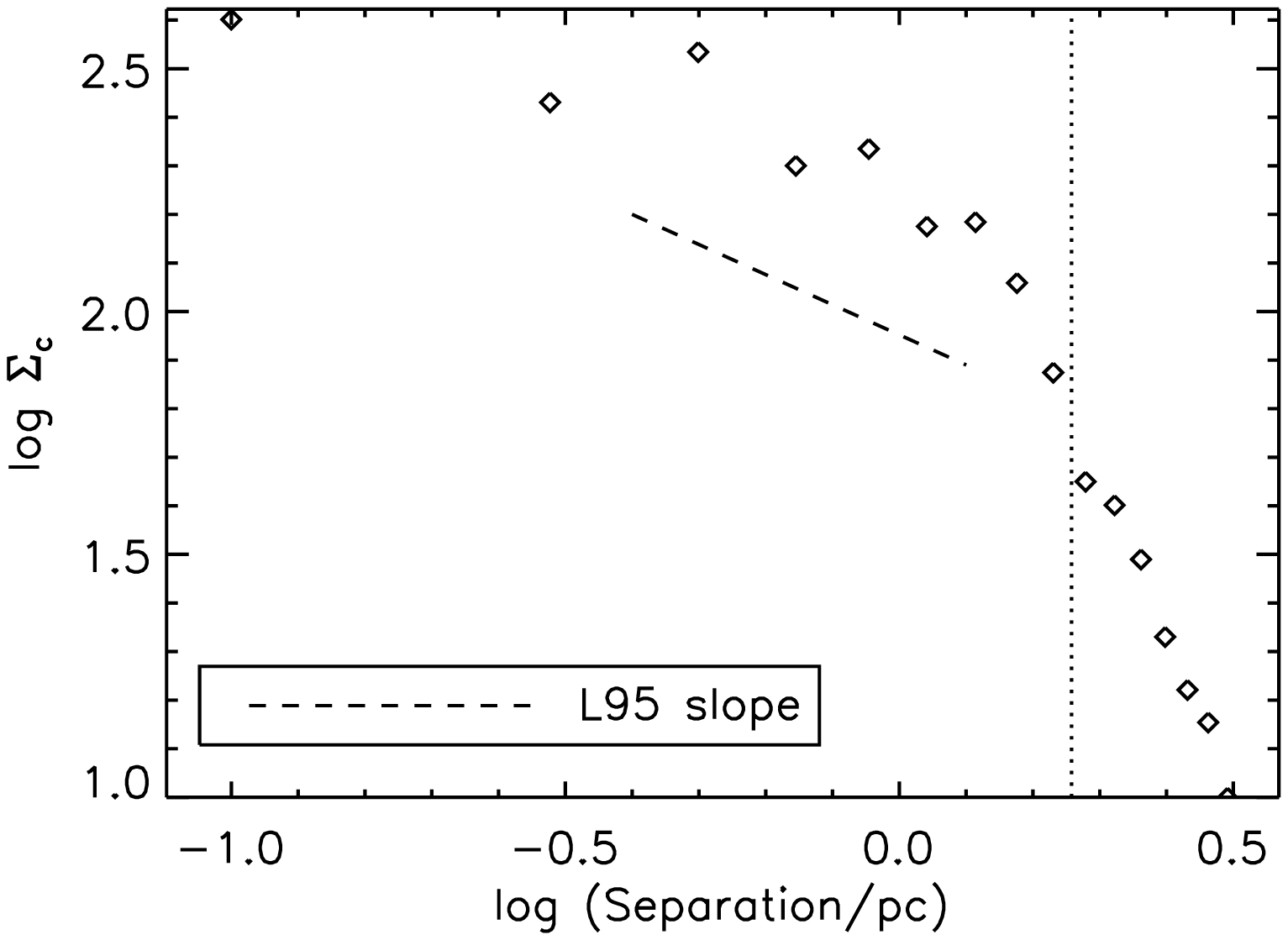} \\
\end{tabular}
\caption{The two-point correlation function (top row) and surface density of companions (bottom
	row) for \None\ (left), \Ntwo\ (middle), and L1622 (right).  The top panel is shown with
	a linear scaling to better illustrate substructure in \Ntwo\ while the bottom panel is 
	plotted in the traditional log-log scale.  The dotted line in the bottom panels shows
	the slope \citet{Larson95} fit at larger separations to the distribution of YSOs
	in Taurus.  Note that the smallest one or two separation bins are likely incomplete
	due to the finite size of dense cores.  In addition, the larger scales may be incomplete
	above a separation of $\sim$15\arcmin\ (1.8~pc) due to the finite area observed. 
	This upper scale is 
	indicated by the vertical dashed line in the bottom panels.} 
\label{fig_twopoint}
\end{figure}

We calculate $w(\theta)$ and $\Sigma(\theta)$ for each of the three regions in Orion~B, and
show the results in Figure~\ref{fig_twopoint}.  Two-point correlation functions effectively
measure the degree to which sources are clustered together beyond what would be expected from
a random distribution.  \citet{Davidge12}, for example, shows that a single cluster
embedded in a lower density halo shows a change in $w$ between the two regimes.  
Generally, two-point correlation functions are calculated for
point sources, which our dense cores are not.  Therefore, caution must be exercised at the
smallest separation bins: clustering cannot be measured effectively 
on scales smaller that the size of 
a dense core.  Since there is a range in core sizes, there is not a single cutoff value
to apply to this distribution, but we estimate that the lowest two separation bins in
Figure~\ref{fig_twopoint} may be affected.  The lowest bin is certainly incomplete, due to the 
finite resolution of the observations.  As with any two-point correlation function, the
upper end of the distribution also becomes incomplete due to the finite areal coverage of the 
observations, although this effect is somewhat accounted for in $w$ by comparison to a random 
distribution over an identical area.  Within these two separation boundaries, we see that
each of the three regions have a two-point correlation function that drops off steeply
with increasing separation.  All three have values of $w(\theta)$ above about 1 for size
scales of $\sim$1.5~pc ($\sim 12$\arcmin), 
which roughly corresponds to the visual width of clusters within
each region.  In \None\ and L1622, $w(\theta)$ drops off fairly regularly with increasing
separation, whereas \Ntwo\ shows several distinct bumps in the distribution.  
A visual
inspection of the \Ntwo\ two-point correlation function indicates that the 
bumps correspond to typical size scales between clusters
in the region.  For example, the bump at 2 to 3~pc separations indicates the projected
distance between neighbouring clusters, while the bump at
4 to 5~pc corresponds to more distant pairings of clusters.  Although the main clusters of
cores appear to be regularly spaced on first glance, a careful examination reveals a 
variation in separations of at least 50\%, suggesting that the cluster-cluster separation
cannot be explained in terms of simple thermal fragmentation analogous
to that suggested on a smaller scale by \citet{Teixeira06} in NGC~2264.  
The clusters in \None\ are even
less regularly spaced, explaining the smoothness in $w$ for that region, while L1622
only includes one obvious cluster.

\citet{Larson95} analyzed the surface density of companions of YSOs, $\Sigma$, in Taurus,
and found two distinct power-law relations, with a break at $\sim 0.04$~pc, and a slope
above this of $\Sigma \propto \theta^{-0.62}$.  \citet{Simon97} found a similar result
in YSOs in Ophiuchus, with a slope of --0.5, and a shallower slope in the Trapezium of
--0.2, but argues that the three distributions are consistent with a value of --0.5 
given the uncertainties.  \citet{Larson95} and \citet{Simon97} interpret a single
power law distribution as corresponding to an underlying fractal dimension of the molecular
cloud of about 1.5 (2 plus the slope). 
Our measurements of $\Sigma$ are shown in the bottom panels of 
Figure~\ref{fig_twopoint}.  At face value, none of the three regions appear to be 
consistent with a single power law relationship.  As mentioned earlier, however, it is
important to consider the area observed properly.  Individual observations used in our
mosaic are well-sampled to a diameter of $\sim$30\arcmin, and in all three regions, this
is a reasonable approximation to the narrowest map dimension observed.  
The surface density of companions,
therefore, may not be well measured for separations above half of this diameter, or 15\arcmin.  
The vertical lines in Figure~\ref{fig_twopoint} show this limiting separation.  Much of the
curvature seen in $\Sigma$ becomes apparent only above the limiting separation, suggesting
that where the data is most reliable, a single straight power law with a slope similar to
\citet{Larson95} or \citet{Simon97} is a reasonable approximation.  
Reassuringly, \citet{Johnstone01} found a similar slope in their analysis of SCUBA
observations of part of the \Ntwo\ region, with their best fit of 
$w(\theta) \propto \theta^{-0.75}$.
It will be interesting
to make this measurement for the larger contiguously observed regions in the GBS, such as
Orion~A, to see whether or not a single power law is representative of dense core clustering at
the largest scales.

Next, we measured the clustering $Q$ parameter following \citet{Cartwright04}
for each of the three regions observed.
$Q$ is calculated as $\bar{m} / \bar{s}$, where $\bar{m}$ is the mean MST branch length,
normalized by the factor $\frac{N-1}{\sqrt{N \times A}}$, $N$ is the number of cluster
members, $A$ is the area of the cluster, and $\bar{s}$ is the mean separation of sources,
normalized by the cluster radius, $R_{circ}$.
\citet{Cartwright04} find that values of $Q$ above 0.8 correspond to clusters described by
a single large-scale density gradient, while values of $Q$ below 0.8 correspond to 
sub-clustered (fractally distributed) sources.  We calculate $A$ as the area contained
within the convex hull\footnote{A convex hull is a shape with the minimum area required
to enclose a set of points under the condition that all neighbouring boundary segments
form $> 90$~degree angles.}
of each cluster, and $R_{circ}$ as the maximum separation between
a source and the cluster centre (calculated as the mean of the convex hull positions) 
following \citet{Gutermuth09}.  We find $Q$ values of 1.18, 0.99, and 0.91 for
L1622, \None, and \Ntwo, respectively.  All three values of $Q$ nominally indicate
a lack of subclustering, which is puzzling for \None\ and \Ntwo\ where subclusters are
clearly visible.  It is reassuring, however, that L1622, with 
no obvious subclusters, has the largest 
value of $Q$. 
We first rule out two possibilities raised by \citet{Cartwright09a} and \citet{Cartwright09b}
for unexpected values of $Q$.  Namely, none of the regions show signs of `anti-clustering', 
as indicated by normalized mean separation of sources above 0.8, or are sufficiently elongated
to require a correction to the observed $Q$, as required for axial ratios greater
than 3 and $ 0.9 < Q < 1 $.
There are several reasons why the values of $Q$ we measure are generally
higher than expected.  First, our observations of the \None\ and \Ntwo\ regions each
cover multiple groupings of cores that are somewhat linearly spaced and not circularly
distributed.  This spatial distribution makes the 
measurements of quantities such as $R_{circ}$ much more 
difficult to define and measure.  We tried calculating $R_{circ}$ as
the maximum separation from the mean source position, as well as half of the maximum
separation between all cores.  The former definition of $R_{circ}$ increases
the estimated $Q$ values (up to 13\% higher), while the latter decreases the estimated
$Q$ values (about 2\% lower).  While none of these variations place the value of
$Q$ measured for \None\ and \Ntwo\
into the `subclustering' regime, they do highlight the fact that such non-circular
core locations can make it difficult to measure $Q$ accurately.
A second possibility is that the application
of a point-source measure like $Q$ to dense cores either introduces a larger 
source of error or changes 
the value at which subclusters versus a single radial density gradient are indicated.
Testing using synthetic distributions of clusters of cores (beyond the scope
of this work) would be necessary to determine if and how $Q$ should be re-calibrated
for this application.

\section{DISCUSSION}
\label{sec_disc}

\subsection{Clustering Measurement Methods}
The existence of
mass segregation in young stellar clusters, and whether it is attributable to a primordial
distribution or dynamics, is an ongoing debate.  \citet{Girichidis12} point out that 
major challenges are the coupled problems of determining whether sub-clustering 
is present and how to define mass segregation itself.  The local 
dynamical timescale for a sub-cluster can be significantly
smaller than the global dynamical timescale, and therefore a region can become mass
segregated in less than a global dynamical timescale due to stellar interactions within
the sub-clusters \citep{Allison09_Orion, Moeckel09, Maschberger11,Girichidis12}. 
Tied to this issue is whether the mass segregation itself
is measured locally or globally.  For example, \citet{Kirk11} find that YSOs within 
Taurus are mass segregated by examining the relative location of the most massive YSO
within each sub-cluster.  At the same time, \citet{Parker11} find that the entire Taurus
complex is {\it inversely} mass segregated: the most massive YSOs are further apart from
each other than typical YSOs are in the complex.  Taurus has an unusual YSO spatial
distribution,
with largely separated sub-clusters (especially those harbouring the most
massive YSOs in the complex), and very few YSOs found anywhere near its geometric 
centre.  In the case of Taurus, the very dispersed distribution of YSOs 
coupled with the
near-thermal velocity dispersion of material \citep[e.g.,][]{Seo15} make it clear that the 
\citet{Kirk11} and \citet{Parker11} 
analyses are tracing very different scales of processes.  The age of the YSOs in Taurus 
is much less than the region crossing-time, and so sub-clusters across Taurus have not had time to
interact.  In other regions, the scale appropriate for measuring a cluster's mass segregation
may be less obvious.

\citet{Parker15} recently compared three different mass segregation analysis
techniques: the \citet{Allison09} $\lambda_{MST}$ technique 
\citep[used in][discussed above]{Parker11}, the \citet{Maschberger11} $M$--$\Sigma$
method discussed in Section~\ref{sec_clusters_msig}, and the \citet{Kirk11} MST-group 
analysis discussed
in Section~\ref{sec_clusters_mst}.  
The $\lambda_{MST}$ method effectively considers a population of sources
as a single cluster, and determines whether or not the more massive sources are located closer
together than a comparable random sample of sources.  The $M$--$\Sigma$ method allows for
the possibility of a monolithic cluster or sub-clustering, and measures whether or not more massive
sources tend to be located in regions of higher than average surface density of sources.  The 
MST-group analysis tends to assume that a system is made up of distinct sub-clusters, and
examines whether or not in each one the most massive source is more centrally-located than is
typical.  

In their study, \citet{Parker15} create a fractal YSO distribution with
several different assignments of mass to the YSOs,
and test the ability of each of the three techniques to measure correctly the
presence or absence of mass segregation.
As expected, the $\lambda_{MST}$ method fares
best when the mass segregation is set up in a way that matches how the method
detects it e.g., closely located massive stars.  Also, the $M$--$\Sigma$ method fares best when
the most massive stars are inserted in regions
of high local surface density.  \citet{Parker15} did not include a third test tuned to 
the strengths of the MST-group method, i.e., centrally-located most massive members in 
sub-clusters, so it is not surprising
that this method fared the worst of the three in their comparisons.  \citet{Parker15}
correctly point out that the MST-group method tends not to measure mass segregation for
all of the $N$ most massive sources (specifically ten in their synthetic distributions),
since larger sub-clusters tend to contain multiple more-massive sources while smaller
sub-clusters may only contain relatively low mass sources.  Instead, the MST-group
method makes measurements based on the relative masses of each sub-cluster.  This situation
highlights the importance of clarity in defining what is meant by mass segregation
and how it is being measured.  

Another point alluded
to by \citet{Parker15} is that the MST-group method can only produce reliable results if
the sub-clusters identified are distinct physical entities.  While subclustering
can be difficult to
determine in practice, one important component lacking in the \citet{Parker15} analysis
is a careful consideration of uncertainties in sub-cluster membership based on the MST-based
criterion for group definition.  If sub-cluster membership is highly uncertain
(due to, say, uncertainties in \Lcrit), 
very different mass segregation estimates result, then the MST-group results should be 
treated as being questionable.
This uncertainty underlines the
importance of ensuring the sub-clusters are well-defined (preferably also 
visually distinct) before embarking on any further analysis.  Conversely, when there are
obviously distinct sub-clusters, the $\lambda_{MST}$ method (unless applied separately
to each sub-cluster) is then better at tracing bulk properties of the region imprinted
at formation and not the present mass segregation in each sub-cluster.  Where 
distinct sub-clusters are present, 
as in the Taurus example discussed above, the typical
separation between the most massive members measured by the $\lambda_{MST}$
technique is more influenced by the spacing between
sub-clusters harbouring massive members than it is to the local mass segregation.  

In \None\ and \Ntwo, there are very visually distinct groupings of dense cores
separated by several parsecs.  HARP CO(3--2) observations show a typical linewidth
of C$^{18}$O / $^{13}$CO of 1 -- 3~km~s$^{-1}$ across the two regions \citep{Buckle10}.
This dispersion is likely to be higher than that of the denser gas, and therefore
can be used to provide a lower limit to the interaction timescale between the groups 
of roughly 1~Myr, i.e., the time to travel several pc at several km~s$^{-1}$.  
Dense cores detectable at submillimetre wavelengths have an estimated lifetime
several times smaller than this, i.e., several tenths of a Myr 
\citep[e.g.,][]{Enoch08,Hatchell07,Kirk05}, although recent {\it Herschel} results
in Aquila suggest that the full population of dense cores there may have a lifetime closer to 
1~Myr, while higher density cores (n$_{H2} > 10^5$~cm$^{-3}$) have lifetimes of 0.4~Myr or
less \citep{Konyves15}.  
While these estimates are necessarily approximate,
they do suggest that there has been insufficient time for many interactions between the
clusters we identify, and hence that it is sensible to characterize mass segregation
on the scale of these clusters, rather than the region as a whole.
The fact that our error analysis (see Appendix~B) shows our results are robust to 
reasonable variations 
in the cluster definitions for the MST-group analysis, and that a similar conclusion
is reached using the independent $M$--$\Sigma$ method supports our findings of mass
segregation within these dense core clusters.

\subsection{Implications}

Dense cores are in a particularly interesting regime to measure clustering properties.
On the one hand, they provide the initial conditions for the formation of (proto)stellar
clusters and should be taken into account in cluster modelling.  On the other hand,
dense cores are themselves the product of the history of substructure formation
within the molecular cloud, and continue to evolve through processes including
mass accretion and gravitational collapse, and their present-day properties
may provide some insight into these.  To the best of our knowledge, predictions
of the clustering properties of dense cores have not yet been made either theoretically
or using numerical simulations.  We therefore suggest that our observations be included
as a benchmark test in future work.  It will also be helpful to expand the measurement 
of dense core clustering properties to additional systems where the local environment
(or even the age of the system) may have influenced the present-day appearance
of dense core clusters in a different way than in Orion.

The study of clustering properties in stellar and protostellar systems
has been well studied both theoretically and observationally.
Numerical simulations have varied predictions on whether or not 
there is primordial mass segregation
in clusters. 
Some simulations find mass segregation of stars / sink cells in clusters from very early times
\citep[e.g.,][]{Maschberger11,Girichidis12,Kirk14,AMyers14}. 
Other simulators find no evidence of mass segregation \citep[e.g.][]{Parker14}, while
\citet{ParkerDale15} find that evidence for or against mass segregation depends on the method
used to measure it.

Observations of mass segregation are somewhat divided, 
although it appears that the majority of studies
do indicate some degree of mass segregation is common.
Young stellar clusters tend to show signs of present-day mass segregation,
which, given the approximate age of the systems, is often interpreted as implying that the 
segregation is at least partially primordial
\citep[e.g.,][]{Carpenter97,Hillenbrand98,BonnellDavies98,Gouliermis04,Stolte06,Kirk11,Gennaro11,Davidge15}. 
Note, however, that in more crowded systems, there are several observational biases which need
to be carefully considered to measure true mass segregation \citep{Ascenso09}.
Several recent studies, however, do not find evidence of primordial (and in some cases, 
even present-day) mass segregation \citep[e.g.,][]{Allison09_Orion,Allison10,Parker12,Wright14}. 

Looking at even younger systems, observations appear to favour mass segregation.
\citet{Megeath05} and \citet{Hunter06} find small, tight groupings of very young, 
high-mass, protostars in W3 IRS-5 and NGC~6334, respectively, which are surrounded by a 
larger cluster of low- and intermediate- mass stars \citep[see, however,][for an example of
a young protocluster system where the most massive member is not centrally located]{Hunter14}.
Similarly, \citet{Kryukova12} and \citet{Elmegreen14} examine young protostars
(primarily or entirely Class~I, respectively), and find that the highest luminosity protostars
tend to be found in regions of higher source surface density.  We note that
the \citet{Kryukova12} protostar sample includes Orion~A and B along with several
other nearby star-forming regions.  Since high protostellar luminosity 
is caused by either high intrinsic source luminosity, implying a high-mass protostar is already
present, or a high accretion rate, implying a high-mass protostar is likely forming, these
observations also suggest that some mass segregation is imprinted at the start of the
protostellar formation phase, as insufficient time has passed for there to be subsequently
significant motions \citep{Elmegreen14}.  Furthermore, \citet{Elmegreen14} finds evidence
of `collaborative accretion', i.e., neighbouring protostars tending to have similar luminosities,
perhaps implying they all are accreting from a locally high density zone of material.
\citet{Zhang15} argue for a similar picture based on high-resolution observations of a 
high-mass star-forming infrared dark cloud.  The cluster-forming gas is densest in the centre,
which allows for the most massive stars to (start) form(ing) there first, while lower 
mass star formation
can continue for longer in the lower-density outskirts \citep[see also the discussion in]
[about star formation timescales as a function of final stellar mass]{Myers11}. 

Our observations that dense core clusters are somewhat mass segregated support the hypothesis 
that the centre of forming clusters provide
a more favourable environment for the most massive protostars to form, with 
more dense gas available at the cluster centre for local accretion.  We emphasize
that this does not rely on a one-to-one mapping between dense core mass and final 
protostellar mass, rather merely that the most massive protostars form out of the 
most massive (and densest) dense cores.  The picture of more massive protostars
forming in a cluster centre due to the favourable conditions there is
consistent with the competitive accretion scenario \citep[e.g.,][]{Bonnell04}.
That model, however, typically assumes initially equal mass protostellar seeds to
track subsequent gas accretion, and it is unclear what the spatial distribution of dense cores
would be in the model.  \citet{Krumholz07} argue that radiative effects, while often
overlooked in simulations, can be of key importance to determine how much fragmentation occurs.  

The one aspect that our current study cannot address is the effect of environment.  Our results,
coupled with those of dense cores in Orion~A (Lane et al., in prep.) show that in the 
denser, more active star-forming environments of our local Gould Belt clouds, dense cores
are mass segregated within their local clustered environments.  The next step is to
determine whether this same property holds true in sparser, more quiescent star-forming
regions such as Taurus, which we are in an excellent position to examine with the full
JCMT GBS dataset available.

\section{CONCLUSION}
\label{sec_conc}
We examined the clustering properties of the dense cores in the L1622, \None, and
\Ntwo\ regions of the Orion~B molecular cloud.  In particular, we
focused on mass segregation under the assumption that generally cores with higher total
fluxes will form the highest mass protostars.  
Using two complementary and independent methods, we find that the
dense cores are mass (flux) segregated in Orion B.  A minimal spanning tree analysis
(Section~\ref{sec_clusters_mst}) 
shows that visually apparent clusters tend to have a centrally located
most massive member, and often a general tendency for more massive members being
located closer to the cluster centre.  A comparison of the core mass and local core-core surface
density shows that more massive cores tend to be found in more highly clustered environments.
If the most massive protostars tend to be born within more massive dense cores, our result
implies that mass segregation in stellar clusters may in part be imprinted already in the
dense gas from which they form.  
An analysis of dense cores in the nearby Orion A molecular cloud
also finds similar results.  If dense core mass segregation holds over a wider variety of 
star-forming environments, these data provide a new observational test for 
simulations and theories 
of clustered star formation.  It would also imply that clustering of stars can only be
understood by studying the causes of (dense) substructure in molecular clouds.

\appendix

\section{MST-based Clusters}
\begin{figure}[p]
\begin{tabular}{cc}
\includegraphics[width=2.2in]{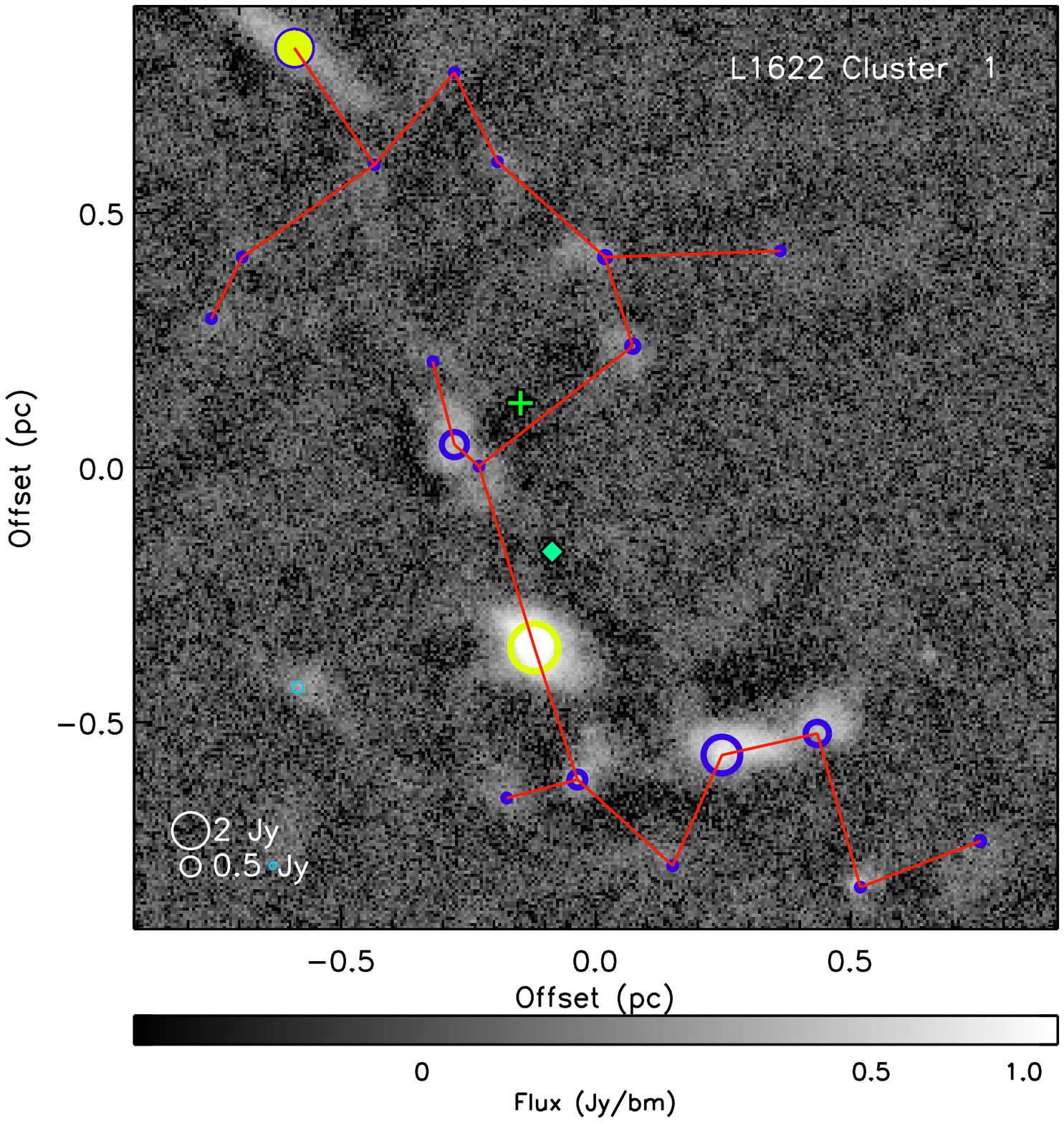} &
\\
\includegraphics[width=2.2in]{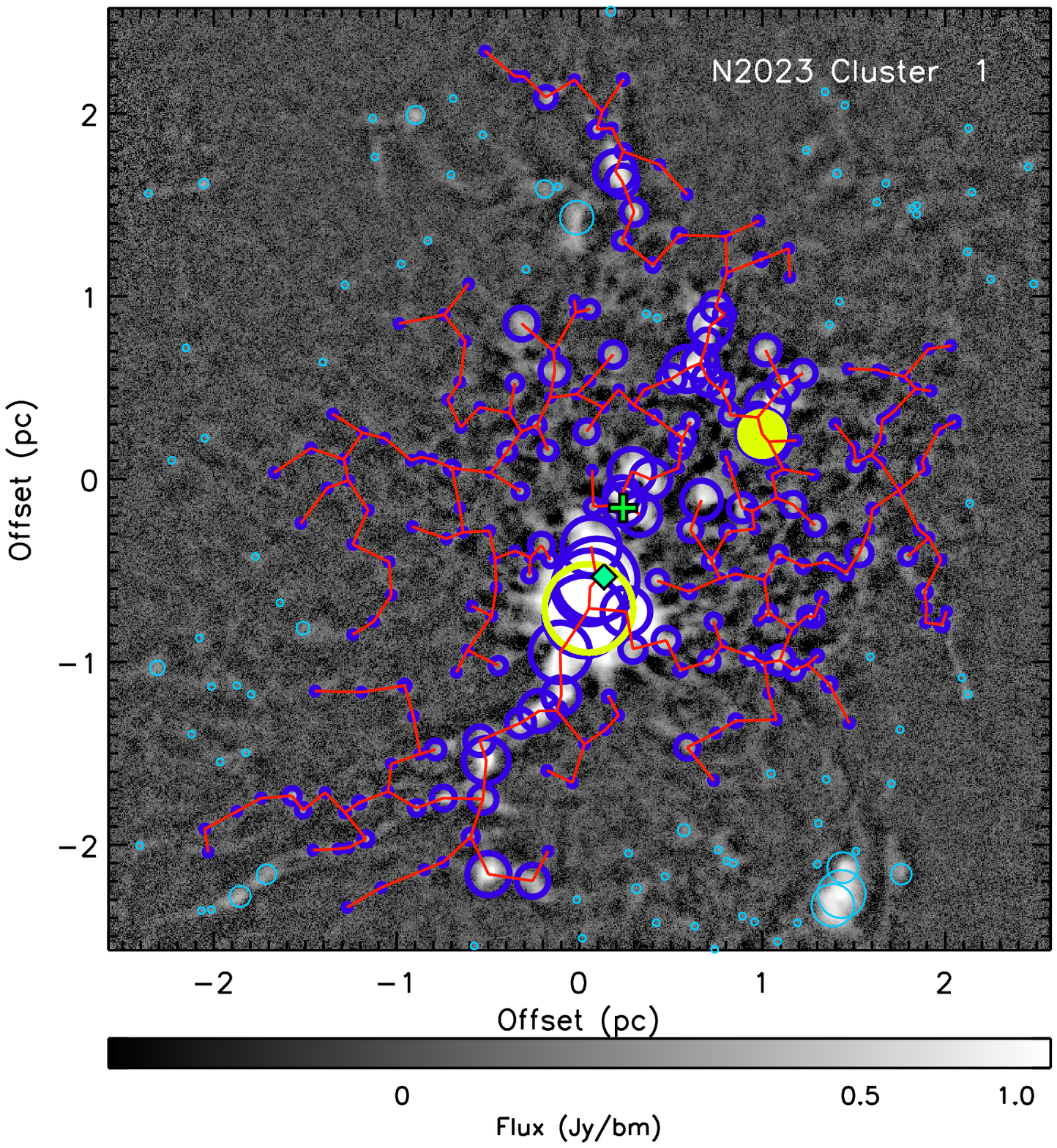} &
\includegraphics[width=2.2in]{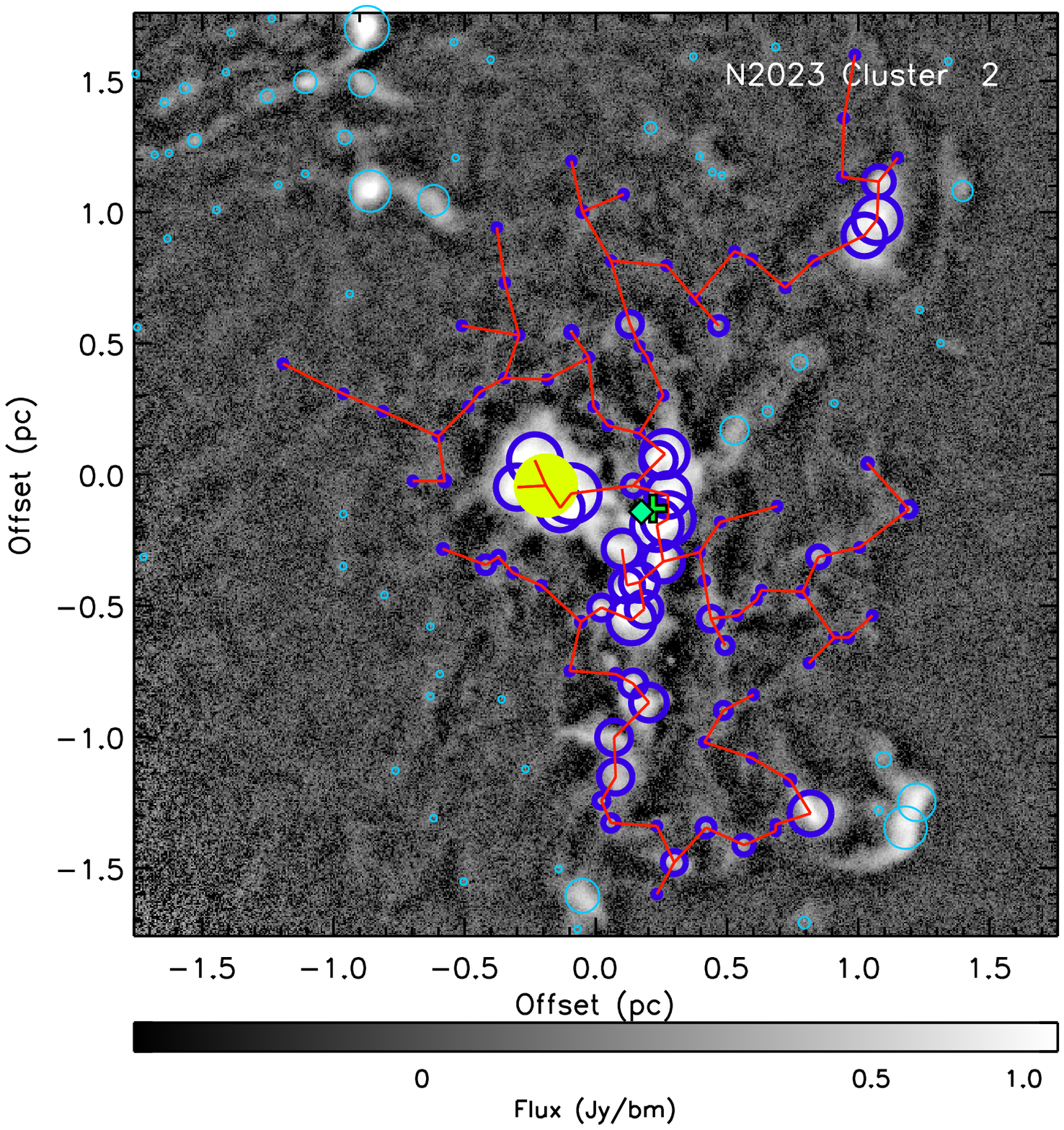} \\
\includegraphics[width=2.2in]{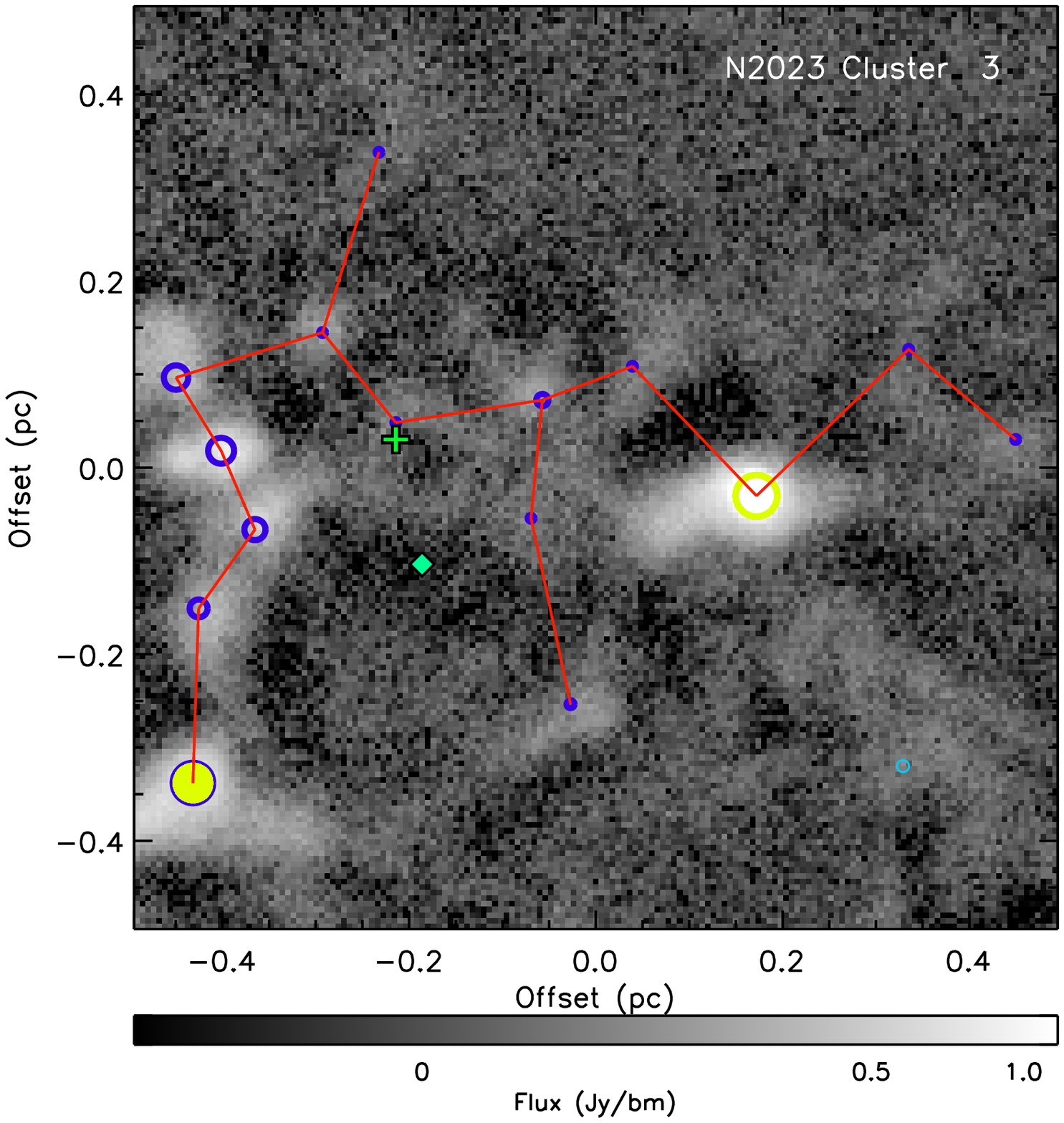} &
\includegraphics[width=2.2in]{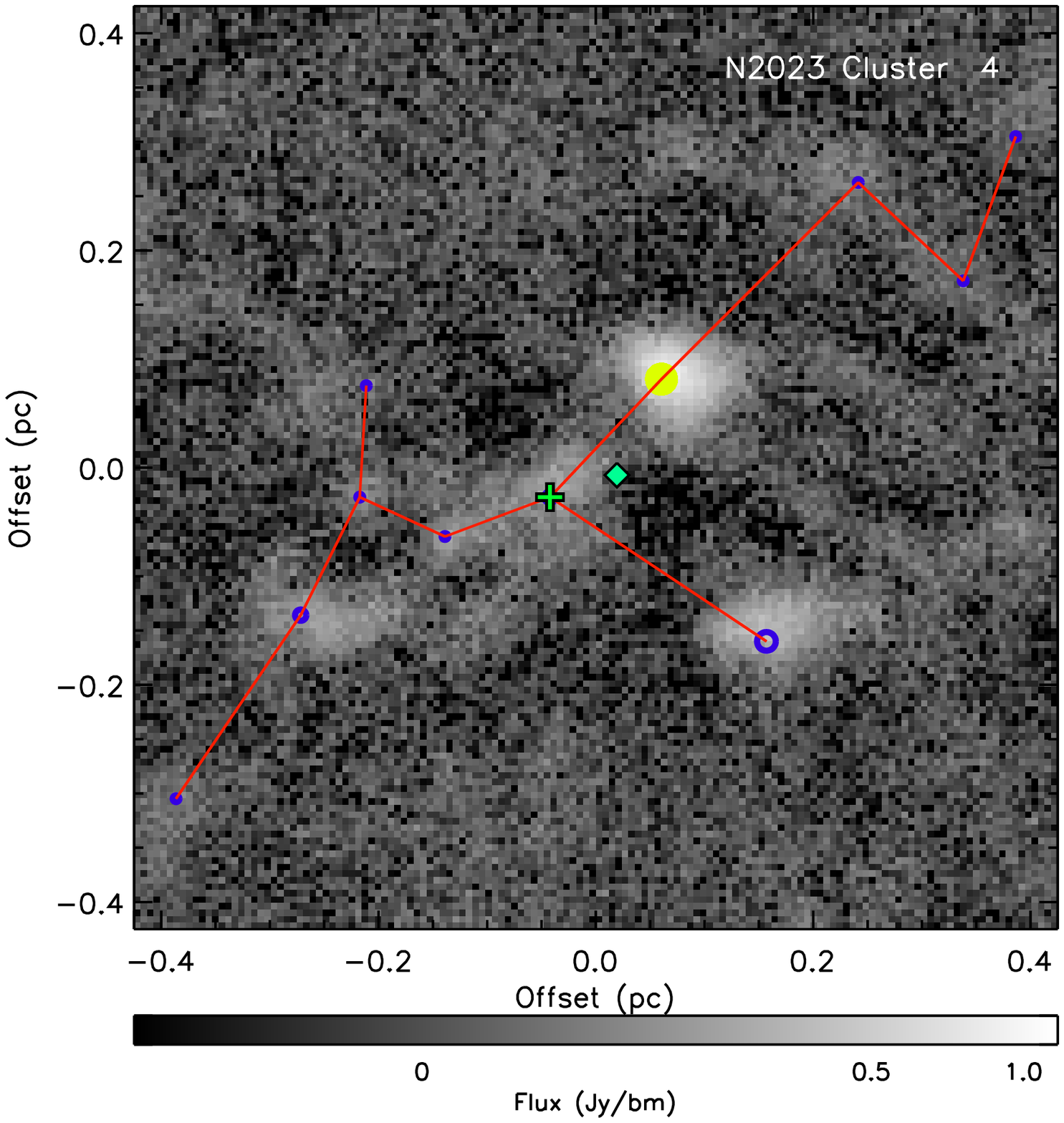} \\
\end{tabular}
\caption{Dense core clusters identified in L1622 (top left panel) and \None\ (remaining
	panels) using MSTs.  The background greyscale images show the SCUBA-2 850~\mum
	map.  The thick blue circles denote the positions of dense cores which
	are members of that cluster, while the thin teal circles denote non-cluster members.
	The size of the circle scales with the total flux (see legend in top left panel).
	The empty and filled yellow circles show the highest flux cluster member and
	starless core cluster member respectively.
	The red line shows the MST structure after branches larger than \Lcrit\
	have been removed.  The green plus symbol shows the cluster centre, calculated as
	the median position of cluster members.  The turquoise diamond symbol shows the
	flux-weighted mean core position.}
\label{fig_mst_clusts1}
\end{figure}

\begin{figure}[p]
\begin{tabular}{cc}
\includegraphics[width=2.7in]{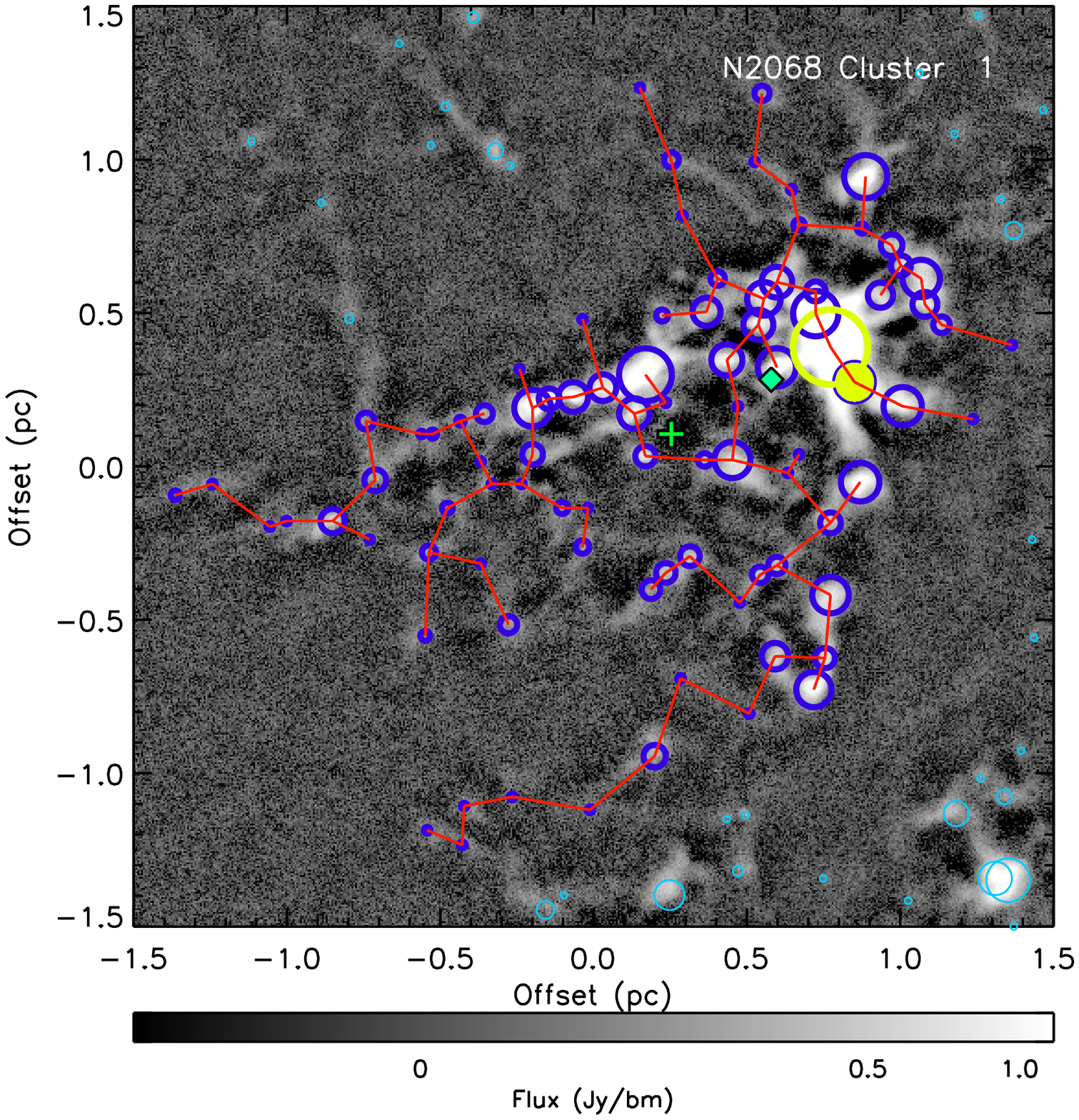} &
\includegraphics[width=2.7in]{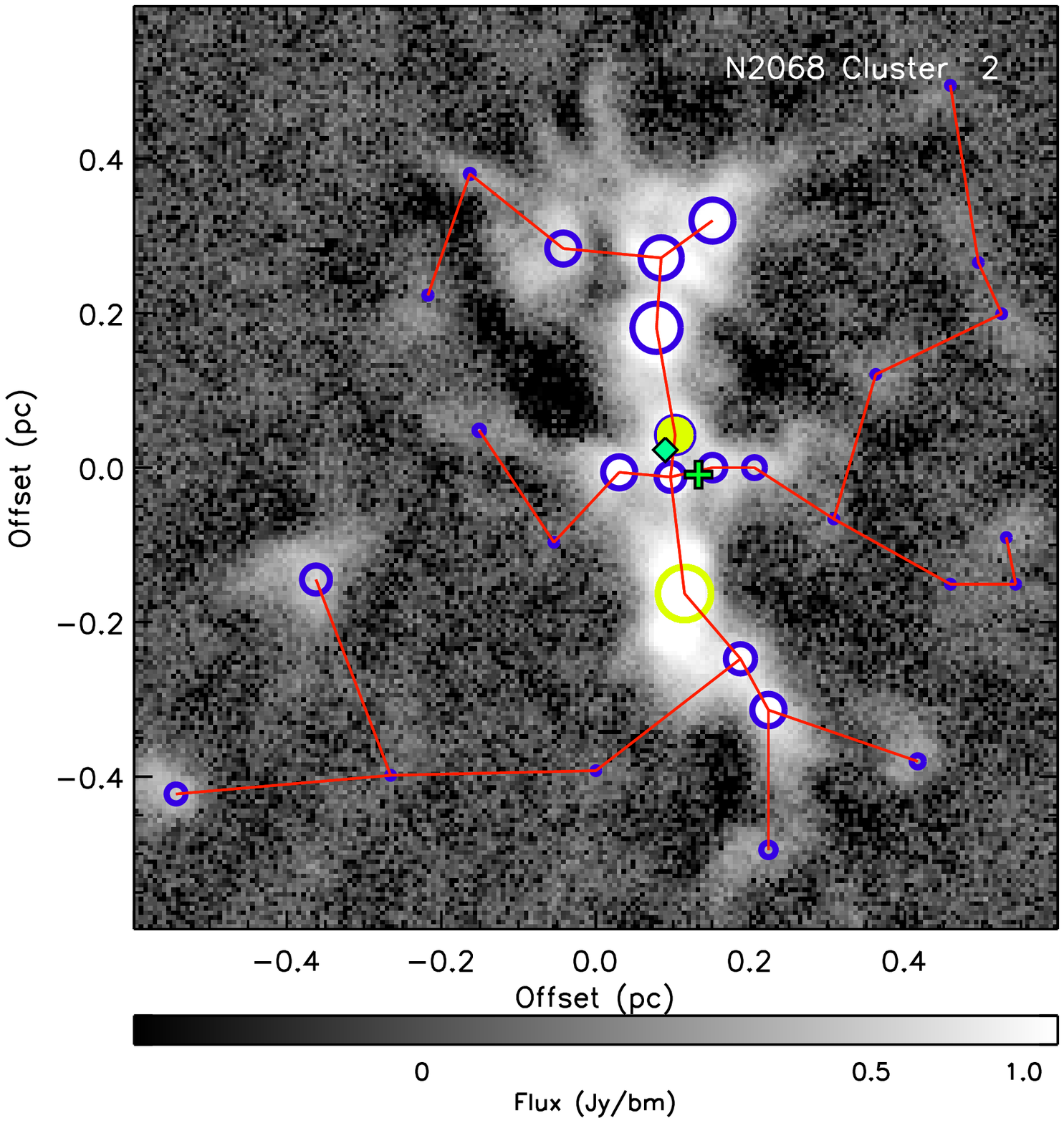} \\
\includegraphics[width=2.7in]{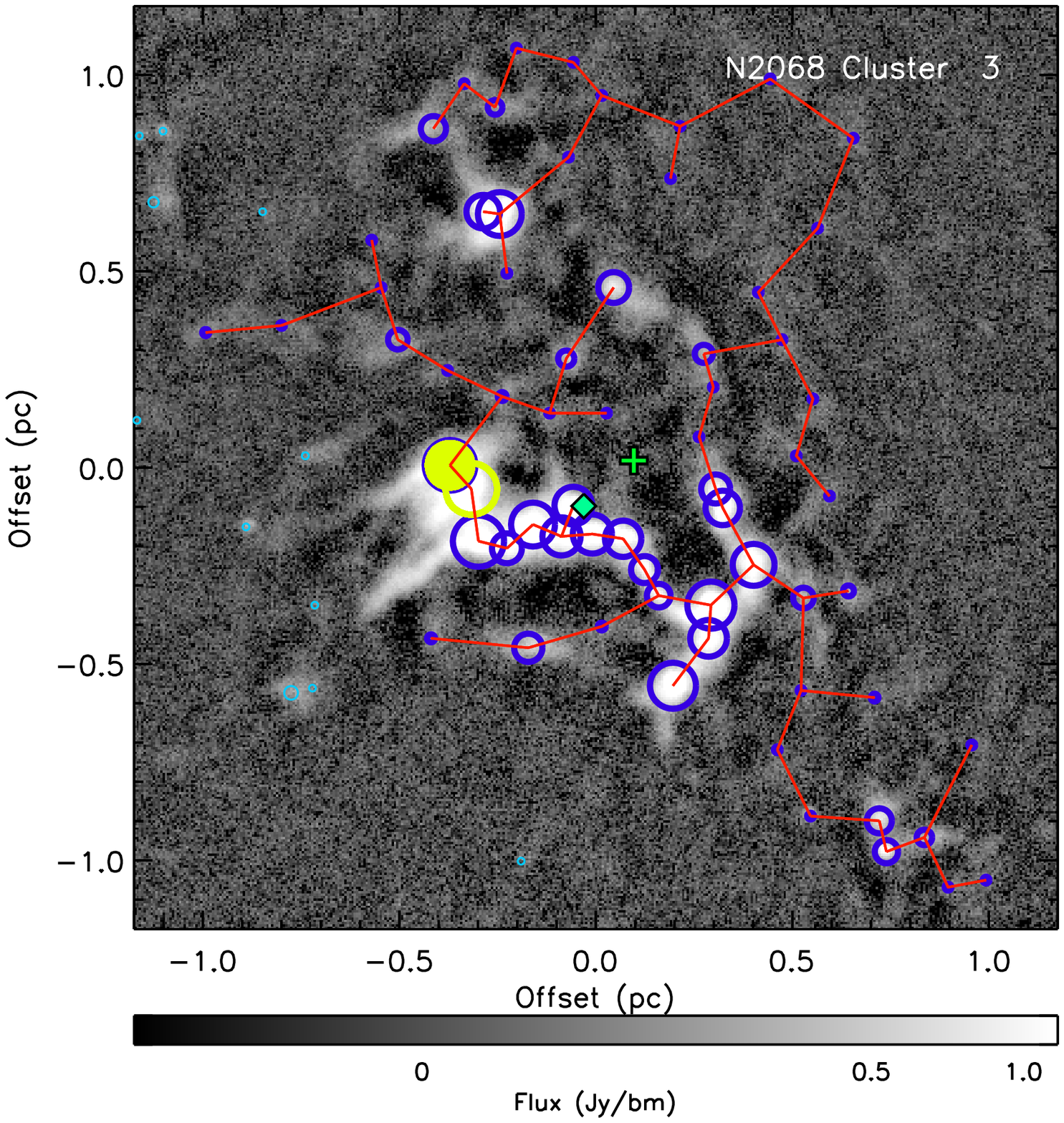} &
\includegraphics[width=2.7in]{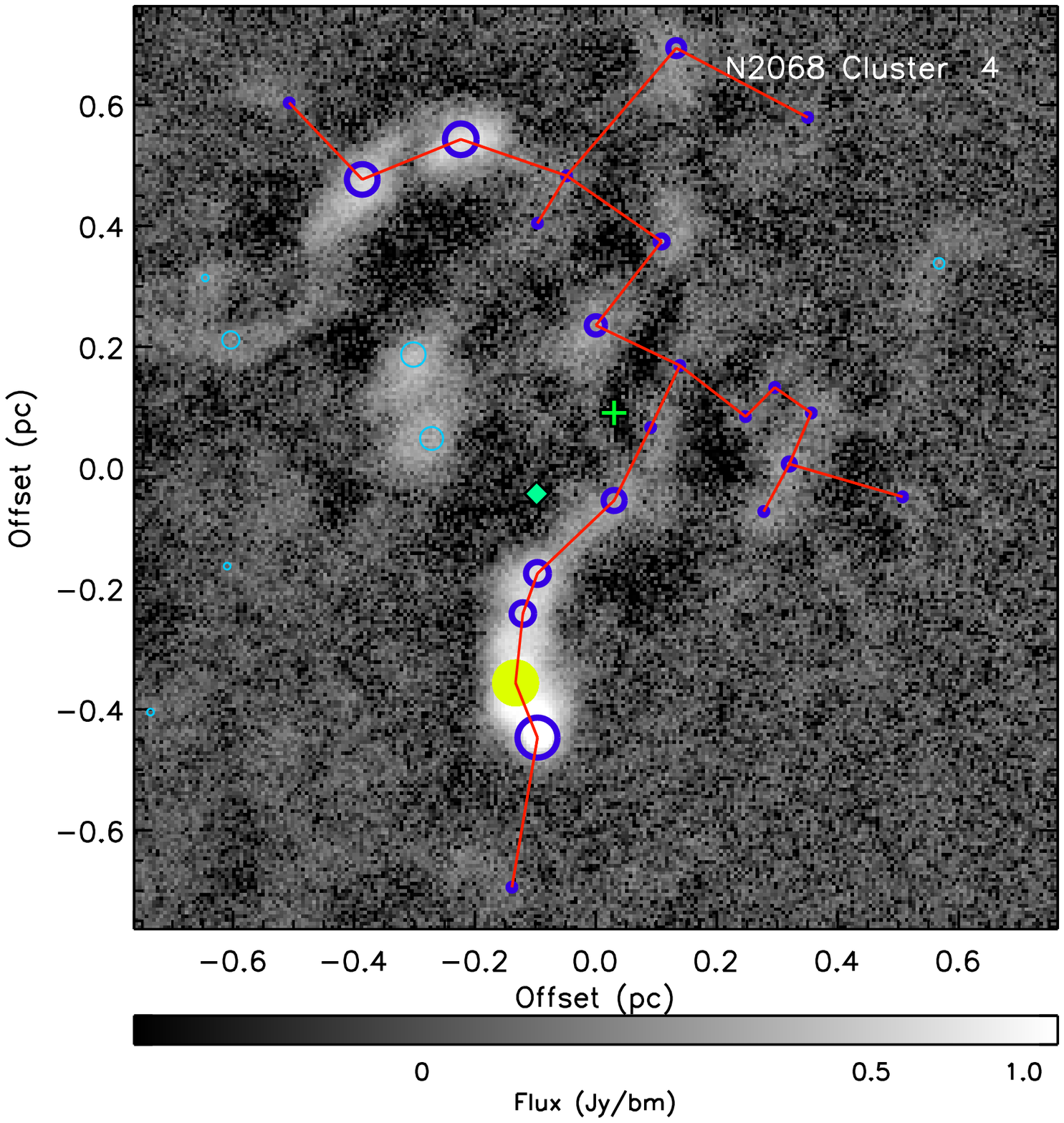} \\
\includegraphics[width=2.7in]{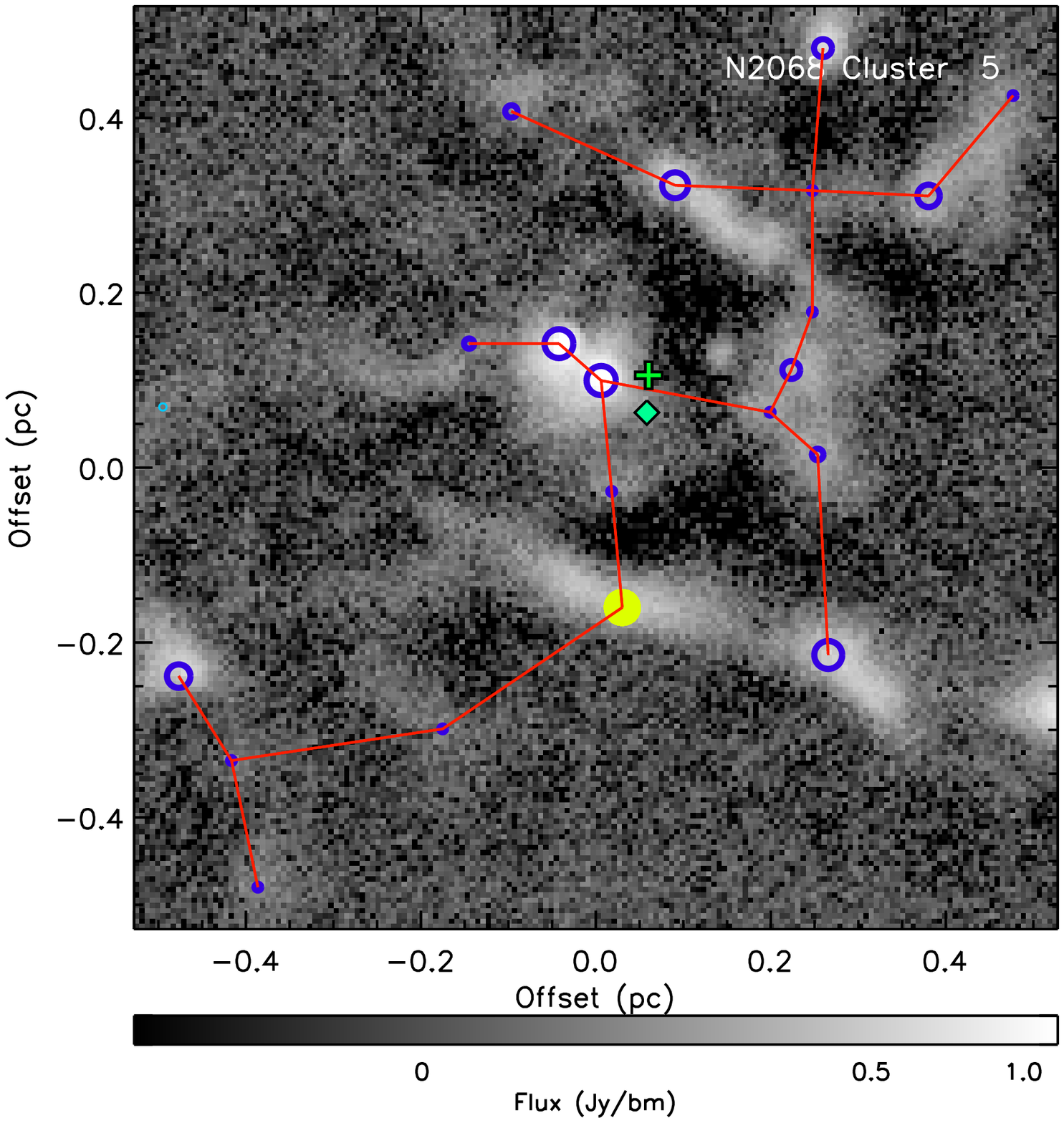} &
\includegraphics[width=2.7in]{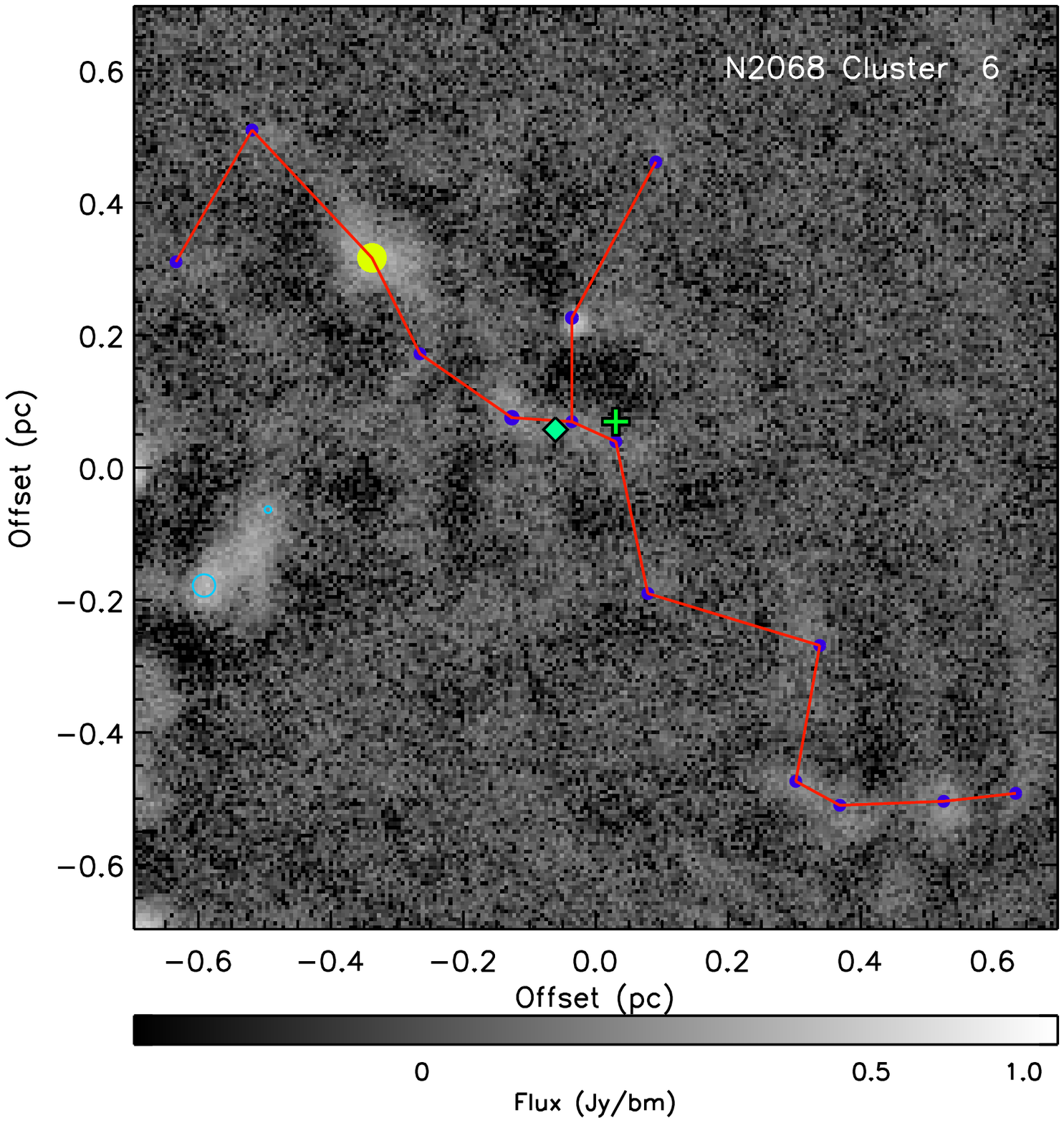} \\
\end{tabular}
\caption{Dense core clusters identified in \Ntwo\
	using MSTs.  See Figure~\ref{fig_mst_clusts1} for the plotting conventions used.}
\label{fig_mst_clusts2}
\end{figure}

Here, we present a series of snapshots of all of the dense core clusters identified using
the MST technique across the three regions.  Figures~\ref{fig_mst_clusts1} and 
\ref{fig_mst_clusts2} show the clusters in L1622 and \None, and \Ntwo, respectively.

\section{MST and Offset Ratio Analysis Uncertainties}

\subsection{Cluster Definition Uncertainties}
As discussed in \citet{Kuhn14}, one important test for any non-parametric cluster
model (such as our MST analysis), is a thorough check on the effect of
the variation in user-specified values on the results.  While Figures~\ref{fig_mst_clusts1} and
\ref{fig_mst_clusts2} show that the clusters we identify are visually reasonable, here,
we explicitly test the impact of varying the \Lcrit\ and $N$ (the minimum number of cluster
members) on our results.

\begin{figure}[htb] 
\begin{tabular}{cc}
\includegraphics[width=2.7in]{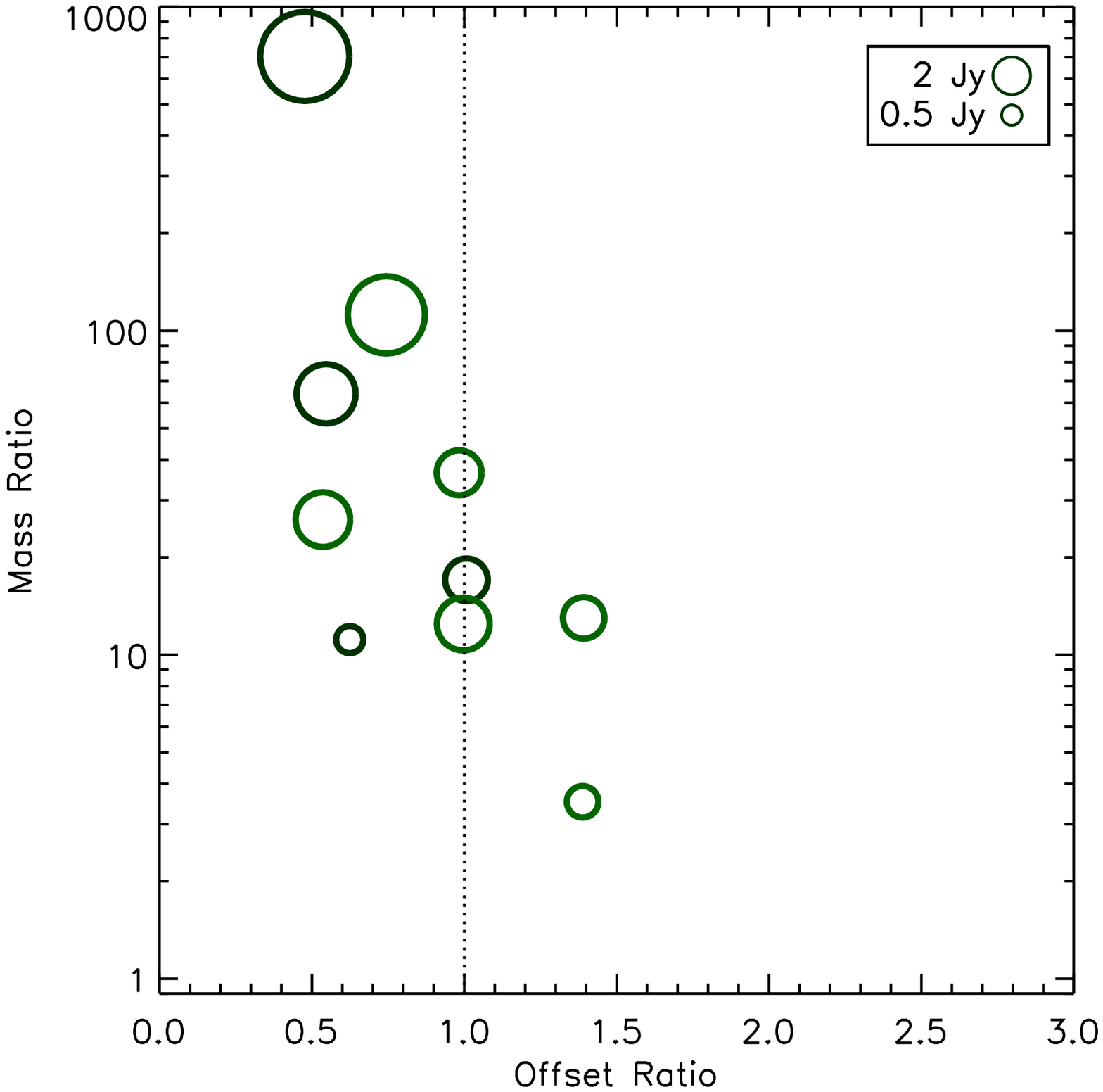} &
\includegraphics[width=2.7in]{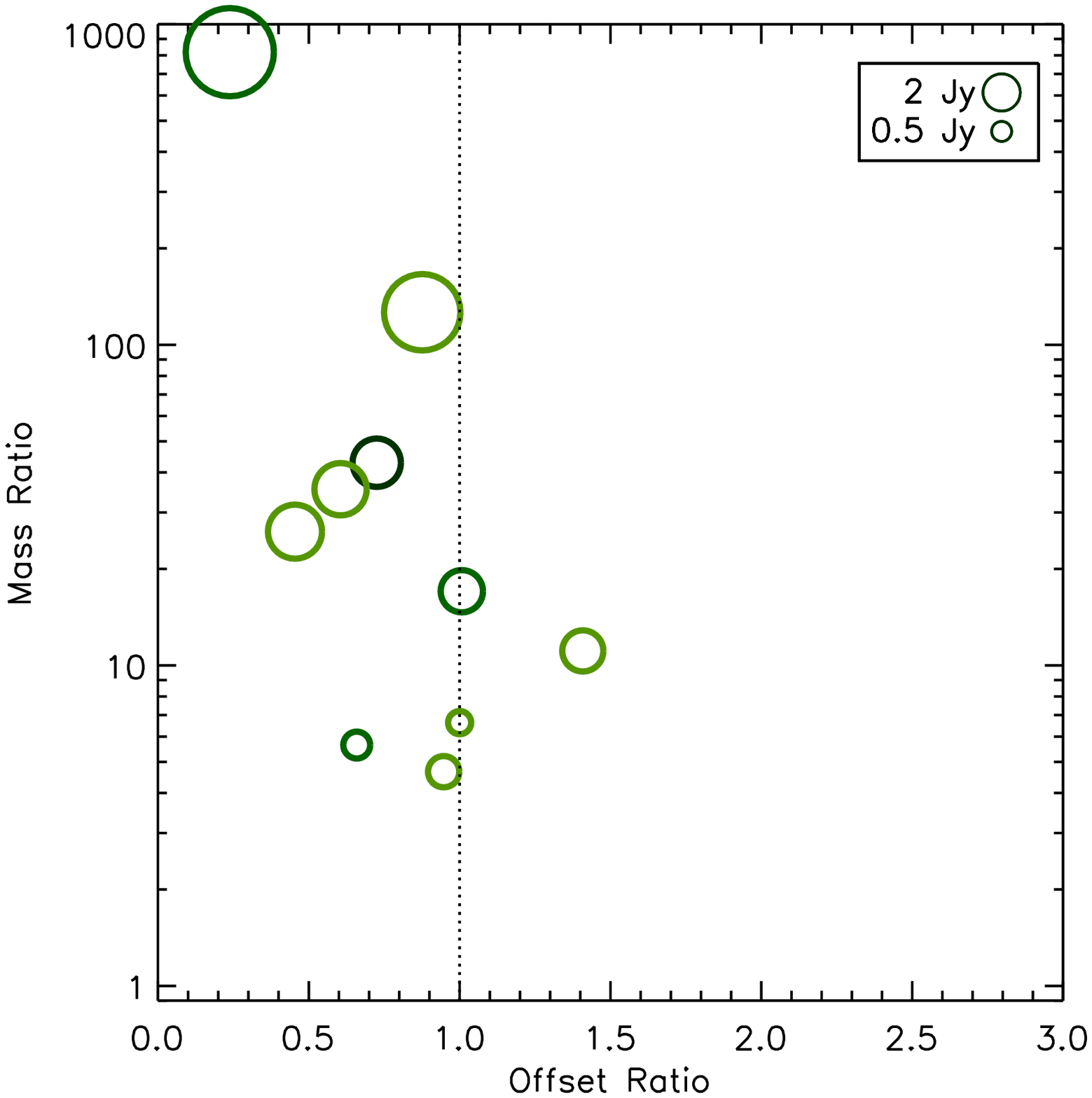} \\
\end{tabular}
\caption{A comparison of the offset ratios measured when extreme values of \Lcrit\ are applied
	to the determination of cluster membership, using the full sample of dense
	cores.  This can be compared to the left panel of Figure~\ref{fig_offset_ratio}, where
	the green circles indicate the offset ratios for the originally-determined
	clusters in Orion~B.  Left: the application of a small
	value of \Lcrit.  
	Right: the application of a large value of \Lcrit.  See Figure~\ref{fig_offset_ratio}
	for the plotting conventions used.
	}
\label{fig_offsets_Lcriterr}
\end{figure}

For each of the three regions, we tested a range
of \Lcrit\ values, corresponding to the intersection of linear fits to the cumulative branch
length distribution somewhat beyond the range of good fits.
We used this range 
to determine the maximal impact of different values of \Lcrit\ on the definition and membership
of the clusters, and how this then changes the results of the offset ratio measurements.
Since it has relatively few cores, L1622 has the poorest best fit to the cumulative branch
length distribution, and the largest range of `okay' fits.  In our tests of this region, we
set the maximum \Lcrit\ to be sufficiently large to include {\it all} cores in the cluster.
In all other cases (minimum \Lcrit\ for L1622 and both minimum and maximum \Lcrit\ for the 
other two regions), we find \Lcrit\ varies less than 18\% 
from the best fit values\footnote{The range of acceptable \Lcrit\ values for each region
is as follows: 0.35~pc to 1.7~pc in L1622, 0.26~pc to 0.30~pc in \None, and 0.23~pc to 0.29~pc 
in \Ntwo.}.
Within this range of \Lcrit\ values, we find that the main properties of most identified clusters
change very little.  In \None, clusters 3 and 4 remain identical under the full
range of \Lcrit\ values, while clusters 1 and 2 are identical for smaller \Lcrit\ values, but
become merged into a single entity for large values of \Lcrit.  In \Ntwo, clusters 1, 2, 4, and 5
remain nearly identical for the full range of \Lcrit\ values, with the occasional loss or gain
of members at their peripheries.  \Ntwo's cluster 3 and 6 are similar for larger values of
\Lcrit\ but for smaller \Lcrit, cluster 3 is
split into two clusters, and cluster 6 no longer has a sufficient number of members
(i.e., more than 10) to be classified as a bona fide cluster.  Interestingly, even with these significant
membership changes, the inner part of cluster 3 and its centre position stay remarkably similar.
L1622, with its wider range of possible \Lcrit\ values, shows a greater variation in membership,
ranging from the inclusion of all cores for the largest \Lcrit, and only 9 cores in the `cluster'
for the smallest \Lcrit.
Despite the changes in some clusters' appearances, however, the overall distribution
of cluster offset ratios remains remarkably similar.  Changes in cluster
membership often tend to be somewhat symmetric, i.e., members being added or subtracted from
multiple sides of the cluster with an increase or decrease in \Lcrit.  As such, 
the cluster centre tends to change very little with
variations in \Lcrit.  Similarly, the median offset tends to vary little if only a few 
large offsets are included or excluded from the calculation, and hence the offset ratio
tends to be similar for small changes to the cluster membership.
We demonstrate the relative invariance in the offset ratios we measure in 
Figure~\ref{fig_offsets_Lcriterr}.  Regardless of the \Lcrit\ applied, the majority
of clusters have centrally located most massive members.

\begin{figure}[htb]
\begin{tabular}{cc}
\includegraphics[width=2.7in]{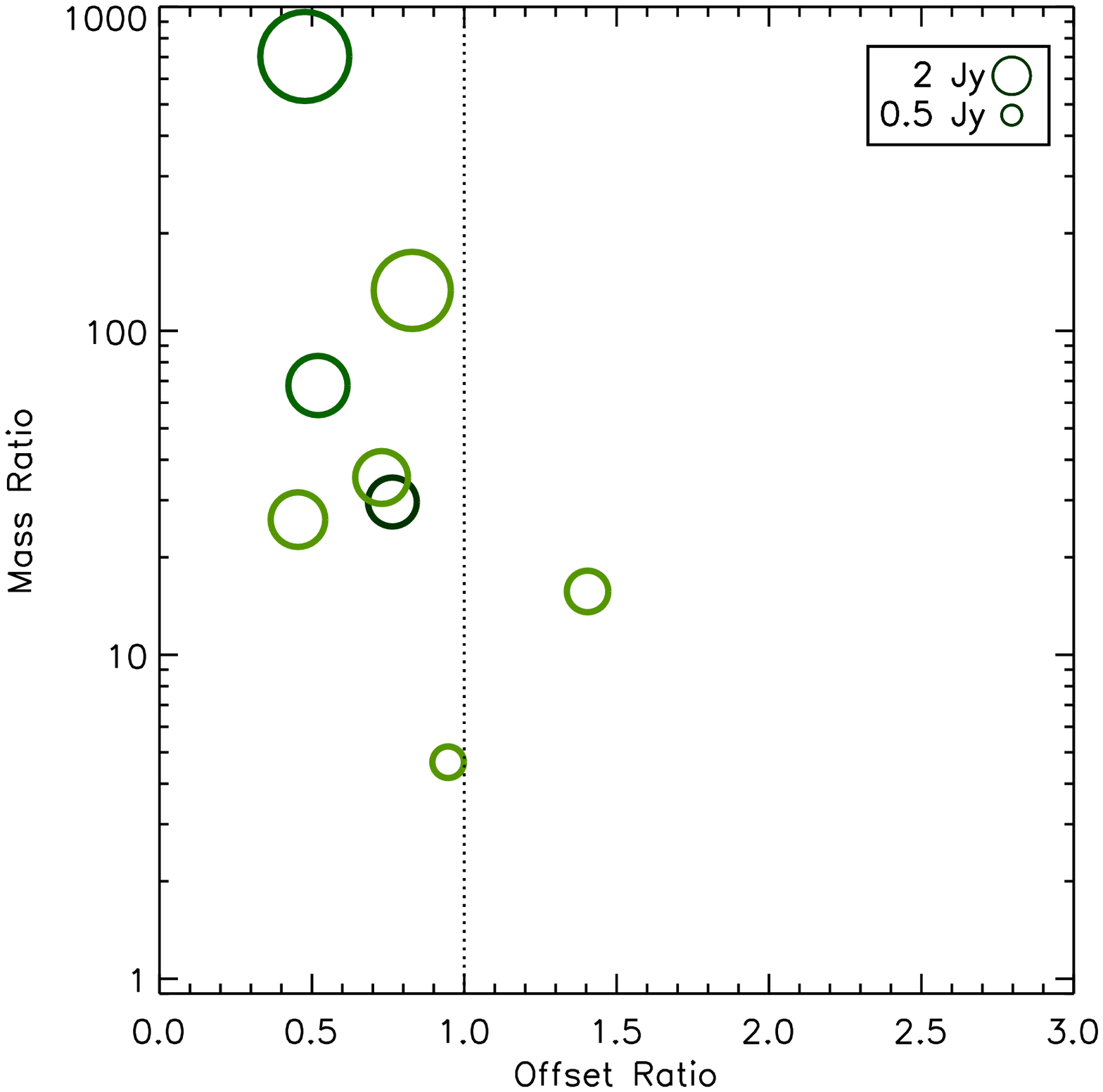} &
\includegraphics[width=2.7in]{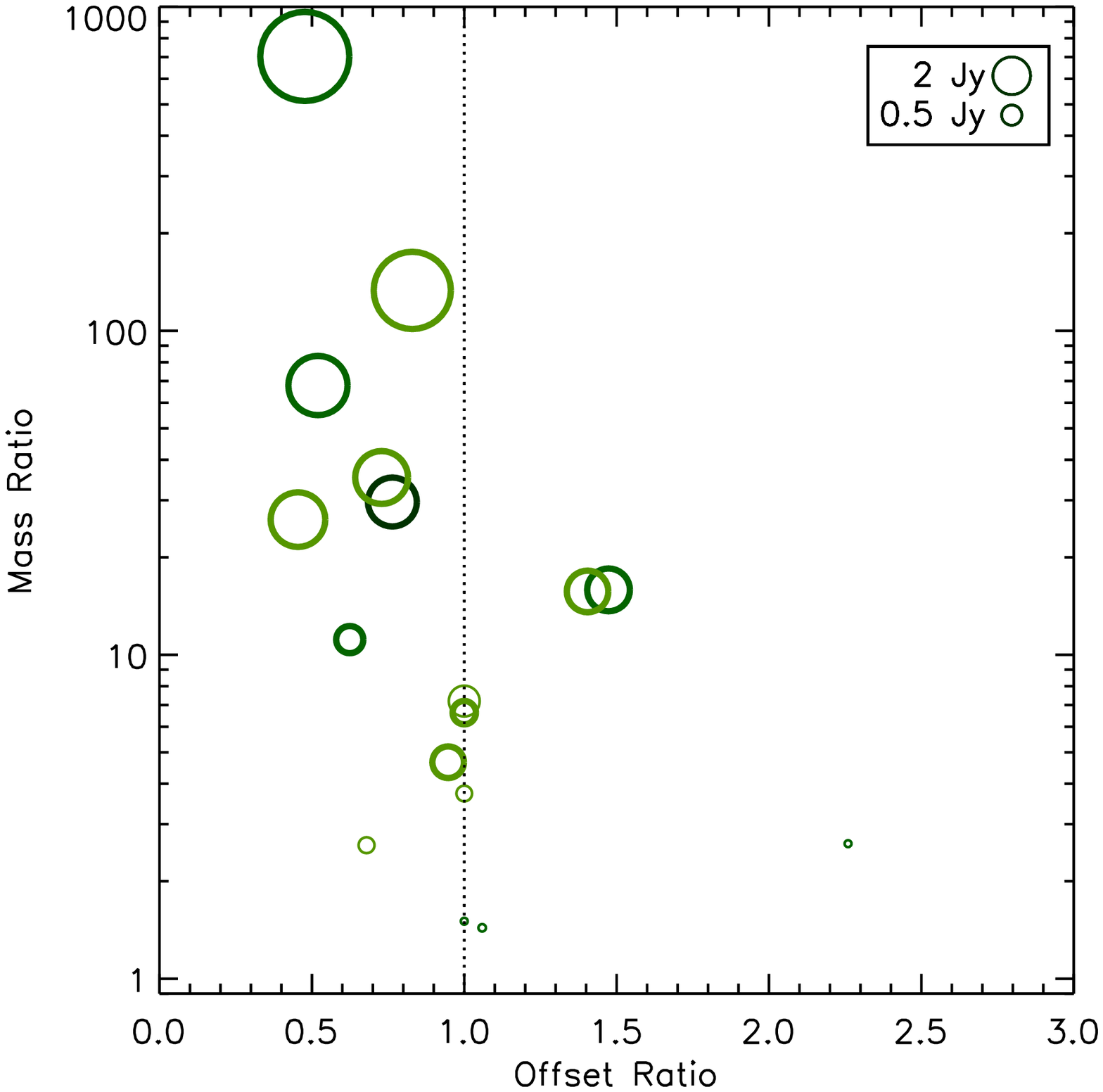}\\
\end{tabular}
\caption{A comparison of the offset ratios measured when different values of $N$ (minimum
	number of cluster members) are applied
	to the MST analysis using the full sample of dense
	cores.  These figures can be compared to the left panel of Figure~\ref{fig_offset_ratio}, 
	where
	the green circles indicate the offset ratios for the originally-determined
	clusters in Orion~B.  Left: $N > 15$ members.
	Right: $N > 5$ members, with members not satisfying the original $N > 10$ criterion
	shown with thinner circles.  See Figure~\ref{fig_offset_ratio}
	for the plotting conventions used.
	}
\label{fig_offsets_Nmin}
\end{figure}

Furthermore, we test the effect of changing the minimum number of cluster members
on our results.  From our standard requirement of $N > 10$, we try $N > 15$ and $N > 5$.
A higher value of $N$ serves to reduce our existing cluster sample.  As can already be
seen in Table~\ref{tab_clusts}, only three of our clusters have $N < 20$, and these 
three clusters are eliminated from our sample with an $N > 15$ requirement.  
Figure~\ref{fig_offsets_Nmin} shows that with these smaller clusters removed, our 
overall results are qualitatively unchanged: most clusters still have offset ratios less
than one, with only one of eight clusters in this restricted sample having an offset
ratio above one.  When we decrease the minimum number of members required to be classified
as a cluster, new clusters are added to the analysis; setting $N > 5$ adds six new 
`clusters', three each in \None\ and \Ntwo, as the right panel of Figure~\ref{fig_offsets_Nmin}
shows.  Of these six additions, one has an offset ratio below one, three have
offset ratios of exactly one, and two have ratios
greater than one.  These small-$N$ `clusters' clearly have less of a tendency for a
centrally-located most massive member than their higher-$N$ bretheren, but this is 
understandable.  With such a small number of members, some of the `clusters' may not
be physically associated.  Even if they are associated, the offset ratio is strongly sensitive to
the location of the cluster centre which becomes ill-determined in the small-$N$ regime.  The small `clusters' identified also have a tendency
to contain only very low mass members ($< 0.2$~Jy for the \None\ `clusters') and the 
most massive member is often not very distinct (four of the six small `clusters' have 
most massive members which are less than three times the median mass).  All of these
factors lead us to expect to see less of a trend in the offset ratio for these very 
small-$N$ clusters and do not take away support from our main findings.

\subsection{Core Definitions and Resolution}
Beyond the definition of each cluster, another source of uncertainty in our 
analysis of offset ratios is our definition of core boundaries and therefore 
their total fluxes.  There are two potential issues which could influence which
core we identify as having the highest flux, and therefore change the offset ratio
measured.  The first issue is substructure on scales smaller than the SCUBA-2 850~\mum
beam.  Namely, depending on how the cores we identify fragment at smaller scales, we might
identify a different highest flux core if using higher resolution observations.  
We test this possibility by examining the SCUBA-2 450~\mum map, whose effective beamsize
is nearly twice as small as the 850~\mum map, and search for signs of fragmentation in
cores near the cluster centres.  Indeed, in the largest, highest density, clusters in \None\
and \Ntwo\ (cluster 1 in \None\ and clusters 1 and 2 in \Ntwo), we see that the core
flagged as the highest flux core shows signs of fragmentation.
A careful examination of these fragmented cores shows one of two behaviours:
1) part of the highest flux core will likely still be the highest flux core even
if the fragmentation is accounted for, or 2) with fragmentation included, a different core
would likely be flagged as the highest flux core, but this new core is
at a comparable (or smaller) separation to the cluster centre.  Although it is possible
that these results would change at even higher resolutions, it is reassuring that 
the results appear to be robust to at least modest improvements in resolution.

The second potential issue is the outer core boundary.  Cores may have a high total
flux due to being a bright compact source or due to having 
a large boundary encompassing a significant
amount of more diffuse emission (or both).  The fraction of material from a given core that will
end up in the stars that form out of it is not clear, and could easily vary from core to
core (and, similarly, the amount of material a protostar accretes from beyond the core
may also vary).  Nonetheless, there is a perhaps naive expectation that the `highest flux core'
of interest in the offset ratio analysis 
is one that has a large amount of compact emission, rather than merely 
a large total flux due to a large areal extent of diffuse emission.  We examine both the
850~\mum and 450~\mum emission maps to identify the sources of strong compact emission
within each cluster, and whether or not these correspond to the cores flagged as the highest
flux cores in our earlier analysis.  In several cases (cluster 2 in \None\ and clusters 3, 4,
5, and 6 in \Ntwo), the core with the largest amount of compact emission 
differs from the core with the highest total flux.  In all but one of these cases
(cluster 4 in \Ntwo),
the position of the brightest compact emission core lies {\it closer} to the 
cluster centre than the core with the highest total flux, which would lower the offset
ratios we measure.

In addition to these two tests, we note that Figures~\ref{fig_mass_offset1} and
\ref{fig_mass_offset2} show a general tendency for more massive cores to have smaller
offsets, which also suggests that our choice of the most massive core has 
a small influence on our results.
We therefore conclude that our core definitions do not appear to bias artificially
the offset ratio lower, due to either the resolution of the telescope or the precise
core boundary adopted.  

\subsection{Cluster Centres}
Finally, the offset ratio analysis depends on where the cluster centre is located.  
Here, as in \citetalias{Kirk11}, we adopt the median position as the cluster centre.
\citetalias{Kirk11} argue that using instead the centre of mass as the cluster centre
has the potential to bias the centre position toward the location of the most massive
cluster member, and hence result in smaller offset ratio measurements.  Our present 
analysis suggests a similar possibility.  In Figures~\ref{fig_mst_clusts1} and 
\ref{fig_mst_clusts2}, the flux-weighted mean core position, i.e., the centre of mass
under the assumption of a constant conversion between flux and mass, is indicated
in addition to median core position adopted as the cluster centre.  There is a clear
tendency for the highest flux core to be located closer to the `centre of mass' 
position than the median core position.  Therefore, our offset ratio measurements 
would decrease with this alternate cluster centre definition, which would serve
to strengthen our overall finding of the central location of high-flux cores.

\acknowledgements{
The authors thank the anonymous referee and Eric Feigelson for comments which 
improved the paper and the presentation of our results.
HK thanks Jonathan Tan for a conversation which helped to inspire some of this analysis,
and Phil Myers for some helpful comments on a draft of this paper.
The authors wish to recognize and acknowledge the very significant cultural role
and reverence that the summit of Maunakea has always had within the indigenous
Hawaiian community.  We are most fortunate to have the opportunity to conduct
observations from this mountain.
The JCMT has historically been operated by the Joint Astronomy Centre on behalf of the
Science and Technology Facilities Council of the United Kingdom, the National Research
Council of Canada and the Netherlands Organisation for Scientific Research. Additional
funds for the construction of SCUBA-2 were provided by the Canada Foundation for
Innovation. 
The authors thank the JCMT staff for their support of
the GBS team in data collection and reduction efforts.
The Starlink software \citep{Currie14} is supported by
the East Asian Observatory.  
This research used the facilities of the Canadian Astronomy Data Centre operated by the
National Research Council of Canada with the support of the Canadian Space Agency.
Figures in this paper were creating using the NASA IDL astronomy library
\citep{idlastro} and the Coyote IDL library ({\tt http://www.idlcoyote.com/index.html}).
}

\facility{JCMT (SCUBA-2)}
\software{Starlink\citep{Currie14}, IDL}

\bibliographystyle{apj}
\bibliography{orionbib}{}


\end{document}